\documentclass[examplefnt,biber]{nowfnt} 

\usepackage[utf8]{inputenc}
\usepackage{rotating,tabularx}

\DeclareUnicodeCharacter{0301}{\'{e}}

\usepackage[ddmmyyyy]{datetime} 
\usepackage{enumitem} 
\newlist{UserSystemUtterance}{description}{2} 
\setlist[UserSystemUtterance]{style=multiline, labelwidth=\widthof{somelabelwidth}, font=\normalfont\textsc , leftmargin=\labelwidth, align=right} 

\usepackage{booktabs} 

\newcommand{\ie}{\emph{i.e.}}
\newcommand{\eg}{\emph{e.g.}}
\newcommand{\etc}{\emph{etc}}

\newcommand{\hamed}[1]{\textcolor{blue}{{\bf [Hamed: }{\em #1}{\bf ]}}}
\newcommand{\johanne}[1]{\textcolor{purple}{{\bf [Johanne: }{\em #1}{\bf ]}}}
\newcommand{\jeff}[1]{\textcolor{magenta}{{\bf [Jeff: }{\em #1}{\bf ]}}}
\newcommand{\filip}[1]{\textcolor{red}{{\bf [Filip: }{\em #1}{\bf ]}}}

\usepackage{xcolor}
\usepackage{mdframed}
\usepackage{wrapfig}

\renewcommand{\hamed}[1]{}
\renewcommand{\jeff}[1]{}
\renewcommand{\filip}[1]{}

\title{Conversational Information Seeking}

\subtitle{An Introduction to Conversational Search, Recommendation, and Question Answering}

\maintitleauthorlist{
Hamed Zamani \\
University of Massachusetts Amherst \\
zamani@cs.umass.edu
\and
Johanne R. Trippas \\
RMIT University \\
j.trippas@rmit.edu.au
\and
Jeff Dalton \\
University of Glasgow \\
jeff.dalton@glasgow.ac.uk
\and
Filip Radlinski \\
Google Research\\
filiprad@google.com
}

\issuesetup
{%
 copyrightowner={H.~Zamani, J.~R.~Trippas, J.~Dalton, F.~Radlinski},
 volume        = xx,
 issue         = xx,
 pubyear       = 2023,
 isbn          = xxx-x-xxxxx-xxx-x,
 eisbn         = xxx-x-xxxxx-xxx-x,
 doi           = 10.1561/XXXXXXXXX,
 firstpage     = 1, 
 lastpage      = 222
 }

\usepackage{mwe}

\author[1]{Zamani, Hamed}
\author[2]{Trippas, Johanne R.}
\author[3]{Dalton, Jeff}
\author[4]{Radlinski, Filip}

\affil[1]{University of Massachusetts Amherst; zamani@cs.umass.edu}
\affil[2]{RMIT University; j.trippas@rmit.edu.au}
\affil[3]{University of Glasgow; jeff.dalton@glasfow.ac.uk}
\affil[4]{Google Research; filiprad@google.com}

\articledatabox{\nowfntstandardcitation}

\begin{document}

\definecolor{highlight}{RGB}{250, 250, 250}
\newmdenv[linewidth=0.05pt,leftline=false,rightline=false,backgroundcolor=highlight,skipabove=20pt,skipbelow=20pt,skipabove=20pt,skipbelow=20pt,innertopmargin=10pt,innerbottommargin=10pt]{topbot}

\newcommand*{\tipbox}[1]{%
  \bgroup
    \setlength\intextsep{-5pt}%
    \begin{topbot}%
      \large%
      \begin{wrapfigure}[2]{l}{0.08\textwidth}%
      \includegraphics[width=0.08\textwidth]{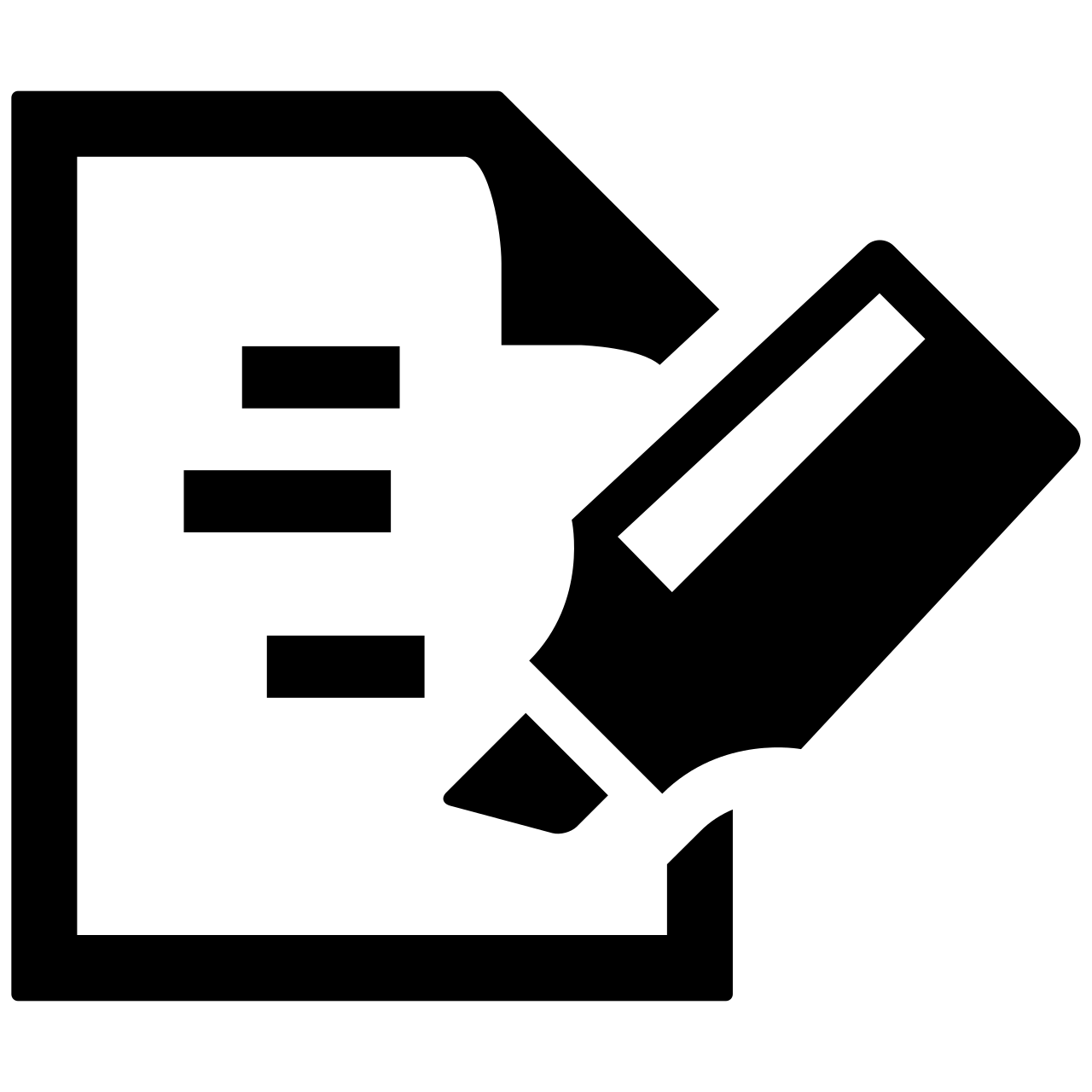}%
      \end{wrapfigure}%
      #1%
    \end{topbot}%
  \egroup
}

\makeabstracttitle

\begin{abstract}
Conversational information seeking (CIS) is concerned with a sequence of interactions between one or more users and an information system. Interactions in CIS are primarily based on natural language dialogue, while they may include other types of interactions, such as click, touch, and body gestures. This monograph provides a thorough overview of CIS definitions, applications, interactions, interfaces, design, implementation, and evaluation. This monograph views CIS applications as including conversational search, conversational question answering, and conversational recommendation. Our aim is to provide an overview of past research related to CIS, introduce the current state-of-the-art in CIS, highlight the challenges still being faced in the community. and suggest future directions.

\end{abstract}

\chapter{Introduction}
\label{chapter1}

\section{Motivation}

Over the years, information retrieval and search systems have become more \emph{conversational}: For instance, techniques have been developed to support queries that refer indirectly to previous queries or previous results; to ask questions back to the user; to record and explicitly reference earlier statements made by the user; to interpret queries issued in fully natural language, and so forth. In fact, systems with multi-turn capabilities, natural language capabilities as well as robust long-term user modeling capabilities have been actively researched for decades. However, the last few years have seen a tremendous acceleration of this evolution.

This has been driven by a few factors. 
Foremost, progress in machine learning, specifically as applied to natural language understanding and spoken language understanding, has recently surged. 
Whereas the possibility of a conversational information seeking (CIS) system robustly understanding conversational input from a person was previously limited, it can now almost be taken for granted. 
In concert with this, consumer hardware that supports and encourages conversation has become common, raising awareness of --- and the expectation of --- conversational support in IR systems. 
From the research community, this has been accompanied by significant progress in defining more natural CIS tasks, metrics, challenges and benchmarks. This has allowed the field to expand rapidly. 
This monograph aims to summarize the current state of the art of conversational information seeking research, and provide an introduction to new researchers as well as a reference for established researchers in this area.

\section{Guide to the Reader}
The intended audience for this survey is computer science researchers in fields related to conversational information seeking, as well as students in this field. We do not assume an existing understanding of conversational systems. 
However, we do assume the reader is familiar with general concepts from information retrieval, such as indexing, querying and evaluation. As this monograph is not a technical presentation of recent machine learning algorithms, we also assume a basic understanding of machine learning and deep learning concepts and familiarity with key algorithms.

The reader will be provided with a summary of the open CIS problems that are currently attracting the most attention, and many promising current results and avenues of investigation. We will also provide an overview of applications attracting interest in the community, and the resources available for addressing these applications.

When discussing the structure of conversations we adopt terminology used in the speech and dialogue research community. The most basic unit is an $utterance$ (analogous to a single query in retrieval). All contiguous utterances from a single speaker form a single $turn$ \citep{Traum:1996:UtteranceUnits}, with a conversation consisting of multiple turns from two or more participants. For the reader we note that somewhat confusingly, a commonly adopted definition in CIS publications defines a turn as the pair of a user turn and a system response turn (a user query and system answer). 

\vspace{2mm}
The focus of this work differs from recent related surveys. We draw the reader's attention to the following most related examples.  \citet{gao:2019:FnTIR} presented an overview focused on specific neural algorithmic solutions for question answering, task-oriented and chat agents. \citet{Freed:2021:ConvAI} also focused on the development of chatbots, often for customer support.  
Our focus is more on characterizing the problem space related to information seeking conversations and providing a broad overview of different problems, metrics and approaches. 
Moreover, the report from the third Strategic Workshop on Information Retrieval in Lorne (SWIRL 2018) \citep{Culpepper:2018:swirl} provided a broader summary of important open challenges in information retrieval, where various challenges associated with CIS were ranked first. That document provides a briefer overview and reading list, more concretely aimed at summarizing open challenges. A more recent report from the Dagstuhl Seminar on Conversational Search~\citep{Anand:2020:Dagstuhl} reiterated these challenges in more detail. Beyond these, more focused recent relevant workshops include SCAI \citep{Penha:2022:SCAI}, KaRS \citep{Anelli:2022:Kars}, Sim4IR \citep{balog:2021:sim4ir}, Future Conversation \citep{Spina:2021:SIGIRForum} and MuCAI \citep{Hauptmann:2020:MuCAI} among others.
Concurrent to this work, \citet{Gao:2022:book} published a book draft on deep learning approaches for conversational information retrieval. This monograph provides a holistic overview of CIS systems, state-of-the-art CIS approaches, and future directions in CIS research. In contrast, Gao \textit{et al.}'s book focuses specifically on deep learning solutions for various subtasks in conversational IR, therefore provides a complementary view to ours.


\section{Scope} 

This monograph focuses on a particular class of conversational systems, namely those that exhibit key attributes of human conversation. We take a cue from \citet{Radlinski:2017:CHIIR}, who propose that a conversational system should incorporate mixed initiative (with both system and user able to take initiative at different times), memory (the ability to reference and incorporate past statements), system revealment (enabling the system to reveal its capabilities and corpus), user revealment (enabling the user to reveal and/or discover their information need), and set retrieval (considering utility over sets of complementary items). 
Here, we study approaches that exhibit at least some of these properties. In particular, we do not delve deeply into \emph{dialogue systems} that restrict themselves largely to identifying slot/value pairs in back and forth exchanges between the system and user.

Additionally, we focus on \emph{information seeking}, which refers to the process of acquiring information through conversation in order to satisfy the users' information needs. This implies that the conversation should exhibit a clear goal or assist the human user in completing a specific task through finding information. 
While significant progress has been recently made on chit-chat systems, with a primary goal of keeping users engaged in realistic conversational exchanges over a prolonged time  (for more information, see~\citep{Yan:2022}), we do not attempt to cover such work in depth.
Our focus thus aligns more with traditional search concepts such as the presence of an information need or user agenda that existed before they engaged with the CIS system, and which can be satisfied through a conversation.

On the other hand, we do not make a strong distinction between \emph{search} and \emph{recommendation} tasks. Rather, we cover both types of conversational information seeking interactions. We see these as strongly related tasks that are becoming more closely related as time passes.
Indeed, we believe that the same task can often be characterized as either. For instance, a query ``hotels in London'' can be seen as either a search task (e.g.~on a desktop interface, for a potential future tourist considering affordability in different areas) or a recommendation task (e.g.~using a smart watch while standing in heavy rain in central London). Clearly device, interface and context play an important role in determining the best next conversational step. 


Finally, we draw attention to three key aspects of CIS that, while having received significant attention, remain largely unsolved. First, the level of natural language understanding in conversational systems remains far from human-level, particularly over long sequences of exchanges. Even over adjacent conversational steps, question/answer interpretation remains challenging. Second, robust evaluation of conversational systems remains a critical research challenge: The highly personalized and adaptive nature of conversations makes test collection construction highly challenging. We will cover many of the common approaches, and their limitations. Third, \emph{conversation} is sometimes taken to imply voice or speech interactions. We do not make this assumption, recognizing that conversations can happen in many types of interfaces and modalities. We discuss research of conversations combining different types of interfaces and presentations in depth.

\tipbox{
Three particularly important aspects of CIS that are very active areas of research include obtaining human-level natural language understanding, robust evaluation of CIS systems, and moving beyond simple text and speech interactions.
}

There are a number of particularly important aspects of conversational information seeking that despite their importance are not covered in depth here, as they apply broadly across many non-conversational search and recommendation tasks. The first is the question of privacy. Clearly this is an essential aspect of all search tasks -- and should be considered in depth in any practical system. We refer readers to \citet{Cooper:2008:Survey,Zhang:2016:Anonymizing} as a starting point for privacy considerations as applied to logging and log analysis.

Similarly, we do not consider the type of information that a user may request or receive -- including information that might be considered offensive or harmful. As this issue is not specific to conversational systems and is heavily studied; A detailed consideration of such information access is thus beyond our scope. We refer readers to \citet{Yenala:2018:Inappropriate,Pradeep:2021:Misinformation} as starting points of recent work in this space.

Along the same lines, fairness is an essential aspect for information seeking and recommendation tasks, yet largely beyond our scope. We note that this includes both fairness in terms of biases that may exist in recommendation to different groups \citep{Ge:2021:Fairness} as well as fairness when considering both consumers of recommendations as well as producers of items being recommended \citep{Abdollahpouri:2020:Multistakeholder}. We refer interested readers to \citet{Ekstrand:2022:FairnessFnTIR} for a complete recent overview. 

\section{Applications}

An alternative way to characterize the scope of this work could be in terms of the relevant \emph{applications} that are addressed. Section \ref{chapter2} will focus on this formulation, starting with a brief introduction on conversational information seeking (Section~\ref{sec:chapter2:Conversational Information Seeking}).
This includes a discussion of different
modalities' (that is, text, speech, or multi-modal) impact on the seeking process, as for instance studied by \citet{deldjoo:2021:SIGIR}.
We then continue with the topic of conversational search and its various proposed definitions (Section~\ref{sec:chapter2:Conversational Search}), culminating with one that relates CIS to many other related settings \citep{Anand:2020:Dagstuhl}.
Section~\ref{sec:chapter2:Conversational Recommendation} introduces conversational recommendation  \citep{Jannach:2021:recommender} followed by conversational question answering in Section~\ref{sec:chapter2:Conversational Question Answering}, where for instance \citet{Qu:2019:sigir} provide a powerful characterization of the relationships between these areas of study.
We continue Section~\ref{chapter2} by explaining how CIS applications can be used in different domains, and focus on e-commerce, enterprise, and health in Section~\ref{sec:chapter2:Conversational Information Seeking in Different Domains}. The section concludes with details on intelligent assistants with relation to~CIS.

\section{A High-Level Architecture for CIS Systems}

To create a structure for the remainder of this work, we follow the general structure of most CIS systems. This choice guides the main body of this monograph: Each section in this part focuses on a core technological competency that is essential to a modern CIS system. In particular, a CIS system must first choose an interface (\S 1.5.1). It must then have an approach to maintain the state of a conversation (\S 1.5.2), and at each system turn determine the system's next utterance (\S 1.5.3). One particular challenge that is attracting attention is when the system should take initiative versus responding passively (\S 1.5.4). 


\tipbox{
Key design considerations of a CIS system include its chosen interface, how it maintains conversational state, and how it selects the system's next utterance. One particular challenge for the latter is that of when the system should take initiative.
}

\subsection{Conversational Interfaces and Result Presentation}

Section~\ref{chapter3} provides an overview of conversational interfaces. We begin with a historical perspective, where we explain differences between existing conversational interfaces such as spoken dialogue systems, voice user interfaces, live chat support, and chatbots. This overview illustrates the use of conversations within closely related CIS applications~\citep{Mctear:2016:conversational}. 
Next, research on result presentation through different mediums (desktop or small device) and modalities (text, voice, multi-modal) are discussed in Section~\ref{sec:chapter3:result presentation}, such as recent work by \citet{Kaushik:2020:interface}. This overview emphasizes the difficulties with highly interactive result presentation and highlights research opportunities.  
Following this, Section~\ref{sec:chapter3:initiative in conversational systems} introduces different kinds of initiative in conversational systems, including system-initiative, mixed-initative, and user-initiative, for instance well-characterized by~\citet{Zue:2000:IEEE,Wadhwa:2021:SystemInitiative}. This section aims to explain the different kinds of initiative, and the consequences on human-machine interactions. 
We finish the section with a discussion of conversational interfaces limitations including, for instance, limitations as experienced by visually impaired searchers \citep{Sahib:2015:Evaluating}.

\subsection{Tracking and Understanding Flow}

The focus of Section~\ref{chapter4} is on the varying approaches that make it possible to follow conversational structure. 
We begin with an overview of how to represent a single turn, such as is done with Transformer models \citep{Raffel:2020:jmlr}, and how turns are often classified into dialogue acts \citep{Reddy:2019:tacl}. 
Section~\ref{sec:chapter4:Topic Tracking in Conversation} then looks at how the different turns of a conversation are usually tied together through state tracking and text resolution across turns. In particular, the structure of longer conversations is looked at in-depth in Section~\ref{sec:chapter4:Modeling Conversation Discourse}, although noting that existing models are often limited in their ability to capture long-distance conversational structure \citep{Chiang:2020:aaai}. We cover work that operates over long-term representation of CIS exchanges in Section~\ref{sec:chapter4:History Understanding Tasks}, followed by recent work that attempts to model longer conversations in the final section, epitomized by work on selecting the right context for understanding each turn \citep{Dinan:2019:convai2}.

\subsection{Determining Next Utterances}

The next step for a canonical conversational system is selecting or generating a relevant response in the conversational context. This is the focus of Section~\ref{sec:chapter5}. We begin with an overview of the different types of responses, including short answers, long answers, and structured entities or attributes. The short answer section presents early Conversational QA (ConvQA) systems then discusses the transition to more recent Transformer architectures based on pre-trained language models. Section~\ref{sec:chapter5:ConvQA Knowledge Graphs} then examines how ConvQA is performed over structured knowledge graphs including systems that use key-value networks \citep{Saha:2018:AAAI}, generative approaches, and logical query representations \citep{Plepi:2021:ESWC}. Following this, we discuss open retrieval from large text corpora as part of the QA process. 
In particular, Section~\ref{sec:chapter5:Long Answer Ranking} goes beyond short answer QA to approaches performing conversational passage retrieval from open text collections including multi-stage neural ranking, for instance recently considered by \citet{Lin:2020:MultiStage}. 
We briefly discuss long answer generation approaches in Section~\ref{sec:chapter5:Long Form Response Generation}  including both extractive and abstractive summarization methods. 
We conclude the section with 
conversational ranking of items in a recommendation context, including models that use multi-armed bandit approaches 
to trade-off between elicitation and item recommendation.     

\subsection{Initiative}

Section \ref{sec:mixed_init} provides a detailed look at mixed-initiative interactions in CIS systems. We start with reviewing the main principles of developing mixed-initiative interactive systems, and describing different levels of mixed-initiative interactions in dialogue systems \citep{Allen:1999:MII,Horvitz:1999:PMU}. We briefly review system-initiative interactions with a focus on information seeking conversations, such as the work of \citet{Wadhwa:2021:SystemInitiative}, in Section~\ref{sec:mixed_init:initiating}. We then delve deeply into intent clarification as an example of important mixed-initiative interactions for CIS in Section~\ref{sec:mixed_init:clarification}. We introduce taxonomy of clarification and review models for generating and selecting clarifying questions, such as those by \citet{Aliannejadi:2019:sigir,Zamani:2020:GCQ}. In presenting the work, we include models that generate clarifying questions trained using maximum likelihood as well as clarification maximization through reinforcement learning. Additionally, Section~\ref{sec:mixed_init:pref_elicit} discusses preference elicitation and its relation with clarification, followed by mixed-initiative feedback (\ie, getting feedback from or giving feedback to users via sub-dialogue initiation) in Section~\ref{sec:mixed_init:feedback}. 

\section{Evaluation}

Beyond the details of how a CIS system functions, fair evaluation is key to assessing the strengths and weaknesses of the solutions developed. Section \ref{chapter7:evaluation} looks at evaluation in CIS holistically. After considering possible ways of studying this broad space, this section breaks down evaluation by the setting that is evaluated. 
Specifically, offline evaluation is treated first, in Section~\ref{sec:chapter7:offline evaluation}. A variety of frequently used offline datasets are presented (such as Multi-WOZ \citep{Budzianowski:2018:Multiwoz}), and strengths and limitations are discussed including the use of simulators to produce more privacy-aware evaluations as well as the use of non-text datasets. Online evaluation is considered next, with Section~\ref{sec:chapter7:online evaluation} contrasting lab studies, crowdsourcing, and real-world evaluations. An example of these is where commercial systems may ask evaluation questions of their users \citep{Park:2020:Hybrid}. 
Finally, the metrics applied in these settings are covered in Section~\ref{sec:chapter7:metrics}. While readers are referred to \citet{Liu:2021:Metrics} for a full treatment, we present an overview of typical turn-level as well as end-to-end evaluation metrics.

\section{Open Research Directions}

Section~\ref{chapter9:Conclusions} provides a brief summary of this monograph and discusses different open research directions. We collate the major themes discussed throughout this manuscript instead of presenting a detailed account of all possible future research problems. We highlight four key areas for future exploration. First, Section~\ref{subsec:chapter8:Interactivity} covers
challenges related to modeling and producing conversational interactions as a way to transfer information between user and system. Second, we highlight the importance of result presentation and its role in CIS research in Section~\ref{subsec:chapter8:Result Presentation}.
Third, we emphasise the importance of different CIS tasks in Section~\ref{subsec:chapter8:Types of Conversational Information Seeking Tasks}. Finally, Section~\ref{subsec:chapter8:Measuring Success} covers measures of success during the highly interactive CIS process and ultimate evaluation of CIS systems.

\section{Further Resources}

Beyond the main body of this work, Appendix~\ref{appendixA} briefly presents a more holistic historical context for this monograph. This appendix mainly includes information about early research on interactive information retrieval, as well as on dialogue-based information retrieval, such as the I\textsuperscript{3}R \citep{Croft:1987:I3R} and THOMAS \citep{oddy1977information} systems (see Section~\ref{appendixA:IIR}). We discuss approaches for theoretical modelling of interactive information retrieval systems, such as game theory-based models~\citep{Zhai:2016:GameTheoryIR} and economic models~\citep{Azzopardi:2011:EconomicsIIR} in Section~\ref{appendixA:formal}. We also include introductory information about existing literature on session search, such as the TREC Session Track, and evaluation methodologies for session search tasks~\citep{Carterette:2016:ERO} in Section~\ref{appendixA:session}. Finally, we briefly cover exploratory search~\citep{White:2017:ES} and discuss its relationship to conversational information seeking in Section~\ref{appendixA:exploratory}, followed by a very brief overview of chit-chat and task-oriented dialogue systems in Section~\ref{appendixA:dialogue}. Newcomers to the field of information retrieval are highly encouraged to review this appendix to develop an understanding of where the core ideas behind CIS originated.


\tipbox{
This monograph has been used in multiple tutorials on conversational information seeking at top-tier conferences, \eg, at the SIGIR 2022 \citep{TheCISTutorial-SIGIR2022} and the Web Conference 2023 \citep{TheCISTutorial-WWW2023}. The materials prepared for these tutorials, \eg, presentation slides, interactive demos, and coding practices, are available at \url{https://cis-tutorial.github.io/}.
}
\chapter{Definitions and Applications}
\label{chapter2}

In this section, we provide relevant concepts from previous work in conversational information seeking (CIS) and its tasks, contexts, and applications illustrating the multi-dimensional nature of CIS. 
This introductory section aims to guide the reader with background knowledge on definitions and basic concepts related to CIS. We cover three CIS subdomains, namely conversational search, conversational recommendation, and conversational question answering. These topics are closely related and their boundaries are often blurred. We also introduce some domain-specific applications of CIS, including e-commerce, enterprise, and health, and illustrate their use cases. Lastly, we cover how CIS can be embedded within the subdomain of intelligent assistants.

\section{Conversation}
\label{sec:chapter2:conversation}
The term ``conversation'' carries different definitions in different contexts. The Merriam-Webster Dictionary defines conversation as ``oral exchange of sentiments, observations, opinions, or ideas''.\footnote{\url{https://www.merriam-webster.com/dictionary/conversation}} This refers to the everyday use of conversation by humans. \citet{Brennan:2012:Conversation} defined conversation as ``a joint activity in which two or more participants use linguistic forms and nonverbal signals to communicate interactively'', highlighting the possible use of nonverbal signals in conversations. In contrast, researchers in dialogue systems consider a more pragmatic definition by identifying a few attributes in human conversations. These attributes include turn, speech acts, grounding, dialogue structure, initiative, inference, and implicature~\citep[Ch. 24]{Jurafsky:2021:SLP}. This monograph provides a new definition of conversation, which we believe is better suited for conversational information seeking research. 

A conversation is often defined as a sequence of interactions between two or more participants, including humans and machines, as a form of interactive communication with the goal of information exchange. Unlike most definitions of conversation in linguistics and dialogue systems that only focus on natural language interactions, we argue that a conversation can also exhibit other types of interactions with different characteristics and modalities, such as click, touch, body gestures, and sensory signals. The reason behind including these interactions is the rich history of using them in search technologies that shape the fundamentals of CIS research. That said, long form natural language is still the dominant interaction type in conversations. Therefore, a conversation can be defined as follows.

\tipbox{
\textbf{{Definition 1.}} \textbf{\textit{Conversation}} is interactive communication for exchanging information between two or more participants (\ie, humans or machines) that involves a sequence of interactions. While natural language is considered a prerequisite for conversational interactions, conversations can also exhibit other types of interaction with different characteristics and modalities (\eg, click, touch, and gestures).}

An important characteristic of conversation is its style: \textit{synchronous} versus \textit{asynchronous}. Synchronous conversations happen in real time, where at least two participants (or agents) exchange information. Most human-machine conversations are expected to be synchronous. Asynchronous conversations, on the other hand, happen when information can be exchanged independently of time. Therefore, asynchronous conversations do not require the participants' immediate attention, allowing them to respond to the message at their convenience. Conversations between humans in forums and email threads are asynchronous. A conversation can also be a mixture of synchronous and asynchronous interactions. For instance, a user can have synchronous interactions with a conversational system. Later, a human representative can reach out to the user to follow up on the conversation and better address the user's needs if the conversational system fails.

Researchers in the area of CIS are interested in \textit{information seeking conversations}: conversations in which at least one participant is seeking information and at least another participant is providing information. Information seeking conversations are mostly either among humans (\eg, the interactions between users and librarians for finding information in a library) or between humans and machines (\eg, the interactions between a user and a CIS system). They can be either synchronous, asynchronous, or a mixture of both. 

\tipbox{
\textbf{{Definition 2.}} \textbf{\textit{Information seeking conversation}} is a conversation (cf. Def. 1) in which the goal of information exchange is satisfying the information needs of one or more participants.}

\section{Interaction Modality and Language in Conversation}
\label{subsec:chapter2:Interaction Modality and Language}
According to the above definition of conversation, a conversational system's input from the users may involve many different input types, such as touch, speech, or body gestures. These signals can be translated through traditional input devices such as a mouse or keyboard. For more modern input devices, users can also input gestures, motion, or touch. The output channels from the conversational system can vary from 2D screens to audio output to potentially even holograms.

Users can interact with a conversational system through a range of input devices, including keyboards for typing,  microphones for speech, smartphones for touch, or through a mixture of these and other input devices~\citep{deldjoo:2021:SIGIR}. 
Using a mixture of modalities offers numerous benefits. The key is accessibility; for example, systems with spoken interfaces may be more accessible to users for whom traditional search interfaces are difficult to use~\citep{Weeratunga:2015:Nethra}.
Even though research in CIS primarily refers to conversation as textual or spoken input, other modalities and the mixture of modalities are receiving increased research attention~\citep{liao:2021:SIGIR, Hauptmann:2020:MuCAI, deldjoo:2021:SIGIR}.


The system output or presentation, similar to the input from the user, can consist of different output channels. Given the user's device, context (e.g., time, location, device), and task complexity, conversational systems need to decide which output modality to use for result presentation~\citep{deldjoo:2021:SIGIR}.

\section{Conversational Information Seeking}
\label{sec:chapter2:Conversational Information Seeking}

CIS, the process of acquiring information through conversations, can be seen as a subset of information seeking~\citep{wilson:1999:models}. In the case of information seeking, \textit{any} interaction that aids the finding of information is considered. Hence, searching for information in a book is considered part of information seeking.
In contrast, CIS specifies the interaction type as conversational in which thoughts, feelings, and ideas are expressed, questions are asked and answered, or information is exchanged. CIS is often partitioned into three subdomains: conversational search, conversational recommendation, and conversational question answering. However, we do not make a strong distinction between these subdomains. The reason is that the boundaries between these subdomains are blurred. For instance, a system that helps a user to find and purchase shoes through a conversational interface can be seen as either a conversational search or conversational recommendation. Or a system that answers a sequence of non-factoid questions by retrieving passages can be seen as either conversational search or conversational question answering. Therefore, this monograph focuses on conversational information seeking in general and describes models, theories, and techniques that can be used across all CIS subdomains. We define CIS systems as follows:

\tipbox{
\textbf{{Definition 3.}} A \textbf{\textit{Conversational Information Seeking (CIS) system}} is a system that satisfies the information needs of one or more users by engaging in information seeking conversations (cf. Def. 2). CIS responses are expected to be concise, fluent, stateful, mixed-initiative, context-aware, and personalized.}

\begin{figure}[htbp]
\centering
\includegraphics[width=.6\textwidth]{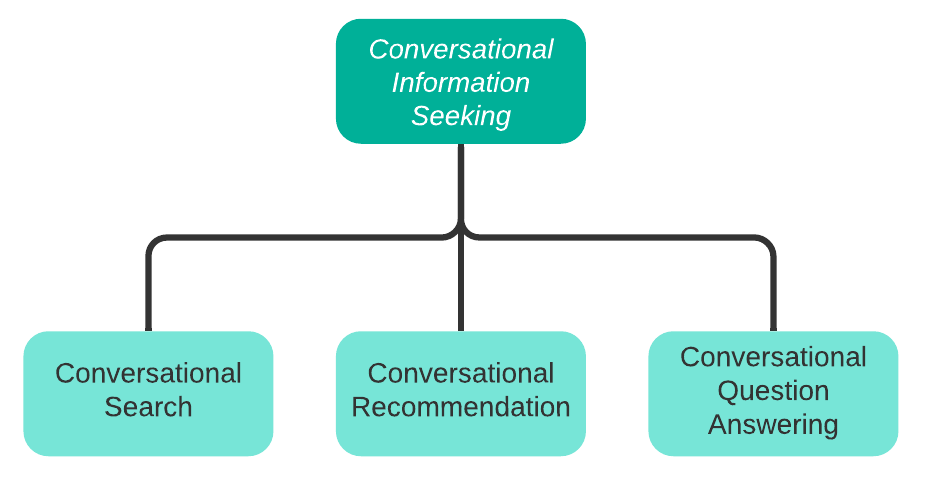}
\caption{Conversational Information Seeking and example subdomains including conversational search, conversational recommendation, and conversational question answering.}
\label{fig:chapter2:CIS_overview}
\end{figure}

In this definition, we provide several properties that are expected from CIS systems. They are explained in the next subsection. Even though we believe that there is no clear distinction between CIS subdomains (depicted in Figure~\ref{fig:chapter2:CIS_overview}), we describe prior work that focused on each of these subdomains in Sections~\ref{sec:chapter2:Conversational Search} -- \ref{sec:chapter2:Conversational Question Answering}.


\section{System Requirements of CIS Systems}
\label{subsec:chapter2:System Requirements of Conversational Search}

To create a truly conversational system, it has been argued that the system should pro-actively participate in the conversation~\citep{Radlinski:2017:CHIIR, Andolina:2018:Investigating, Avula:2020:Wizard, Tabassum:2019:Investigating, Trippas:2018:Informing, Vuong:2018:Naturalistic,Wadhwa:2021:SystemInitiative}. This requires \textit{mixed-initiative}, which implies that the system both responds to utterances, but also at times drives the conversation.
Furthermore, the user-system interactions should create a \textit{multi-turn} dialogue where each participant takes multiple turns to state their information need, clarify this need, or maintain communication functions such as discourse management~\citep{Aliannejadi:2019:sigir, deits2013clarifying, trippas2020towards, Zamani:2020:GCQ}. Indeed, systems can utilize interactive feedback signals such as clarifying questions to optimize the advantages of the conversational technique~\citep{Aliannejadi:2019:sigir, Vtyurina:2017:exploring, Zamani:2020:GCQ}. Mixed-initiative interactions, and in particular clarifying questions, are thoroughly reviewed in Section~\ref{sec:mixed_init}.


The requirements of a system to support the users' interactions are multiple. For example, the interaction history (\eg, queries, relevance feedback, type of interaction device) has to be saved and, where necessary, retrieved by the system~\citep{Reichman:1985:getting,Vtyurina:2017:exploring,Zamani:2020:MACAW}. 
The interaction history as well as user-specific and contextual information can be adopted to provide \textit{personalized} and \textit{context-aware} access to information.
A system should also be able to adapt the results presentation strategies depending on the users' needs. It could be that a user is cognitively engaged, in which case the system can present the results \textit{concisely and fluently with a high comprehensibility}. We note that conciseness and fluency are not specific to natural language and it should be extended to multi-modal conversations. For instance, in speech-only setting, the CIS outputs are expected to be ``listenable''~\citep{Trippas:2019:Thesis}. 

Due to the interactive, adaptive, and conversational nature of these user-system interactions, both user and system turn-time can be less predictable. For example, if the users' input is natural language-based, it can increase the time needed to convey their information need versus a query-based information need. Simultaneously, a system can ask clarifying questions to overcome errors and thus engage with the user through multiple interactions~\citep{skantze2007error}.

One system requirement which is particularly relevant to a speech-only setting is the system's ability to assist the user when speech recognition errors have occurred~\citep{Trippas:2018:Informing}. 
These errors may occur due to background noise, speaker accents, disfluency, spoken language ability, or out-of-vocabulary words, among other reasons. Speakers often compensate with hyper-articulation or restarting voice inputs~\citep{Jiang:2013:SIGIR, Myers:2018:CHI}.
It has been suggested that systems should design in ways to handle the myriad of possible errors and use meta-communication to overcome them~\citep{Trippas:2019:Thesis}.


Existing open-source software to create a CIS system is available. Even though many of these systems cannot be seen as truly conversational, they are updated frequently. For instance, RASA\footnote{\url{https://rasa.com/}} provides flexible conversational software for building text and voice-based assistants but, at the time of writing, lacks mixed-initiative functions. Other conversational systems include Amazon Lex\footnote{\url{https://aws.amazon.com/lex/}} or botpress\footnote{\url{https://botpress.com/}}. Macaw \citep{Zamani:2020:MACAW} provides an extensible framework for conversational information seeking research and supports both mixed-initiative and multi-modal interactions.

\tipbox{Overall, a CIS system is concerned with dialogue-like information seeking exchanges between users and system. Furthermore, the system is pro-actively involved with eliciting, displaying, and supporting the user to satisfy their information need through multi-turn transactions, which can be over one or more sessions.} 

We note that given the complexity of the system and properties listed in Definition 3, most research articles make several simplifying assumptions. For instance, TREC Conversational Assistance Tracks 2019 - 2022 \citep{Dalton:2019:TREC,Dalton:2020:TREC,Dalton:2021:TREC,owoicho:2023:trec} do not consider some of these properties, including personalization.

\section{Conversational Search}
\label{sec:chapter2:Conversational Search}

Conversational search, or the process of interacting with a conversational system through natural conversations to search for information, is an increasingly popular research area and has been recognized as an important new frontier within IR~\citep{Anand:2020:Dagstuhl,Culpepper:2018:swirl}. Furthermore, mobile devices and commercial intelligent assistants such as Amazon Alexa, Apple's Siri, and Google Assistant, in which users interact with a system to search for information, are becoming accepted. Among many other use cases, users can use these systems to receive weather updates, directions, calendar items, and information on any topic covered on the Internet by stating our needs in natural language.

Information seeking, or the process by which people locate information, has traditionally been viewed as a highly interactive process~\citep{oddy1977information, Croft:1987:I3R}. More specifically, searching has been approached as an interactive user-system activity for many years. Furthermore, with the rise in machine learning (ML), natural language processing (NLP), and spoken language comprehension, understanding many users' natural language statements has become more feasible.
Simultaneously, with ever-growing computing power, it has been easier to
comprehend, categorize, or analyze major datasets, which helped to develop genuinely interactive systems that go beyond the conventional ``query box'' action-reaction search model~\citep{Trippas:2018:Informing}. For example, instead of posing a query word in which the user needs to filter through a search engine results page, the user can describe their information need. In addition, the system could inform the user in a more conversational style which documents might be relevant to the query and thus have the system actively involved in the search process. As described by~\citet{Radlinski:2017:CHIIR}, the system could reason about the retrieved documents and actively help the user sift through the information.
Intuitively, conversational search opens up many possibilities as a new interaction paradigm. For example, we may learn how to optimize traditional browser-based ``query box'' searching, improve information accessibility, and decrease information access barriers by incorporating search into everyday dialogues~\citep{Balasuriya:2018:OzCHI, Trippas:2021:CUI:Accessing}.

\begin{figure}[htbp]
\centering
\includegraphics[width=.6\textwidth]{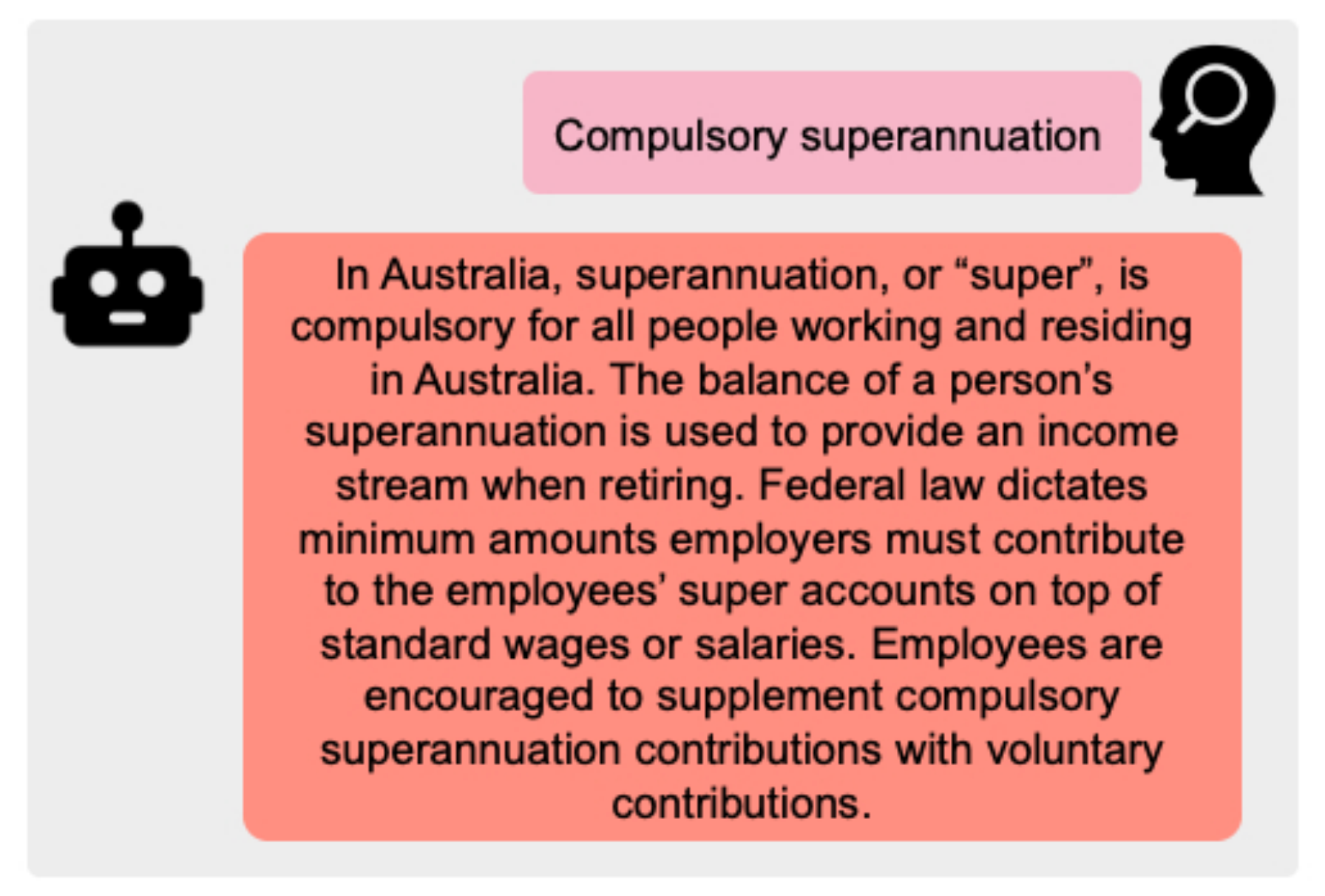}
\caption{Example information seeking task where someone inquires whether superannuation is compulsory in Australia. The user asks a keyword-style query and the system response is an information-dense passage.}
\label{fig:chapter2:example1}
\end{figure}

Consider Figure~\ref{fig:chapter2:example1}, where the statement from a user is short and resembles a keyword-style query and the system response is a long and information-dense passage that is likely hard for the user to consume. In addition, the presentation of the result is not interactive, instead, all the information is presented in one turn, inhibiting the strength of interactivity as an interaction paradigm. Furthermore, the user cannot interact with the content through query refinements or clarifications. This also reinforces the perceived importance of the initial user query, requiring them to formulate ``excellent'' queries from the beginning~\citep{Sahib:2015:Evaluating}. 

\begin{figure}[htbp]
\centering
\includegraphics[width=.6\textwidth]{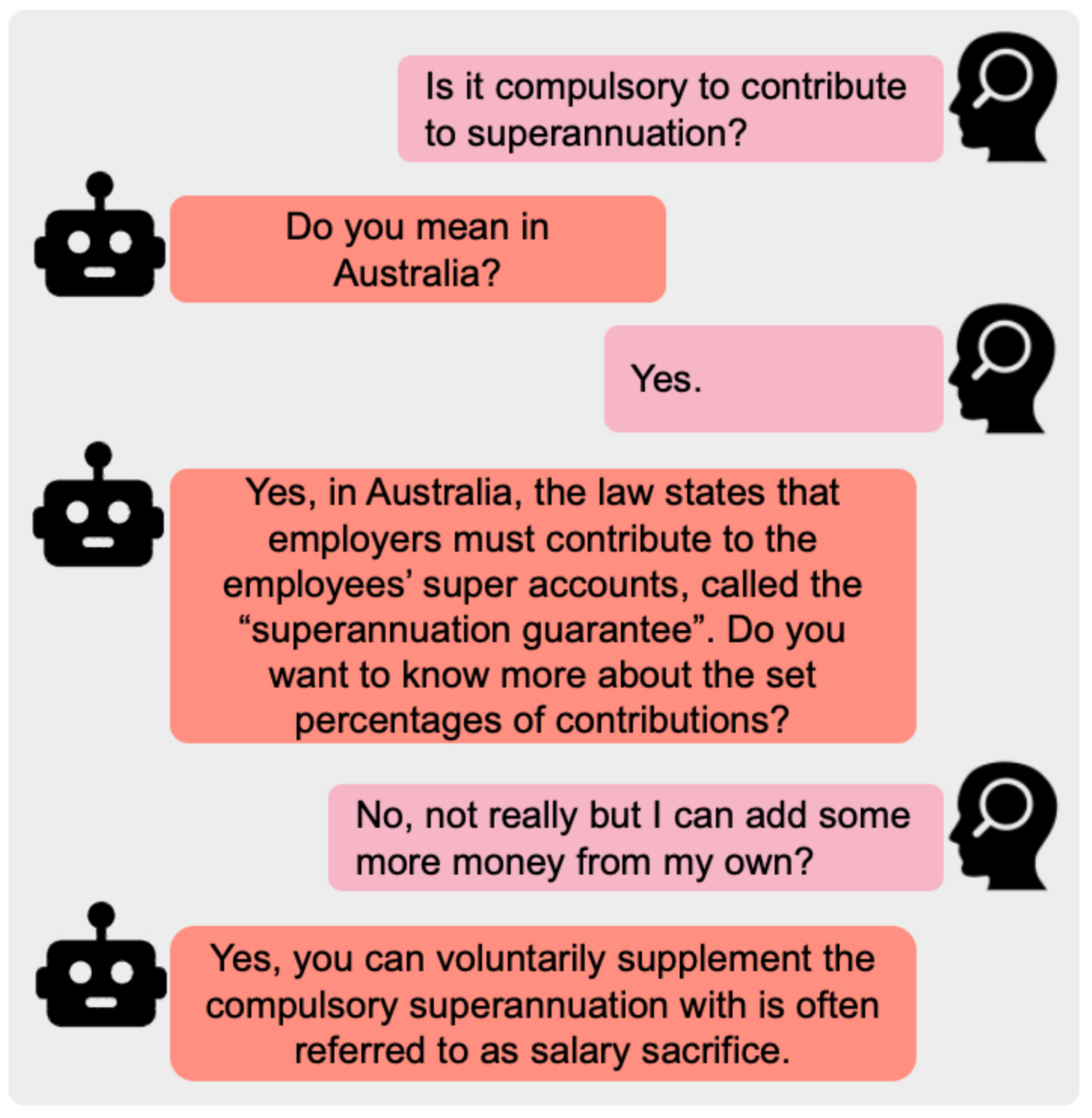}
\caption{Example conversation when someone inquires whether superannuation is compulsory in Australia within a more ideal dialogue.}
\label{fig:chapter2:example2}
\end{figure}

In contrast to the Figure~\ref{fig:chapter2:example1} example, the example in Figure~\ref{fig:chapter2:example2} shows a conversational search dialogue that enables the user to provide their query in a more natural style.
The dialogue is more natural and involves greater natural language exchanges. The dialogue is intuitively divided into pieces to minimise information overload. Furthermore, the system recognizes the user's context, assisting them in refining their inquiry, and maintains an account of prior encounters, eliminating the need for repetition. In addition, the system creates a model of the user and their information needs through problem elicitation. All of these user-system interactions are made possible by both sides conversing in a human-like manner. 

As part of CIS, several definitions of conversational search have been proposed~\citep{Anand:2020:Dagstuhl, Radlinski:2017:CHIIR, Azzopardi:2018:ConceptualizingAI, Trippas:2019:CHIIR}, which are all inline with the CIS definition provided earlier in this section. 
For example, researchers who attended the Dagstuhl seminar on Conversational Search created a typology based on existing systems as a definition~\citep{Anand:2020:Dagstuhl}.
\citet{Radlinski:2017:CHIIR, Azzopardi:2018:ConceptualizingAI} viewed the process mainly from a theoretical and system perspective, 
while~\citet{Trippas:2019:Thesis} viewed it from a cognitive, user-system, and empirical perspective.

As seen in Figure~\ref{fig:chapter2:dagstuhl}, the Dagstuhl typology aimed to position conversational search with respect to other disciplines and research areas. For instance, they drew the lines from IR systems and added properties such as statefulness to derive IIR systems and thus specify conversational search as:

\begin{figure}[t]
\centering
\includegraphics[width=.7\textwidth]{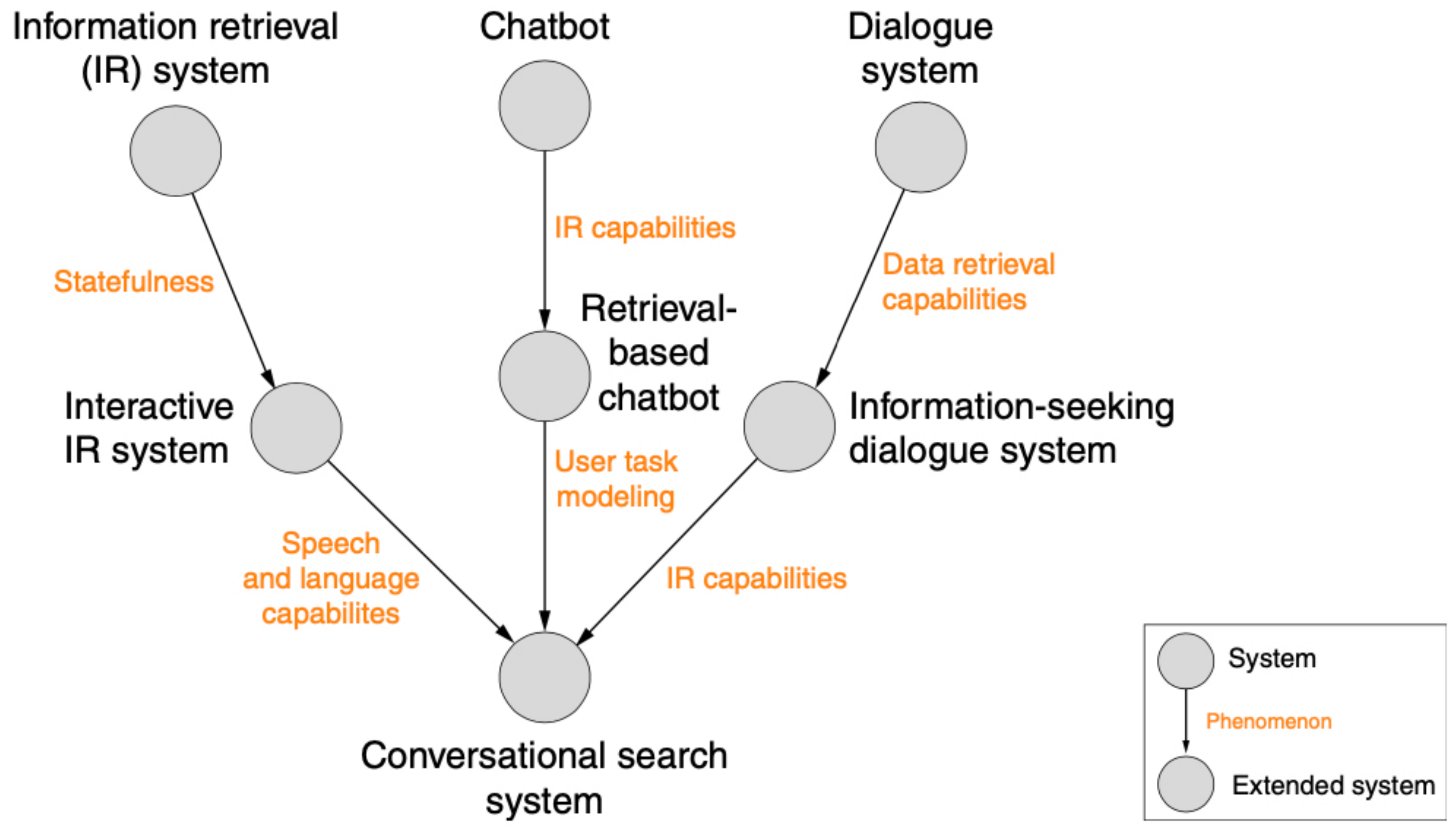}
\caption{The Dagstuhl Conversational Search Typology defines the systems via functional extensions of IR systems, chatbots, and dialogue systems~\citep{Anand:2020:Dagstuhl}.}
\label{fig:chapter2:dagstuhl}
\end{figure}

\begin{quotation}
  \textit{``A conversational search system is either an interactive information retrieval system with speech and language processing capabilities, a retrieval-based chatbot with user task modeling, or an information seeking dialogue system with information retrieval capabilities.''}~\citep[p. 52]{Anand:2020:Dagstuhl}
\end{quotation}

Meanwhile~\citet{Radlinski:2017:CHIIR} define conversational search systems with a more focused and applied view on which properties need to be met.
\begin{quotation}
  \textit{``A conversational search system is a system for retrieving information that permits a mixed-initiative back and forth between a user and agent, where the agent’s actions are chosen in response to a model of current user needs within the current conversation, using both short- and long-term knowledge of the user.''}~\citep[p. 120]{Radlinski:2017:CHIIR}
\end{quotation}

Lastly, \citet{Trippas:2019:Thesis} expanded on~\citeauthor{Radlinski:2017:CHIIR}'s definition and stated that for spoken conversational search:

\begin{quotation}
  \textit{``A spoken conversational system supports the users’ input which can include multiple actions in one utterance and is more semantically complex. Moreover, the conversational system helps users navigate an information space and can overcome standstill-conversations due to communication breakdown by including meta-communication as part of the interactions. Ultimately, the conversational system multi-turn exchanges are mixed-initiative, meaning that systems also can take action or drive the conversation. The system also keeps track of the context of particular questions, ensuring a natural flow to the conversation (i.e., no need to repeat previous statements). Thus the user’s information need can be expressed, formalised, or elicited through natural language conversational interactions.''}~\citep[p. 142]{Trippas:2019:Thesis}
\end{quotation}

All of these definitions look at the provided CIS definition from a search perspective by focusing on retrieving/selecting information items. 




\section{Conversational Recommendation}
\label{sec:chapter2:Conversational Recommendation}

Recommender systems can be seen as information seeking systems that provide users with potentially relevant items based on historical interactions. Unlike a conventional search engine that takes a query as input, most recommender systems use past user-item interactions to produce relevant recommendations~\citep{Konstan:2012:recommender}. 
%
As such, traditional recommender systems aim to help users filter and select items for their information need, often in a closed domain such as books, restaurants, or movies. These systems select possible items from an extensive database and filter them to present the user with the best suitable option~\citep{Resnick:1997:ACM,Thompson:2004:personalised}.

Recently, two survey papers on conversational recommender systems have proposed definitions of this research area as:

\begin{quotation}
  \textit{``A conversational recommender system is a software system that supports its users in achieving recommendation-related goals through a multi-turn dialogue.''}~\citep[p. 105]{Jannach:2021:recommender}
\end{quotation}

and

\begin{quotation}
  \textit{``A recommendation system [ed. conversational recommender system] that can elicit the dynamic preferences of users and take actions based on their current needs through real-time multi-turn interactions.''}~\citep[p. 101]{gao:2021:AIOpen}
\end{quotation}

Based on the above definitions and similar to conversational search, conversational recommender systems ultimately should be \textit{multi-turn}, meaning that there is more than one interaction or two utterances (\ie, one utterance from the user and one from the system). Current conversational recommender systems can answer recommendation requests reasonably well, but often have difficulties maintaining multi-turn conversations~\citep{Jannach:2021:recommender}.




Even though the usage of multi-turn interactions could imply some kind of memory that can keep track of the communication and current state, most previous definitions fail to mention this fundamental requirement for conversational recommendation. Indeed, some form of user-system interaction history with conversational recommender systems is necessary for a system to be able to provide recommendations based on those previous interactions. Thus, storing past interactions to refer to is a key component, similarly to conversational search. At the same time, it is important to simultaneously consider privacy implications of such an interaction history: What exactly is being retained, how it may be used in future, and how people can control what is stored. This is currently an open area of research.

\begin{figure}[h]
\centering
\includegraphics[width=.75\textwidth]{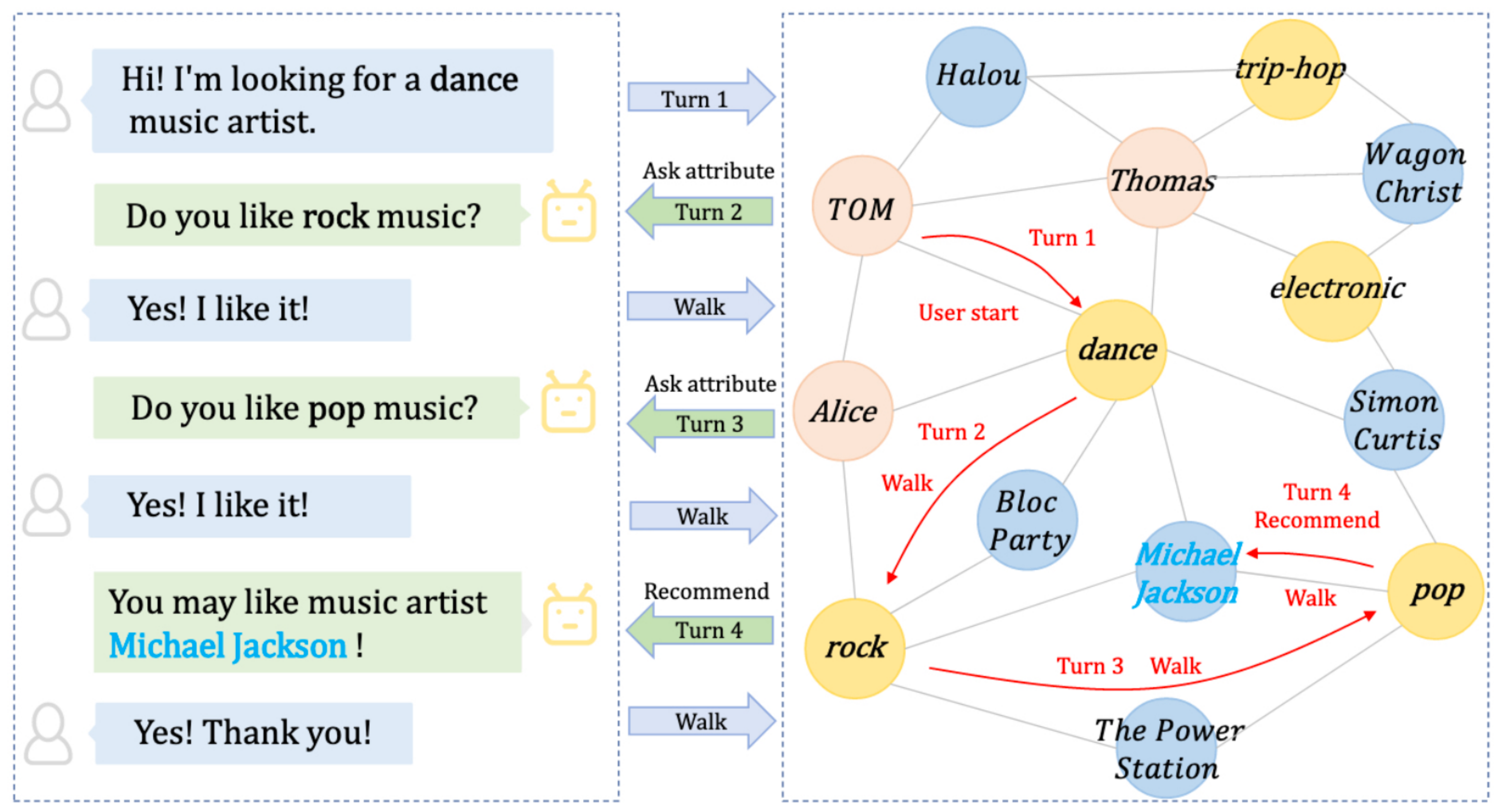}
\caption{An example of user interactions with conversational recommender systems from~\citet{Lei:2020:KDD} with each interaction demonstrating the reasoning.}
\label{fig:chapter2:chapter2_recommender}
\end{figure}

Conversational recommender systems are sometimes referred to as a ``systems ask, users answer'' paradigm~\citep{Sun:2018:ConvRecSys,Zhang:2018:cikm}. This means that only the recommender system could ask questions to elicit users' preferences. Furthermore, this one-way elicitation approach can have difficulties thoroughly capturing the users' needs.
However, more recent work in conversational recommender systems has investigated this rigid paradigm, introducing the \textit{mixed-initiative} approach~\citep{REN:2020:TOIS}.
Indeed, a conversational recommender system should be able to elicit, acquire, store, and utilize user preferences through implicit (\eg, clicking) or explicit (\eg, rating) user feedback~\citep{Pommeranz:2012:elicitation, Christakopoulou:2016:KDD}. This implies that conversational recommender systems should be capable of taking the initiative and thus support mixed-initiative interactions. An example of acquiring the user's preference can be seen in Figure~\ref{fig:chapter2:chapter2_recommender}.

A fundamental characteristic of conversational recommender systems is that they support specific tasks and goals. The system should suggest recommendations while the user interacts with that system to help them find relevant information and thus support the user's decision making process.

Another way to elicit user preferences is through product reviews. However, one drawback of this method is that the user must have reviewed for the system to create a user profile~\citep{Chen:2015:recommender}. Conversational interactions may overcome this issue by simply engaging the user in a conversation about products they liked or disliked in the past or the most important features of products for them~\citep{Iovine:2020:RecSys}, or asking users questions based on others' reviews~\citep{Kostric:2021:Questions}. 
Another advantage of the conversational format for recommendations is to explain why (and/or how) particular items are retrieved~\citep{laban:2020:IVA}. 
In conversational search, users submit a query and explain their information need, which means there can be some transparency on why the system retrieves the given results. However, the decision-making process in recommender systems is much less visible to the users since it is based on prior interactions~\citep{paraschakis:2016:recommender}. Further research on systems that reason and explain through natural language and conversational actions why particular results are retrieved, how they yield ethically sourced recommendations that are culturally relevant, and respect laws and societal norms are warranted~\citep{krebs:2019:tell, Dinoia:2022:recommender}. By providing explanations, conversational systems will enhance human decision-making and will also be improved from an ethical standpoint.

\tipbox{Conversational search and conversational recommender systems share many commonalities. Essentially, both tasks aim to provide users with relevant items based on a ranking, either through a query (search) or user preference (recommendation). This point has been raised in the 1990s by \citet{Belkin:1992} and has recently been revisited in \citep{Zamani:2018:DESIRES,Zamani:2020:JSR}. Furthermore, both systems will interact through \textit{conversations} with the system and share the same characteristics of interaction modality (see Section~\ref{subsec:chapter2:Interaction Modality and Language}).
}

In conclusion, as been repeatedly mentioned, the boundaries between these CIS applications are often blurred, mainly because many comparable technological and computing methods are applied. Using the strengths and advances from each CIS subdomain will move the area of conversational systems forward.

\section{Conversational Question Answering}
\label{sec:chapter2:Conversational Question Answering}


Question answering (QA), the task of providing one of more answer(s) to a given question, has been a longstanding information seeking task within the IR and NLP communities~\citep{Dwivedi:2013:research, Kolomiyets:2011:QAsurvey, Warren:1982:CHATQA, winograd:1974:five}. 
Early QA systems were created in the 1960s and 70s, such as BASEBALL~\citep{Green:1961:baseball} and LUNAR~\citep{woods1972lunar}. Both interfaced  a structured database that could be accessed through very restricted natural language questions. The subject domain was also very restricted, so the user query could be processed and parsed through a manually created domain-specific vocabulary.

Other early systems, usually created for a specific domain, include SHRDLU by~\citet{winograd:1974:five} and CHAT-80 QA by~\citet{Warren:1982:CHATQA}. The SHRDLU system was designed as an interactive dialogue interface to give commands, ask questions, or make statements while the system could react by carrying out the commands, answering questions, and taking in new information.
However, this early system had limited capabilities. For example, as~\citet{winograd:1974:five} explained, the system was narrow and only accepted a limited range of information, specifically in understanding human language and the reasoning behind these interactions.

QA is a specific form of information seeking where the users' needs are expressed in a form of (natural language) question. 
For example, ``Which country has the longest period without a government?''. QA questions are also frequently classified by common properties and can often be classified as factoid, list, definition, relationship, procedural, and conformation questions~\citep{Kolomiyets:2011:QAsurvey}. These particular question types have specific characteristics, such as a factoid question often starts with WH-interrogated words (what, when, where, who) and list questions often start with \textit{List/Name [me] [all/at least NUMBER/some]}~\citep{Kolomiyets:2011:QAsurvey}.

In contrast to classical IR, in which full documents are considered relevant to the user's need, QA is concerned about finding and presenting relatively short pieces of information to answer the queries.
Therefore, QA uses NLP and IR techniques to retrieve small text snippets containing the exact answer to a query instead of the document lists traditionally returned by text retrieval systems~\citep{Voorhees:1999:TREC, gao:2019:FnTIR}. 
The short answers are often retrieved and presented as short text passages, phrases, sentences, or knowledge graph entities~\citep{Lu:2019:SIGIR}. 

With the developments around conversational systems, QA work has received increased attention in the context of CIS~\citep{Christmann:2019:CIKM, Qu:2019:CIKM, Kaiser:2020:conversational}. 
Conversational QA (ConvQA) can be seen as a subsection of CIS but with a narrower focus than conversational search. Even though ConvQA is a popular research topic, we are unaware of any comprehensive definition of ConvQA. The main reason is likely that it is difficult to distinguish it from many conversational search tasks. 



Traditionally, QA has focused on a single question, meaning no historical interaction data is kept. However, it could be argued that conversations should be composed of more than one interaction. Thus, in conversational QA, the user may pose more than one question. Furthermore, 
as explained in earlier sections, conversational interactions imply that the history of previous dialogues is kept and used to answer the user's questions enabling follow-up questions or references to earlier concepts. 
Using the advantage of the conversational aspect, users can query the system interactively without having to compose complicated queries~\citep{gao:2019:FnTIR}. However, to correctly answer the user's question, ConvQA systems need to handle more complex linguistic characteristics of conversations, such as anaphoras (words that explicitly refer to previous conversational turns) or ellipsis (words that are redundant in the conversation)~\citep{Vakulenko:2021:QuestionRe}.

\begin{figure}[h]
\centering
\includegraphics[width=.75\textwidth]{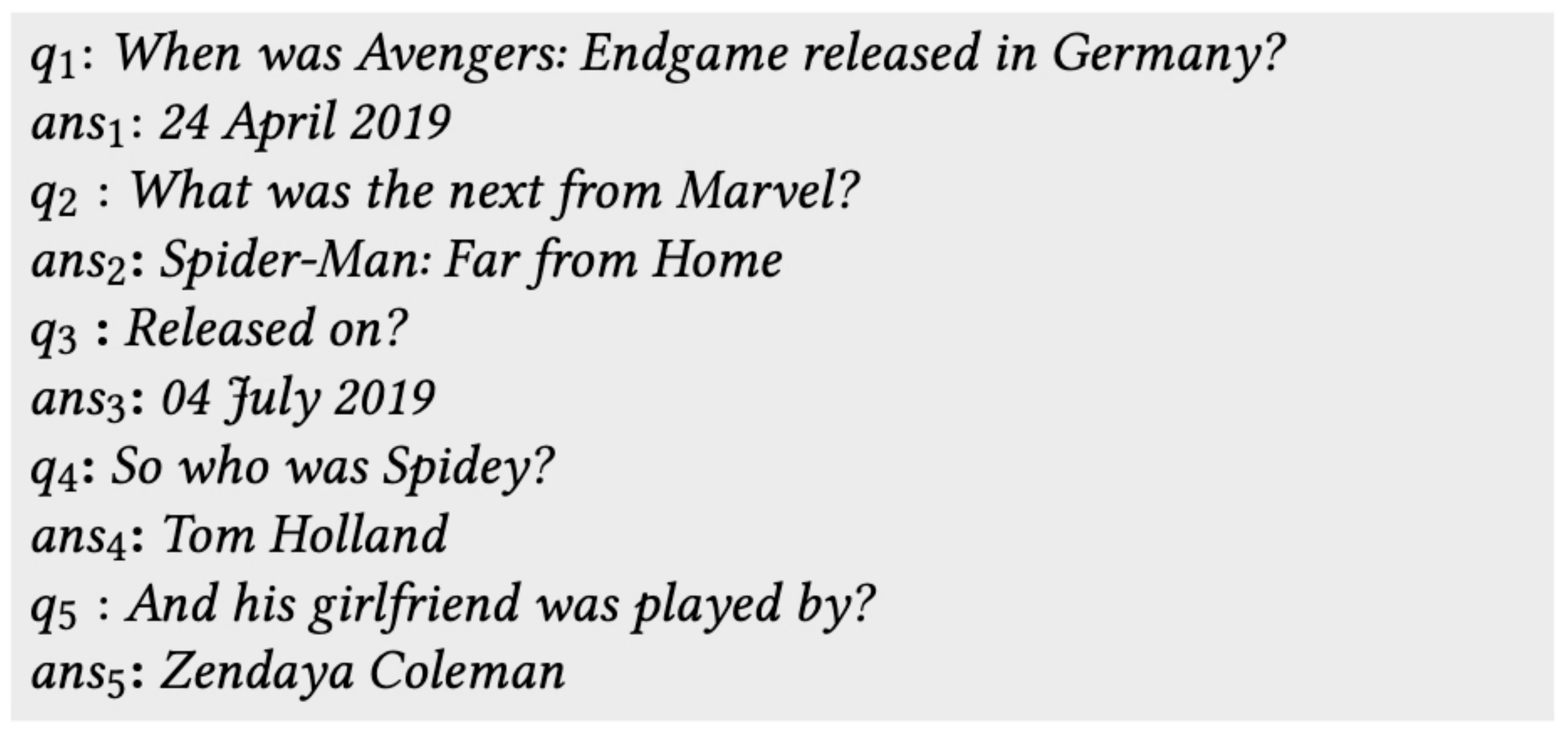}
\caption{An ideal conversational QA interaction example with five turns from~\citet{Kaiser:2021:SIGIR} where $q_i$ and $ans_i$ are questions and answers at turn $i$, respectively.}
\label{fig:chapter2:CoQA_example}
\end{figure}

An example of a series of ConvQA interactions is seen in Figure~\ref{fig:chapter2:CoQA_example}. 
Furthermore, ConvQA is often seen in relation to machine comprehension~\citep{Yang:2018:ecommerce}, which is often based on questions about a given passage of text. The main difference is that machine comprehension organizes the questions into conversations~\citep{Qu:2019:sigir}. This means that leveraging the history is crucial to creating robust and effective ConvQA systems. For example, history can help map the state and changes of the information need to inform current or future responses.
Recent work from~\citet{Kaiser:2020:conversational} also mentions the importance of dialogue context to improve ConvQA.
That is, the user in later interactions can refer to the implicit context of previous utterances.

\section{Conversational Information Seeking in Different Domains}
\label{sec:chapter2:Conversational Information Seeking in Different Domains}

As a paradigm to interact with information, CIS can find items on the web, databases, or knowledge graphs. Conversational information access can also be applied to specific domains such as the financial industry, hospitality, or cooking recipes. This section expands different domains where CIS can be applied in addition to their unique properties. These domains include e-commerce, enterprise, and health.

\subsection{Conversational Information Seeking in E-Commerce}
\label{subsec:chapter2:Conversational Information Seeking in E-commerce}

Finding and buying products through conversational interactions is becoming popular~\citep{Papenmeier:2021:CHIIR,Papenmeier:2022:CHIIR}. E-commerce transactions, the process of buying and selling goods and services online, are steadily increasing.\footnote{\url{https://www.forbes.com/sites/joanverdon/2021/04/27/global-ecommerce-sales-to-hit-42-trillion-as-online-surge-continues-adobe-reports/}} 
Simultaneously, with the uptake of CIS systems with consumers (\eg, Amazon Alexa or Google Assistant), it becomes increasingly easier to identify consumers' context (\eg,~a user searching for washing instructions or re-ordering washing powder may be located in the laundry), resulting in more accurate context-aware responses.


It has been suggested that conversational e-commerce (also referred to as conversational commerce~\citep{vaneeuwen2017mobile}) search and task-oriented dialogues share commonalities. For example, the dialogue for flight reservation and e-commerce will elicit user preferences such as flight destinations akin to an e-commerce product~\citep{Yang:2018:ecommerce}. However, differences between task-oriented dialogue systems and e-commerce queries have also been observed, making e-commerce information need expression much more complex~\citep{Yang:2018:ecommerce}. For instance, e-commerce products often have different facets, such as brand, color, size, or style, resulting in different preference slot combinations or shopping schema. Thus, e-commerce schemas can be complex. It is even suggested that they can be incomplete due to the extended range of product facets. 
\citet{Zhang:2018:cikm} suggested that
user-system interactions in e-commerce CIS systems can be classified into three stages: initiation, conversation, and display.
In their proposed paradigm, the system will loop through questions to understand all user preferences of the product's facets before presenting the user's query results.

Some advantages of using CIS in e-commerce include accessing products through conversational-enabled devices such as mobile phones or smart devices~\citep{vaneeuwen2017mobile}. Furthermore, instead of going to a shop for support, customers can access help instantly through these devices~\citep{lim2022alexa}. In addition, when users are logged in to their shopping profile, personalization and shopping history can optimize shopping experiences. Conversely, CIS systems embedded in an intelligent assistant have the potential to be virtual shopping assistants. Future conversational commerce systems can also be embedded into other emerging technologies, such as augmented reality~\citep{buschel2018here}.

\subsection{Conversational Information Seeking in Enterprise}
\label{subsec:chapter2:Conversational Information Seeking in Enterprise}
An application of CIS which has not received as much attention is searching through conversational interactions in an enterprise setting~\citep{Teevan:2020:dagstuhl}.
CIS enterprise systems aim to help people in a work environment such as meeting rooms and at desks, with predictions that by 2025, 50\% of knowledge workers would use a virtual assistant daily. This prediction is up from 2\% in 2019.\footnote{\url{https://blogs.gartner.com/anthony_bradley/2020/08/10/brace-yourself-for-an-explosion-of-virtual-assistants/}}
Even though there has been an increased interest in workplace-oriented digital assistants in general (\eg, 
Alexa for Business\footnote{\url{https://aws.amazon.com/alexaforbusiness/}} 
or Cortana Skills Kit for Enterprise\footnote{\url{https://blogs.microsoft.com/ai/cortana-for-enterprise/}}), the uptake has been limited.

It is well known that enterprise search applications have different needs than a traditional web search engine, including challenges such as searching over enterprise Intranets or multiple internal sources~\citep{Hawking:2004:Enterprise}. Furthermore, besides using CIS systems in a traditional office environment, many different applications of more varied and complex environments, such as airplane pilots, create an extra layer of complexity~\citep{Arnold:2020:percieved, Gosper:2021:Understanding}.
Many open problems in the intersection of CIS applications and enterprise need further investigation. In particular, issues such as defining appropriate test collections, effective conversational search over distributed information sources, identifying tasks that lend themselves to use a CIS application, and understanding the way employees interact with these systems need to be investigated.

\subsection{Conversational Information Seeking in Health}
\label{subsec:chapter2:Conversational Information Seeking in Health}

Searching for health information is another application for CIS.
Many people already search for health advice online. For example, people will go to symptom checkers to understand if they have an underlying health condition or to identify whether they need professional advice~\citep{Cross:2021:WWW}. Furthermore, a recent study of a CIS application to enable patients to search for cancer-related clinical trials suggest that CIS could help to make health information more accessible for people with low health or computer literacy skills~\citep{bickmore:2016:improving}.

A recent survey suggests that the main areas of CIS applications are located in areas for patients such as treatment and monitoring, health care service support, and education~\citep{car:2020:conversational}. However, user groups such as carers and other health professionals can benefit from these systems besides patients. For example, in a study where physicians used an information seeking chatbot, they reported that the advantages of CIS include diagnostic decision-making~\citep{koman:2020:physicians}.

Even though CIS has major potential, some concerns about implementing these systems in the health domain need to be addressed. For example, these systems may not have sufficient expertise to answer all questions and may even misinterpret or misunderstand these questions, potentially providing a wrong answer~\citep{su2021role}. Although a common challenge to all search systems, this may be exacerbated in a CIS setting if a system were to naively present health \emph{mis}information in a way that reinforces it. Furthermore, these systems can deal with sensitive patient data and thus need to be safeguarded. Voice-only CIS systems may also encounter issues with speech recognition, especially when people are distressed or are in noisy environments~\citep{Spina:2021:SIGIRForum}.

\section{Intelligent Assistants}
\label{sec:chapter2:intelligent assistants}
Intelligent assistants are often associated with CIS and are rising in popularity. 
The number of intelligent voice assistants worldwide is predicted to double between 2020 and 2024, from 4.2 billion to 8.4 billion.\footnote{\url{https://www.statista.com/statistics/973815/worldwide-digital-voice-assistant-in-use}}  
Intelligent assistants are frequently embedded in existing phones, laptops, mobile devices or smart speakers. For instance, assistants such as Google Assistant, Amazon's Alexa, AliMe, or Apple's Siri enable users to receive assistance on everyday tasks with a specific goal (\eg,~turning on or off appliances) or conduct simple question-answering tasks such as asking for weather forecasts or the news.
With the increase in mobile devices and mobile internet connections, users instantly have access to powerful computational and digital intelligent assistants. These may even be designed to access the user's situation or context through GPS locations, the people around them through Bluetooth scans, and previous interactions with their electronic devices~\citep{Liono:2020:intelligent,Trippas:2019:CHIIR} when enabled on the mobile device. However, more research is needed to use all the contextual signals to optimize CIS responsibly and with user privacy in mind.

Different CIS tasks may require access to different knowledge sources and databases. Intelligent assistants need to disambiguate which knowledge source they need to retrieve the information from. For instance, \citet{Aliannejadi:2018:UnifiedMobileSearch} introduced the problem of unified mobile search, in which intelligent assistants identify the target mobile apps for each search query, route the query to the selected apps, and aggregate the search results. In follow-up work, the authors demonstrated the impact of user context and app usage patterns on unified mobile search~\citep{Aliannejadi:2018:InSituUMS,Aliannejadi:2021:UMSTOIS}. Identifying knowledge sources was also used in the Ninth Dialog System Technology Challenge (DSTC9) with a track called ``Beyond domain APIs - Tasks-oriented conversational modeling with unstructured knowledge access''. This track aimed to expand different task-oriented dialog systems by incorporating external unstructured knowledge sources~\citep{gunasekara:2020:overview}. The track's purpose was to investigate how to support \textit{frictionless} task-oriented situations so that the flow of the conversation does not break when users have questions that are out of the scope of APIs/DB but possibly are available in external knowledge sources.

Other applications incorporating CIS systems are embodied robots, \eg, the Multi-Modal Mall Entertainment Robot (MuMMER)~\citep{Foster:2016:MuMMER}. MuMMER was a collaborative challenge in which a robot was made to behave appropriately to human social norms and engage through speech-based interactions. Similarly, social bots enable users to search and engage in information dialogues. This has been thoroughly studied in the context of Alexa Prize Socialbot Challenge~\citep{Ram:2018:Alexa}. Although these interactions involving search for information may differ from a focused CIS system, embedding CIS enables a wider variety of use-cases.


\section{Summary}
\label{sec:chapter2:section summary}
This section provided a high-level overview of CIS and its applications. We first started by providing definitions for conversation, information seeking conversation, and CIS systems. Under these definitions, conversational search, conversational question answering, and conversational recommendation are seen as the subdomains of conversational information seeking tasks. This section also included several system requirements that are expected from CIS systems. 

We later reviewed previous work that characterizes the three subdomains of CIS and discussed their connections.  We lastly provided an overview of how CIS can be used in particular domains and compared CIS to intelligent assistants.
CIS is still being developed and is rapidly expanding as a multi-dimensional and multi-disciplinary research area. 
Overall, this section summarized prior work in conversational information seeking applications to provide an overview.



\chapter{Conversational Interfaces and Result Presentation}
\label{chapter3}

The emergence of conversational systems has empowered the development of a new kind of human--computer interface supporting users to converse with the interface through spoken interactions. In this section, we introduce different kinds of conversational interfaces, set out the limitations, how they support the entire interaction from the users' speech input to the system's output, and investigate the latest research in the presentation of results.

A conversational interface, also identified as conversational user interface (CUI), presents the front-end to a chatbot or virtual personal assistant, enabling the user to interact with the application through various input and output modalities such as speech, text, or touch~\citep{Mctear:2016:conversational,mctear2017rise}. Besides being the system's front-end, the conversational interface integrates or glues together all the underlying system components, represented in a usable application~\citep{Zue:2000:IEEE}. Even though all the recent developments of the separate components have made conversational interfaces more functional, they act as the orchestrator of all the information with their challenges. 

Overall, this section introduces the different conversational interfaces and illustrates the limitation of transferring information in a conversational style for different interfaces. We discuss \textit{initiative} as a critical element in conversational interactions, including the interface limitations with regards to CIS.

\section{Conversational Interfaces}
\label{sec:chapter3:conversational interfaces}

Interfaces that provide users with the ability to interact \textit{conversationally} with systems through different modalities such as speech, gesture, text, or touch are commonly referred to as CUIs. Many additional terms refer to these systems that enable conversational interactions, including chatbots, intelligent assistants, or conversational agents. 

An interface is often referred to be \textit{conversational} when it covers two basic attributes (1) natural language and (2) conversational interaction style~\citep{mctear2017rise}. The \textit{natural language} attribute means that the system and user can use language as in naturally occurring conversations between two or more participants; this contrasts to restricted commands, mouse clicks, or phrases in a graphical user interface (GUI). Furthermore, natural language is more flexible, permitting input to be expressed in many different ways versus one fixed expression. 
Intuitively, allowing the user to input natural language contributes to a more complex system. In addition, \textit{conversational interaction style} is often referred to as basic turn-taking behavior in which the user and system converse one after another. This contrasts with clicking or swiping on GUI elements such as buttons or drop-down menus.
Furthermore, to make an interface even more conversational, the usage of \textit{mixed-initiative} is introduced. Mixed-initiative is more human-like and flexible because both actors can independently contribute to the conversation. 
Lastly, a more advanced system could include \textit{context} tracking enabling follow-up questions and persistent tracking of the topic. Even though many dialogue systems are seen as conversational, they may not be tracking the context and therefore never refer back to a previous question or answer. Instead, they attend to every input individually.

\tipbox{
Basic conversational interfaces often consist of two primary attributes and sub-attributes: natural language which does not consist of fixed expressions, and conversational interaction style which could support turn-taking, mixed-initiative, and context tracing.
}

Even though various forms of conversational interfaces have been around for a long time, we have recently seen a revival of the topic, mostly due to the advances in automatic speech recognition (ASR), natural language processing (NLP), and machine learning in general. Nevertheless, much fundamental research dates back to the 1960s with the first well-known chatbot, ELIZA, having simulated a Rogerian psychologist~\citep{Weizenbaum:1966:ELIZA}. In the following, we provide some historical context for four distinctive groups of conversational interfaces, (1) spoken dialogue systems (SDSs), (2) voice user interfaces (VUIs), (3) live chat support, and (4) chatbots.

\subsection{Spoken Dialogue Systems}
\label{subsec:chapter3:Spoken Dialogue Systems}
Spoken dialogue systems (SDSs) enable users to interact with a system in spoken natural language on a turn-by-turn basis and are an instance of a conversational interface. Many of these systems are used for task-oriented issues with clear task boundaries, such as travel planning. In the 1960s and 70s, the earliest SDSs were mainly text-based. However, once technologies improved in the 80s, more complex components were added, such as more advanced ASR or components that helped recover from conversational breakdowns.
Much government funding from Europe and the U.S. supported research in SDS, which resulted in the European SUNDIAL (Speech Understanding and DIALog) project~\citep{peckham1991speech} and the DARPA spoken language system in the U.S.~\citep{Clark:1988:DARPA}. The SUNDIAL project aimed to design systems that could be used by the public, while the DARPA program focused on the technical aspects. Many of the early research outcomes are still applicable today, such as the Information State Update Theory~\citep{traum2003ISU}, information presentation techniques~\citep{gibbon1997handbook}, or the CSLU toolkit~\citep{Sutton:1997:CSLU}.

\begin{figure}[htbp]
\centering
\includegraphics[width=.6\textwidth]{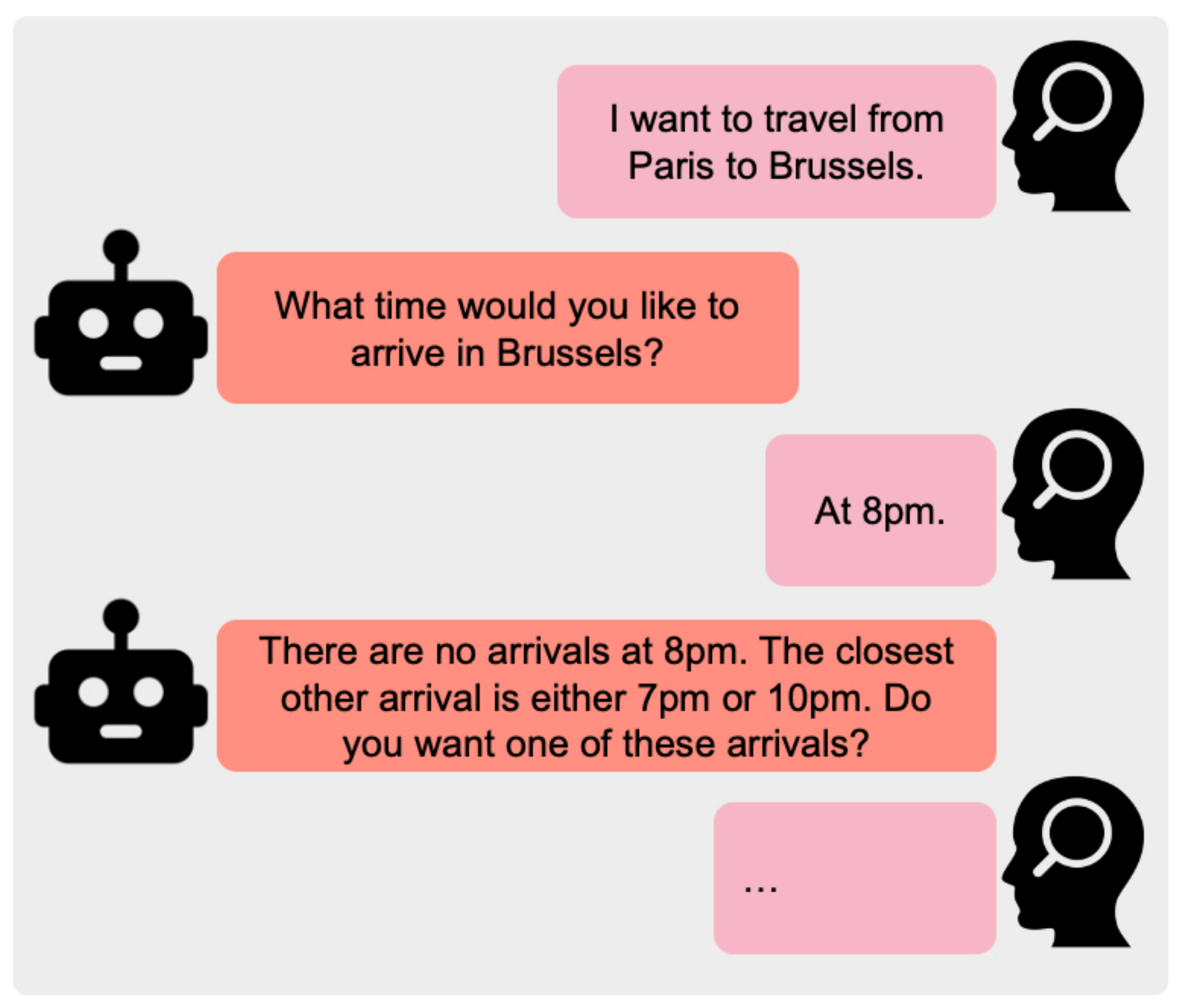}
\caption{Example conversation where the user wants to book a travel and the system provides options.}
\label{fig:chapter3:example1}
\end{figure}

A frequent example task for SDSs is time-tabling for travel services, providing the interface between the user and a database~\citep{fraser1998assessment}. In the Figure~\ref{fig:chapter3:example1} example, the user has the need of finding a reasonable travel plan.

As seen in the first utterance from the system, it is narrowing down the information need by adding a refinement or clarification question. These back and forth interactions are part of the elicitation process for the system to understand and specify the information need.

\subsection{Voice User Interfaces}
\label{subsec:chapter3:Voice User Interfaces}

\begin{figure}[htbp]
\centering
\includegraphics[width=.6\textwidth]{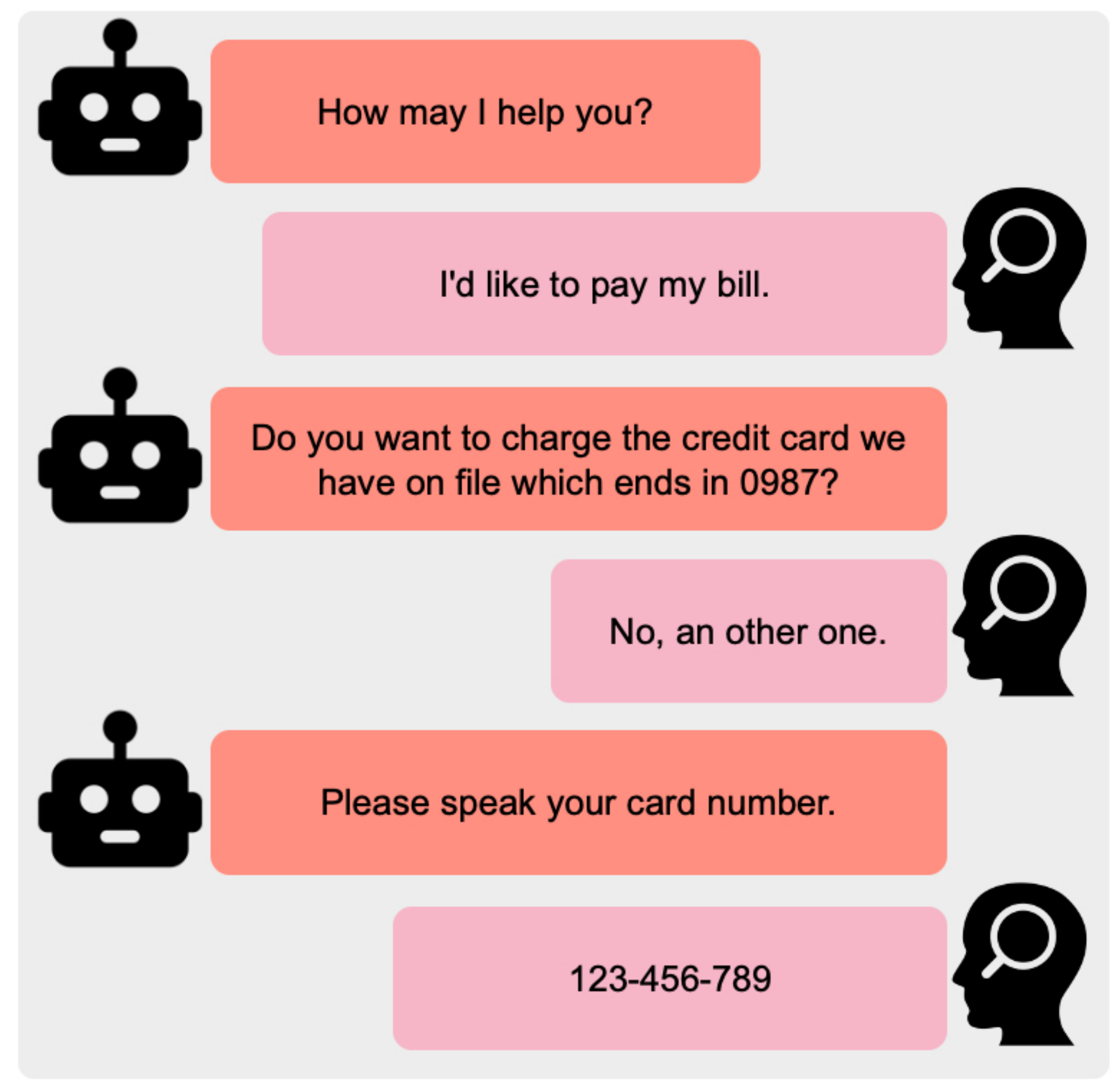}
\caption{Example conversation with a VUI in which the system is eliciting how they can help the user before possibly routing them to a human operator for more complex interactions.}
\label{fig:chapter3:example2}
\end{figure}

Companies have traditionally developed VUIs for commercial benefits, in contrast with SDS that has been created mainly by academic and research labs. For example, AT\&T created an early VUI called \textit{How May I Help You?} which supported call routing~\citep{Gorin:1997:HMIHY}. The automated customer self-service systems are task-oriented and engage in conversation to help the client, thus being classified as a conversational interface. Instead of helping the customer with their problem, such VUIs typically aim to understand the customer's problem sufficiently, after which the user can be routed to the appropriate (human) call taker to help with their problem further. Thus, these call routing services only need to elicit the general problem to refer the call to someone or a specific system module. The system responses are pre-recorded, which is possible for highly structured domain-specific settings. For example, a scenario where a user wants to pay for a service might follow a scripted interaction as shown in Figure~\ref{fig:chapter3:example2}.

In these systems, when none of the options are relevant to the user, the system will narrow down the problem to re-route the call to an appropriate human agent. The connection with CIS is the human-like interactions, eliciting information needs, and narrowing down the relevant answers or services.

The VUI community has involved with the development of W3C standards for scripting spoken dialogues such as VoiceXML,\footnote{\url{https://www.w3.org/TR/voicexml21/}} VoiceXML-based toolkits,\footnote{\url{http://evolution.voxeo.com/}} and the development for speech analtyics.


\subsection{Live Chat Support}
\label{subsec:chapter3:Live Chat Support}
The above interfaces (\ie, SDS and VUI) are mainly used with an underlying automated system. However, many support systems are powered by humans in which the interface is the connection between a user and a service provider. Live chat support is real-time communication between a customer and a support person via instant messaging, often through a pop-up dialogue box.
The service providers can include librarians on a library website~\citep{Matteson:2011:systematic}, technical or sales support on e-commerce websites~\citep{Goes:2021:live}, or health assistance~\citep{Stephen:2014:talking}. Such chat support interfaces are often embedded as web widgets in websites or as an extra feature within an application. The main advantage of live chat support interfaces is that the chat history is persistent and can be referred to by the users. Furthermore, these chats can support asynchronous and synchronous interactions~\citep{Fono:2006:Structuring}.



Some recent work by~\citet{Vakulenko:2021:TOIS} investigated virtual reference interviews of professional librarians. They suggest major differences between librarian interviews and existing datasets used to investigate, analyze, and train CIS topics. For example, they suggested that professional intermediaries are more proactive, write more extended responses, ask follow-up questions, and actively steer the topic of conversation. Further research efforts are needed to understand the impact of different conversational styles of CIS systems~\citep{thomas:2018:CHIIR}.

A ``live chat support'' provider (\eg, the call taker or customer provider) is often synchronous, meaning that the support person answers questions from the user in real-time. However, many support providers are required to answer multiple customers simultaneously, creating a one-to-many relationship. The importance of the support provider's interface, which could support decision making by ranking response suggestions on the information-seeking process or incorporating machine reading to track the conversation, has not been studied extensively~\citep{Xu:2021:customerservice, Yang:2018:ecommerce}.
Furthermore, research on how the support providers deal with task-switching and interruptions could suggest future conversational interface optimisations~\citep{pajukoski:2018:impact}.

\subsection{Chatbots}
\label{subsec:chapter3:Chatbots}

The interactions with chatbots are often based on social engagement through chit-chat (\ie, small talk), in contrast to the task-oriented interactions with SDSs and VUIs. Traditionally, chatbots are mainly text-based. However, more recent chatbots incorporate spoken interactions, images, and avatars to create a more human-like persona.\footnote{Note that a chatbot is different from a bot~\citep{mctear2017rise}. A chatbot is a software application that can perform automated tasks while engaging in conversations with the user. This contrasts with bots, which complete repetitive and mundane automated tasks such as crawling the web or harvesting email addresses from social networks. }

\medskip

All the above systems aim to support users to interact with datasets or databases. Due to the conversational aspect of the interaction, no technical expertise is required to interact with these databases, making them more accessible.
As illustrated with the different CUIs (\ie, SDS, VUI, live chat support, and chatbots), these systems cover a large range of applications and tasks (\eg, from travel booking to chit-chat). 
Although all these CUIs may be considered conversational, they still differ in the degree that people are searching for information, the system maintains control, and flexibility allowed by the user to ask for what they want to find or how they want to have the information presented.
In contrast, searching for information on the web over documents is much less predictable and cannot be implemented by pre-set refinement options. Due to the vast amount of information, more advanced techniques are needed to support users' information needs.
Other questions such as the ambiguity in people knowing when they are talking to a human or machine (\eg, chatbot),\footnote{\url{https://botor.no/}} the trust of people have in these systems, appropriateness of these systems, or transparency around the usage of artificial intelligence in general\footnote{\url{https://digital-strategy.ec.europa.eu/en/policies/european-approach-artificial-intelligence}} are relevant~\citep{Mori:2012:Uncanny, zamora:2017:HAI,gupta:2022:www}.

\section{Result Presentation: From Search Boxes to Speech Bubbles}
\label{sec:chapter3:result presentation}

Result presentation in CIS is tightly coupled with decades of research on interface development for search engines and other information retrieval systems. In this section, we draw the connection between conversational user interfaces required in CIS and past research on result presentation in search engines.

Results presentation, the way search results are communicated, has been a major research area for many years \citep{Croft:2010:Search}.
The general approach to presenting search results is a vertical list of information summarizing the retrieved documents. These results should not only return relevant results but also display them so that users can recognize them as relevant to their information need.

\begin{figure}[htbp]
\centering
\includegraphics[width=.9\textwidth]{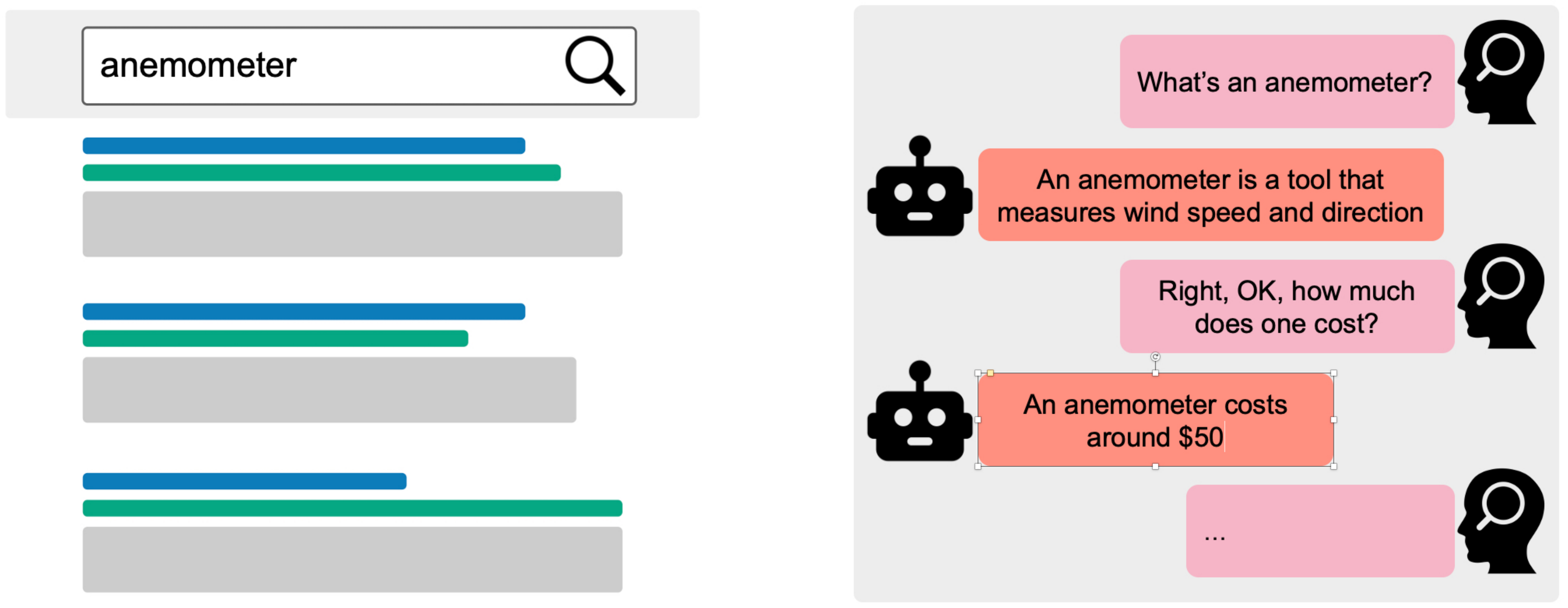}
\caption{Traditional SERP example versus a conversational style interaction \hamed{maybe replace with something with text?}}
\label{fig:chapter3:SERP}
\end{figure}

Even though many people have become accustomed to searching through these search boxes, finding information can still be a demanding task with much information to filter through. Traditionally, a user would submit an information need through keywords in a search engine search box. In return, search engines 
present a ranked list with potential relevant documents for that query, also referred to as the search engine result page (SERP). This SERP consists of the traditional ``ten blue links'' in which each item or result consists of a document title, a short summary (\ie, snippet), URL, and often other meta-data such as date or author (see Figure~\ref{fig:chapter3:SERP})~\citep{hearst2009search, Paek:2004:wavelens}.

The user would then review this returned ranked list and select an item they think would satisfy their information need. However, the first clicked item will often not satisfy the users' information need. Instead, the user will go back and forth between inspecting SERPs, looking at the contents of documents and submitting new queries. These interactions mimic a limited or one-sided conversation driven by the user.
In this instance, the user has ``control'' over the actions taken and the system has limited capabilities to interact with the user. These systems are sometimes referred to as \textit{passive}~\citep{Avula:2020:Thesis, Trippas:2018:Informing}.

The alternative interaction paradigm of CIS aims to overcome the limitations of the results presentation strategies of existing search engines by becoming more active. That is, instead of presenting a ranked list, these CIS systems can be more flexible with their information presentation strategies by adapting to the user's needs.

Even though research has shown that different presentation techniques and answer organization are needed for different modalities, limited research has been conducted in \textit{how} (the content expression) and \textit{what} (the content response) to present in conversational search~\citep{Chuklin:2018:prosody, Trippas:2015:sigir, Vtyurina:2020:ICTIR}. 
Furthermore, not only the retrieved information needs to be presented but depending on the modality of the results presentation, other interactions such as meta-conversations (\ie, information about the information, for example, information about a document or page), need to be presented~\citep{kiesel:2021:TOIS, Trippas:2018:Informing}.


\tipbox{People search differently depending on the device (\eg, desktop versus mobile) and modality (\eg, text versus audio). }

Some of these differences are highlighted in the following subsections.

\subsection{Text-Only Result Presentation on Desktops}
\label{subsec:chapter3:Text-Only Result Presentation on Desktops}

Much research has been conducted on the appearance of SERPs in browsers~\citep{hearst2009search}. In a visual setting, researchers have investigated features such as snippet length~\citep{Cutrell:2007:CHI, kaisser2008improving, Maxwell:2017:SIGIR, Rose:2007:summary}, snippet attractiveness~\citep{Clarke:2007:Influence, he2012bridging}, or the use of thumbnails~\citep{Teevan:2009:snippets, Woodruff:2002:Comparison}.

\begin{figure}[htbp]
\centering
\includegraphics[width=.5\textwidth]{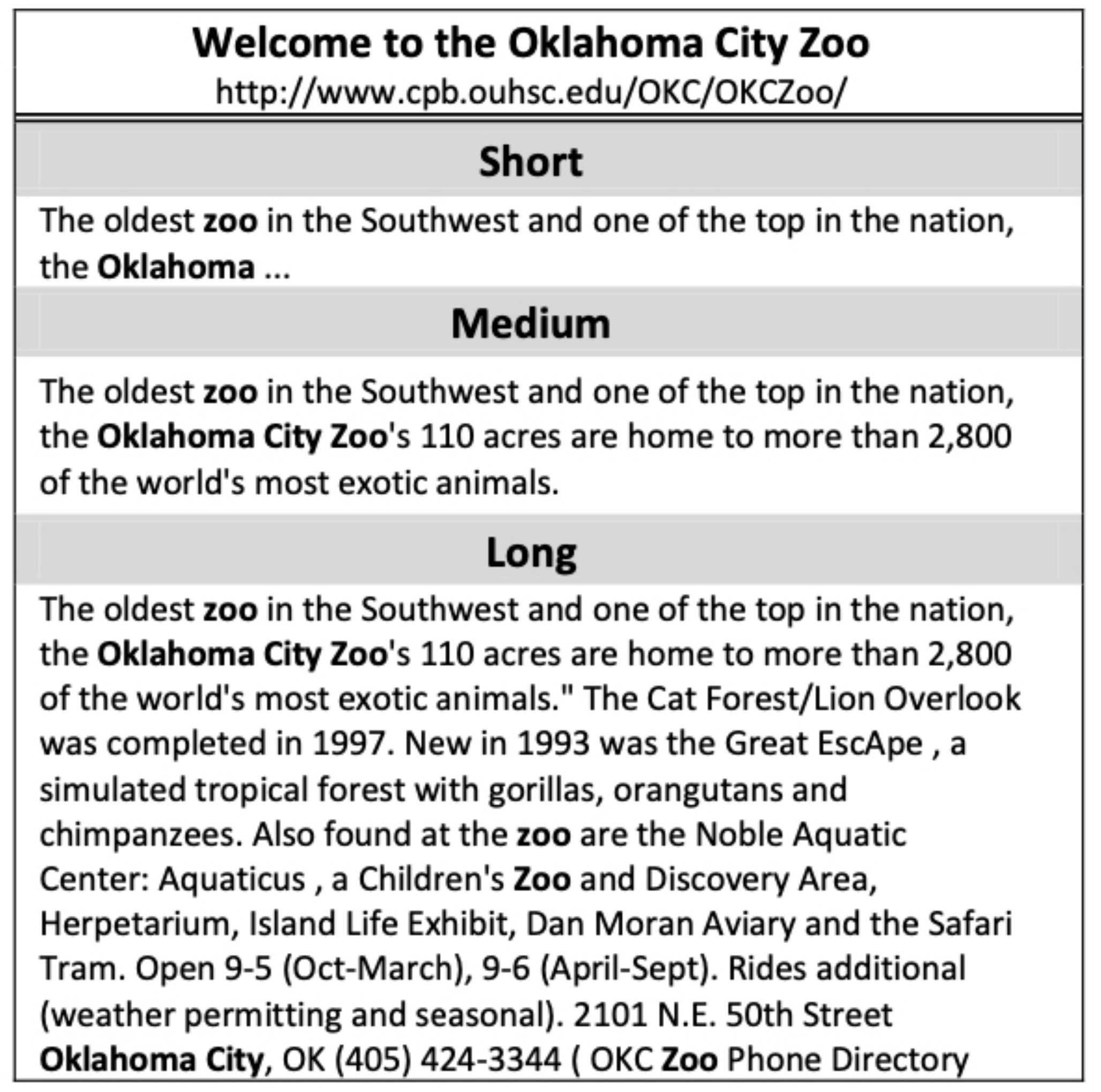}
\caption{Snippet length differences~\citep{Cutrell:2007:CHI}.}
\label{fig:chapter3:snippet_lenght}
\end{figure}

Research on results presentation has suggested that the presentation has an impact on the usability of the system.
For instance, \citet{Clarke:2007:Influence} investigated the influence of SERP features, such as the title, snippets, and URLs on user behavior. They suggested that missing or short snippets, missing query terms in the snippets, and complex URLs negatively impacted click-through behavior.
In addition, \citet{Cutrell:2007:CHI} used an eye-tracking study to explore the effects of changes in the presented search results. They manipulated the snippet length with three different lengths (short [1 text line], medium [2-3 lines], and long snippets [6-7 lines]) as shown in Figure~\ref{fig:chapter3:snippet_lenght}. Their results suggested that depending on the search task (\ie, navigational or informational), the performance improved with changing the length of the snippet. 
For navigational queries, optimal performance happened with short snippet lengths, while extended snippets helped the most for informational tasks.

Further research into snippet summary length confirmed the findings that different snippet lengths were preferred depending on the task~\citep{kaisser2008improving}. A more recent study by~\citet{Maxwell:2017:SIGIR}, re-investigated the varying snippet length and the information content within the snippets. Their results suggest that users preferred more informative and extended summaries, which they perceived as more informative. However, even though participants felt that longer snippets were more informative, they did not always help the users to identify relevant documents.

Techniques in which visual changes to the text are made, such as clustering, highlighting, or ``bolding'' query words in their context, sentence fragments, or query-biased summaries have been extensively investigated for traditional results presentation \citep{hearst2009search}. Furthermore, besides only showing text in the SERP, search engine companies have added more techniques to display results through feature snippets, knowledge cards, query suggestions, or knowledge panels. 
More research on these presentation styles in CIS is needed to understand the impact of these techniques in a conversational setting.

\begin{figure}[htbp]
\centering
\includegraphics[width=1\textwidth]{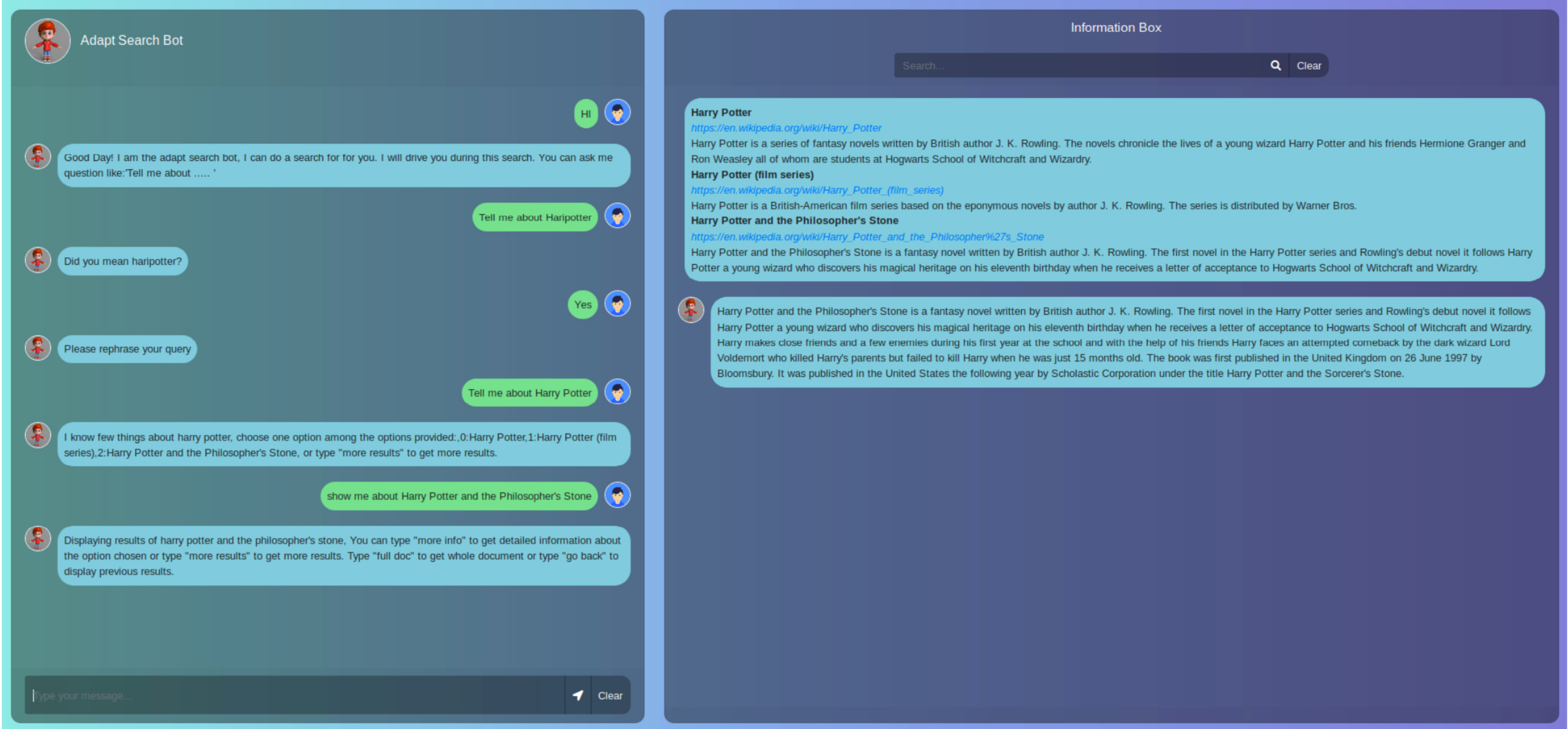}
\caption{A visual example of the Conversational Agent by~\citet{Kaushik:2020:interface}. The agent exist of a conversational search assistant (left) with a more traditional search interface (right).}
\label{fig:chapter3:conversational_agent_kaushik}
\end{figure}

Limited research has been conducted into conversational results presentation for desktop. A recent prototype for text-only chat-based search by 
\citet{Kaushik:2020:interface} 
combined a conversational search assistant (\ie, Adapt Search Bot), with a more traditional search interface (\ie, Information Box), see Figure~\ref{fig:chapter3:conversational_agent_kaushik}. The user can either interact with the assistant on the left side of the application or with the retrieved information on the right panel. The authors described this design as flexible for users to interact with the agent and the search engine itself. Furthermore, their design supported users interacting with the search engine with the agent initiating dialogues to support the search process.
However, further research could help understand the impact of different presentation techniques, chat-based search, and distributed results presentation (\eg, results on both left and right panels).

\begin{figure}[t]
\centering
\includegraphics[width=0.8\textwidth]{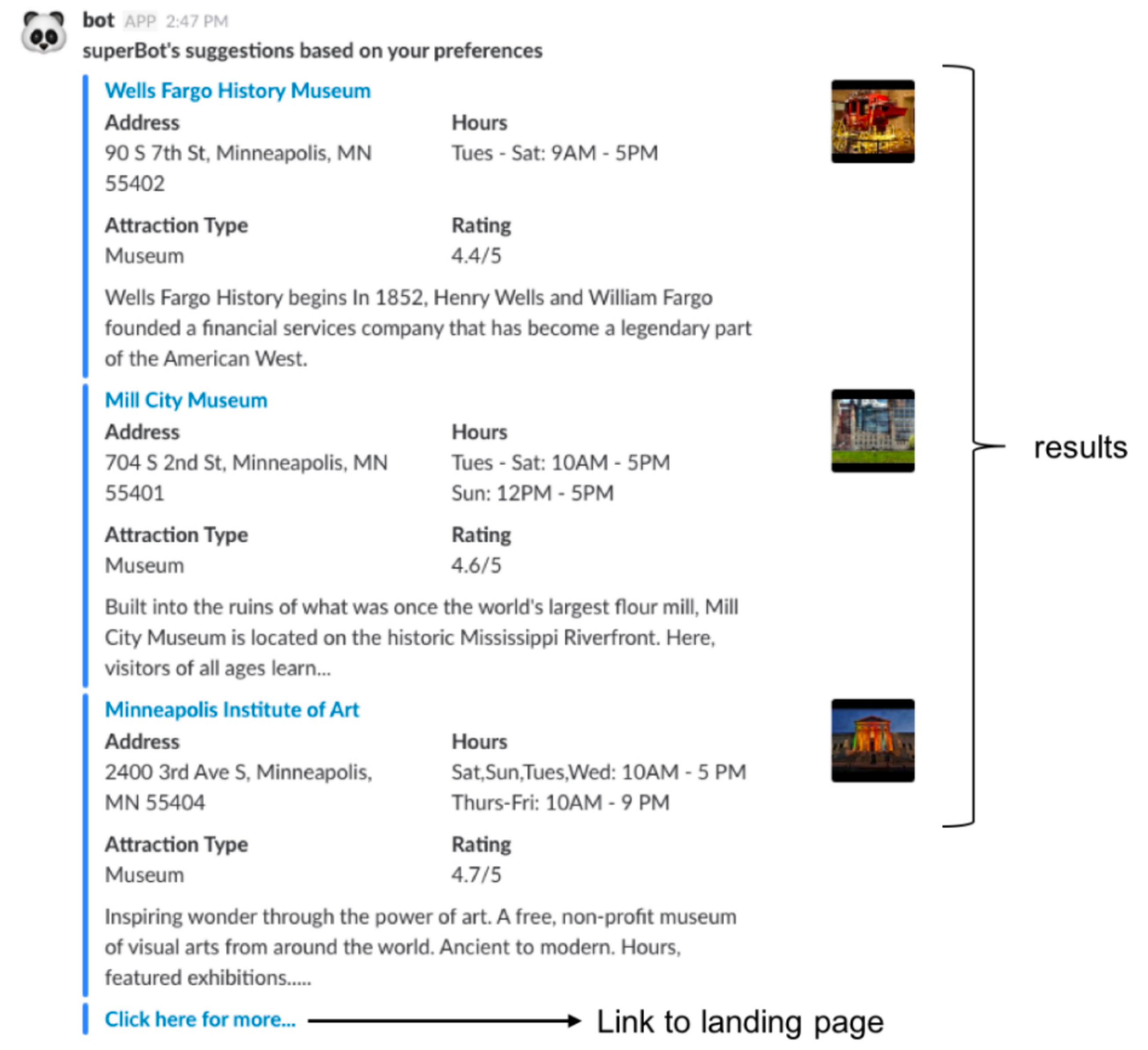}
\caption{A \textit{searchbot} results presentation example inside Slack on a desktop~\citep{Avula:2018:SearchBot}.}
\label{fig:chapter3:searchbot_avula}
\end{figure}

Another alternative for searching through conversational interactions on a desktop was presented by embedding a \textit{searchbot} directly into an existing messaging platform (\ie, Slack) by~\citet{Avula:2018:SearchBot}. 
The searchbot interfered in a collaborative setting (i.e., a search interaction with more than one searcher)
by injecting information relevant to the conversation between the two users. 
An example of a searchbot results page within Slack is presented in Figure~\ref{fig:chapter3:searchbot_avula}. As seen in the figure, the results were always followed by a ``click here for more'' option, redirecting the users to a different SERP.
The results of this study suggest that dynamically injected information can enhance users' collaborative experience. Further research into the presentation of the results in such a collaborative CIS setting is needed to enhance our understanding of optimizing this search experience.

\subsection{Text-Only Result Presentation on Small Screens}
\label{subsec:chapter3:text-only result presentation on small screens}
People interact differently when searching for information on a mobile or desktop device~\citep{Jones:1999:improving, Church:2011:understanding, Ong:2017:SIGIR}.
Researchers have suggested that the shift to mobile search has also been a paradigm shift in web search~\citep{Ong:2017:SIGIR}. Differences in screen size and being able to access search engines in different contexts or ``on-the-go'' have impacted how we search.

With the increasing use of mobile devices such as smartphones, researchers have also investigated the results presentation on different screen sizes~\citep{Ong:2017:SIGIR, Kim:2015:eyetracking}.
Because of the smaller screen sizes on mobile devices, it is important to investigate the result presentation and optimize for the screen real-estate. For example, an average-sized snippet for a desktop site may not be appropriate for a smaller screen since it may involve more scrolling and swiping.

\begin{figure}[htbp]
\centering
\includegraphics[width=.65\textwidth]{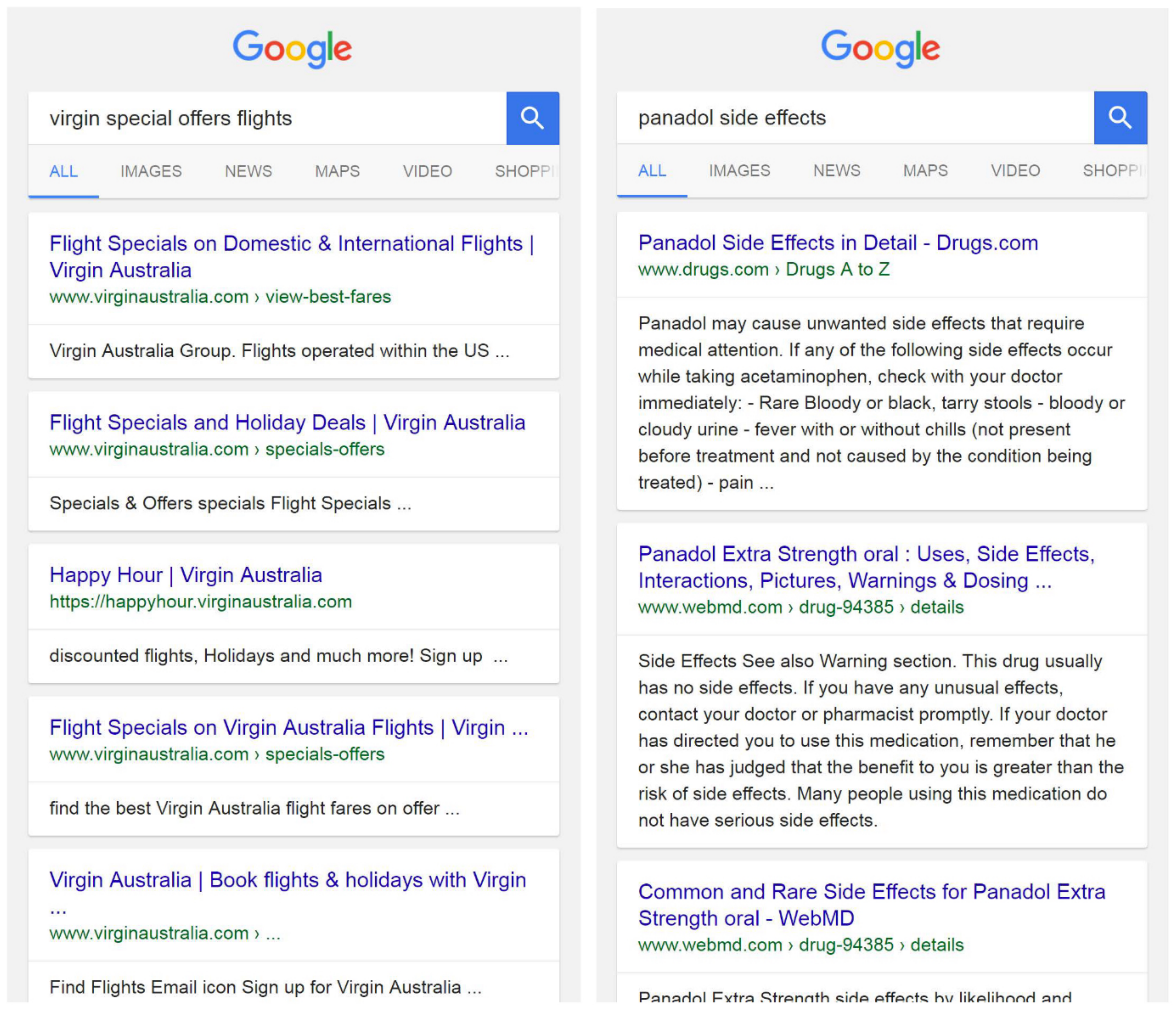}
\caption{Examples of SERPs with short (left) and long (right) snippets by~\citet{Kim:2017:length}.}
\label{fig:chapter3:snippet-length-mobile}
\end{figure}

\citet{Kim:2017:length} studied different snippet lengths on mobile devices. An example of varying snippet length on a small screen is presented in Figure~\ref{fig:chapter3:snippet-length-mobile}.
They demonstrated that participants who were using more extended snippets took longer to search because it took them longer to read the snippets. They suggested that unlike previous work on the effect of snippet length, the extended snippets did not seem that useful for mobile devices and that snippets of two to three lines were most appropriate. Furthermore, it has been suggested that short snippets may provide too little information about the underlying document, which can have an adverse effect on the search performance~\citep{sachse2019influence}. In general, depending on the information need, different snippet lengths could be used to optimize the user experience.

\begin{figure}[htbp]
\centering
\includegraphics[width=.3\textwidth]{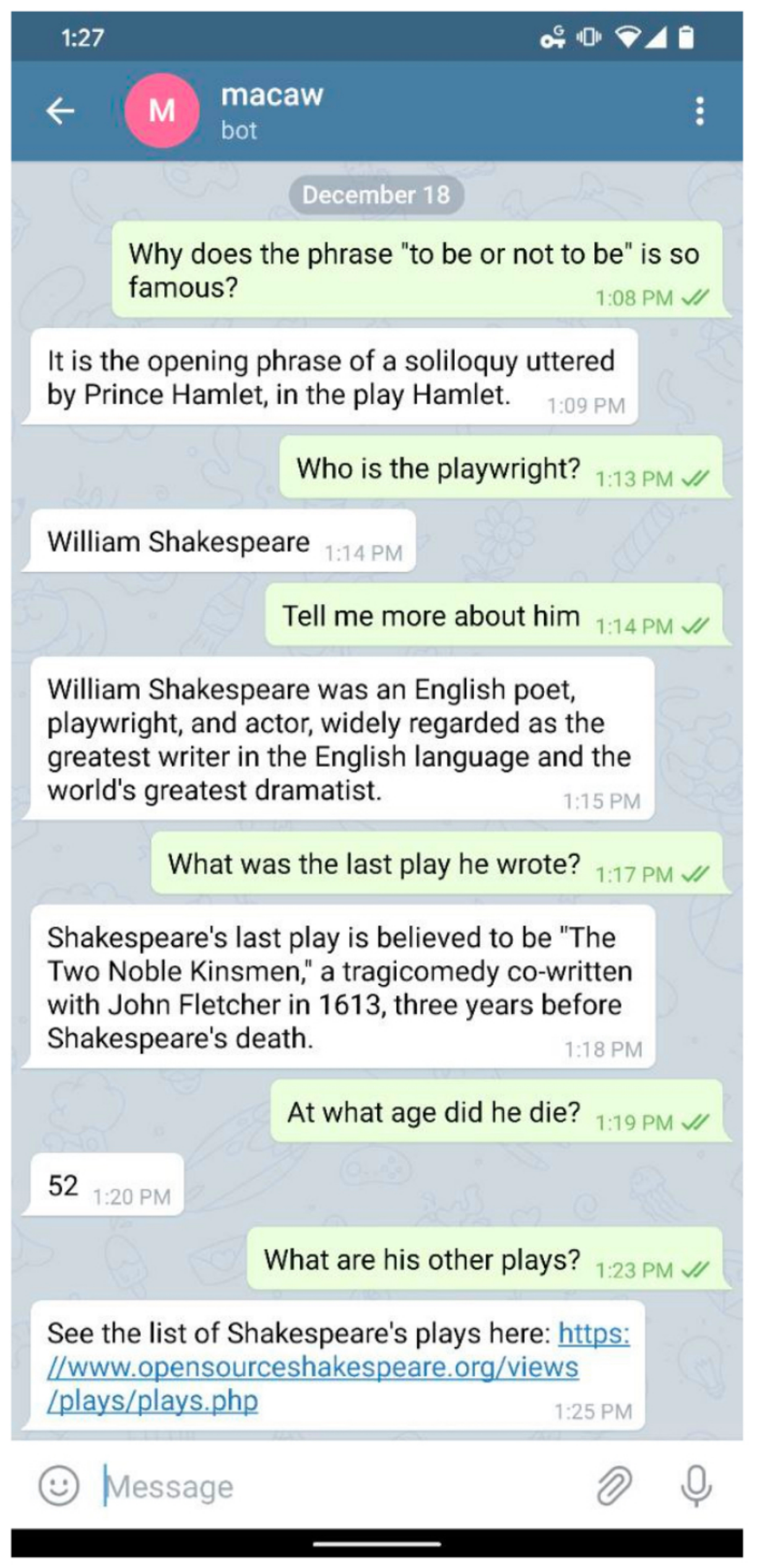}
\caption{Example screenshot of results presentations with Macaw~\citep{Zamani:2020:MACAW} using Telegram.}
\label{fig:chapter3:macaw}
\end{figure}

Even though results presentation has not been fully explored in a CIS context, CIS systems can be developed and deployed on already installed mobile messaging applications such as Telegram (see Figure~\ref{fig:chapter3:macaw})~\citep{Zamani:2020:MACAW}. This means that people are already familiar with the application and it can often be deployed and accessed over multiple devices and platforms. Furthermore, embedding these CIS systems within existing messaging applications means the user does not need to download and install new apps for every service.

However, further research is needed to understand how users interact with information through such messaging applications. For example, little is known about how to display multi-modal information on small screens (\ie, how much information should be displayed versus the trade-off from screen real-estate).

\subsection{Speech-Only Result Presentation}
\label{subsec:chapter3:voice-only result presentation}

Result presentation research has traditionally been focused on visual representation. However, with the ongoing trend of CIS and the improvement of speech recognition, researchers have started investigating how to present results in a speech-only setting.\footnote{We use speech-only, which is the structural act or mechanism to speak. However, some of the studies described use \textit{audio} as a sound or \textit{voice}.} 
It has been suggested that using speech to search is a natural extension of the visual search engines, potentially changing how we access information~\citep{Trippas:2019:Thesis}. However, several researchers have also suggested that simply translating a SERP from a visual to a speech setting is not desirable~\citep{lai2009conversational, Trippas:2019:Thesis, Vtyurina:2020:ICTIR}. 
For instance, \citet{Vtyurina:2020:ICTIR} found that simply translating text results into audio impacts the user experience negatively and requires higher cognition. Thus, it has been suggested to steer away from the ``ten blue link'' paradigm and instead re-think the interactions with search systems. Furthermore, due to the temporal nature of speech, results can be adapted on the fly to change the presentation, thus supporting the user in their information need more actively.

Similarly, as in traditional web search versus searching on smaller screens, it has been suggested that snippet length should be altered depending on the information need. In a study by~\citet{Trippas:2015:sigir}, the preference for summary length was investigated with a crowdsourcing setup. Specifically, they studied the summary length by comparing user preference between text and speech-only results. They observed that users preferred longer, more informative summaries in a text setting, than with audio summaries. Furthermore, different results were observed depending on the query style (single- or multi-faceted): users preferred shorter audios for single-faceted queries, although for more ambiguous queries, this preference was not clear. 

More recent work by~\citet{Vtyurina:2020:ICTIR} also compared results presented over text versus speech.
They used a mixed-methods study with a crowdsourcing and laboratory component, finding that user preferences differ depending on the presentation mode (text or speech). However, they also found that users can still identify relevant results even if presented in a more cognitively demanding speech format. The authors suggested that further improvements to the snippets can help optimize and guide the use of speech-based search interfaces. As part of this study, the authors provided the following presentation guidelines for speech-only results presentation:
        \begin{itemize}
            \item Use prosody to avoid monotone voice
            \item Avoid abbreviations in the spoken results
            \item Avoid truncation of sentences
            \item Avoid repetitive terms in spoken results
        \end{itemize}

Research on using prosody for results presentation was conducted by~\citet{Chuklin:2018:prosody, Chuklin:2019:CLEF}.
They investigated audio manipulation as an alternative to ``highlighting'' or ``bolding'', which is frequently done in a visual interface. They used a crowdsourcing study by modifying speech prosodies such as pitch, pauses, and speech rate in readout snippets. They found that some emphasis features help users identify relevant documents and also increase snippet informativeness.

Many open problems related to the support and guiding of searchers through results presentation exist. For example, 
presentation order bias~\citep{azzopardi:2021:CHIIR, kiesel:2021:CUI}, 
interaction with tabular data~\citep{Zhang:2020:SIGIR},
personas of the conversational system~\citep{Nass:2005:Wired}, 
persuasiveness of synthetic speech~\citep{Dubiel:2020:Persuasive}, 
meta-communication to support communication breakdowns~\citep{Trippas:2018:Informing},
or using non-speech sounds to increase user engagement with search results~\citep{Winters:2019:strategies, arons1997speechskimmer}. 
For example, order bias has been suggested to affect which results summaries receive the most attention from users in a visual setting~\citep{Joachims:2005:SIGIR}. Work has suggested a possible bias towards first and last readout search results depending on the kinds of information need, single- versus multi-faceted~\citep{Trippas:2015:nwsearch}. This example of a \textit{serial-position effect} (\ie, the tendency to recall the first and last items best and the middle items worst) are open problems.

\subsection{Multi-Modal Results Presentation}
\label{subsec:chapter3:multi-modal result presentation}
Past research on CIS primarily focuses on uni-modal interactions and information items. That is, all information is generally either exchanged in text or speech-only format within one turn. However, more recently, researchers have started investigating in more detail the advantages of multi-modal CIS (MMCIS), in which multiple input and output approaches are used~\citep{deldjoo:2021:SIGIR, liao:2021:SIGIR}. Presenting search engine results over a multi-modal channel aims to increase the knowledge transfer of different modalities, enhancing the search experience~\citep{Schaffer:2019:CUI}.

A multi-modal interface can process two or more user input modes in one turn, for instance, speech, images, gestures, or touch~\citep{Furth:2008:multimodal}. Multi-modal systems try to recognize human language, expressions, or behaviors which then can be translated with a recognition-based system. These multi-modal interfaces are often seen as a paradigm shift away from the conventional graphical interface~\citep{Oviatt:2015:Morgan}. 
Similar to a multi-modal dialogue system, MMCIS systems aim to provide completeness to the unimodal counterpart by providing information through multiple modalities~\citep{Firdaus:2021:Aspect}.
Furthermore, the theoretical advantage of these different inputs is that they are very close to human expression and thus are an efficient way of human-computer interaction.
Thus, multi-modal interfaces enable humans to input signals to machines naturally through a mixture of interactions to convey the intended meaning~\citep{Rudnicky:2005:multimodal} and it is often suggested that multi-modality increases the intuitiveness of an interface.

By coupling the intuitiveness of conversations with human conversation, which is inherently multi-modal, the strengths of human communication can be combined, enabling a natural form of information seeking. In addition to the system trying to elicit information from the user to satisfy the information need and perform queries in the background, the system also needs to decide \textit{which}, \textit{what}, \textit{how}, and \textit{when} to present information.

\citet{Rousseau:2006:Framework} created a conceptual model, called WWHT, describing four main concepts of multi-modal information presentation, based on four concepts ``What'', ``Which'', ``How'', and ``Then'':
\begin{itemize}
    \item \textbf{What} is the information to present?
    \item \textbf{Which} modality(ies) should we use to present this information?
    \item \textbf{How} to present the information using this(ese) modality(ies)?
    \item and \textbf{Then}, how to handle the evolution of the resulting presentation?
\end{itemize}

When designing multi-modal CIS interactions, a fundamental problem is the option, combination, or sequence of different outputs of ``displaying'' results. For example, it is logical that relying only on a speech-only result presentation in a loud environment will be undesirable. Instead, using a combination of modalities to present the results in such an environment may be advantageous. Furthermore, as identified and demonstrated by~\citet{deldjoo:2021:SIGIR}, MMCIS, and therefore the information presentation problem, is suitable in the following conditions: 

\begin{itemize}
    \item the person who is searching has \textbf{device(s)} available which allows for more than one interaction mode (multi-device and multi-modal),
    \item when the task’s \textbf{context} is important and can be captured with a device in a suitable modality enhancing personalization,
    \item when task \textbf{complexity} can be supported by the mode of device interaction,
    \item when the results can be returned in an appropriate \textbf{output modality} given the device, context, and complexity.
\end{itemize}

Many open challenges for CIS results presentation in a multi-modal domain exist. Problems include selecting the optimal output modality depending on the context or the user's ability, adapting or changing the output modality to be different from the retrieved modality, or fusing the response to present the results in multiple modalities~\citep{deldjoo:2021:SIGIR}. New tools like Task Multimodal Agent Dialogue (TaskMAD) support wizard-of-oz data collection and experimentation with multiple modalities \citep{Speggiorin:2022:TaskMAD} to support research in these future directions.

\section{Initiative in Conversational Systems}
\label{sec:chapter3:initiative in conversational systems}

The demand to access information rapidly in a natural way has substantially increased due to the proliferation of reliable mobile internet, mobile devices, and conversational systems. Humans create and collect more information than ever before\footnote{\url{https://www.forbes.com/sites/nicolemartin1/2019/08/07/how-much-data-is-collected-every-minute-of-the-day/}} through blog posts, social media, emails, news articles, or videos while using them for education, entertainment, finance decisions, or other decision making~\citep{Zue:2000:IEEE}. In addition, querying this information has become omnipresent, with an estimated 75,000 Google searches per second in 2019.\footnote{\url{https://www.domo.com/learn/infographic/data-never-sleeps-7}} DuckDuckGo, a privacy-focused search engine with an estimated market share of 0.18\% of global searches, received 23.65 billion search queries in 2020,\footnote{\url{https://www.theverge.com/2018/10/12/17967224/duckduckgo-daily-searches-privacy-30-million-2018}} illustrating the scale of search in our daily life.

Furthermore, with the rise of smartphones and mobile internet, we have been accustomed to accessing this information on the go and while multitasking.\footnote{\url{https://www.thinkwithgoogle.com/intl/en-aunz/marketing-strategies/app-and-mobile/device-use-marketer-tips/}} However, accessing information through a small screen and on-screen keyboard while travelling can be cumbersome. Therefore, conversational interfaces in which natural language can be used to interact with information have great promise. Indeed, spoken human language is attractive since it is the most intuitive way of conversation. Furthermore, it is often seen as a very efficient, flexible, and inexpensive means of communication~\citep{Zue:2000:IEEE, Trippas:2019:Thesis}. 
In addition to human language, additional support input can be given through gestures as part of the multi-modal input (see Section~\ref{subsec:chapter3:multi-modal result presentation}).


\begin{figure}[h]
\centering
\includegraphics[width=.7\textwidth]{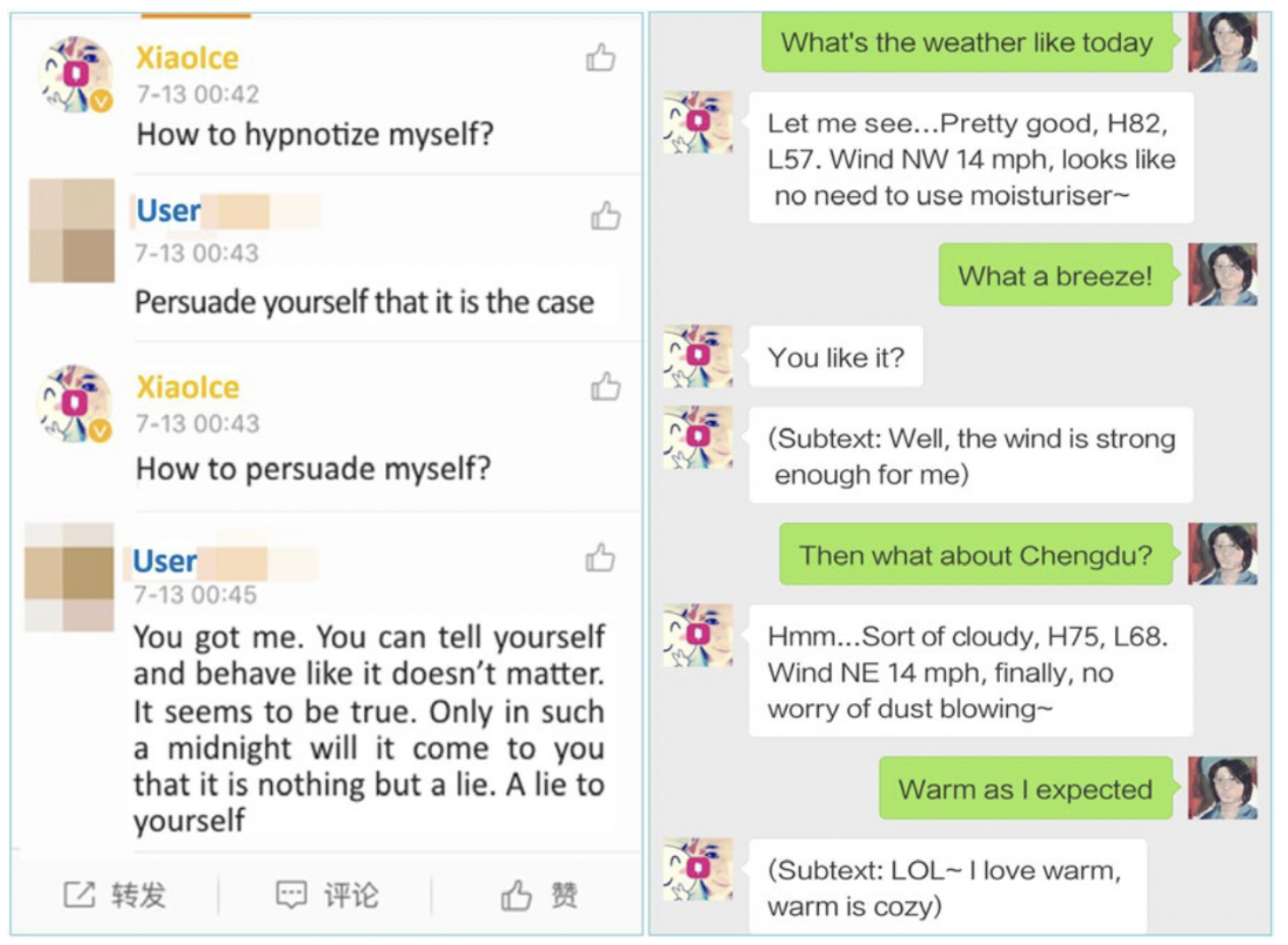}
\caption{Two example chats with XiaoIce~\citep{shum:2018:eliza}.}
\label{fig:chapter3:xiaoice}
\end{figure}

\begin{figure}[htbp]
\centering
\includegraphics[width=.6\textwidth]{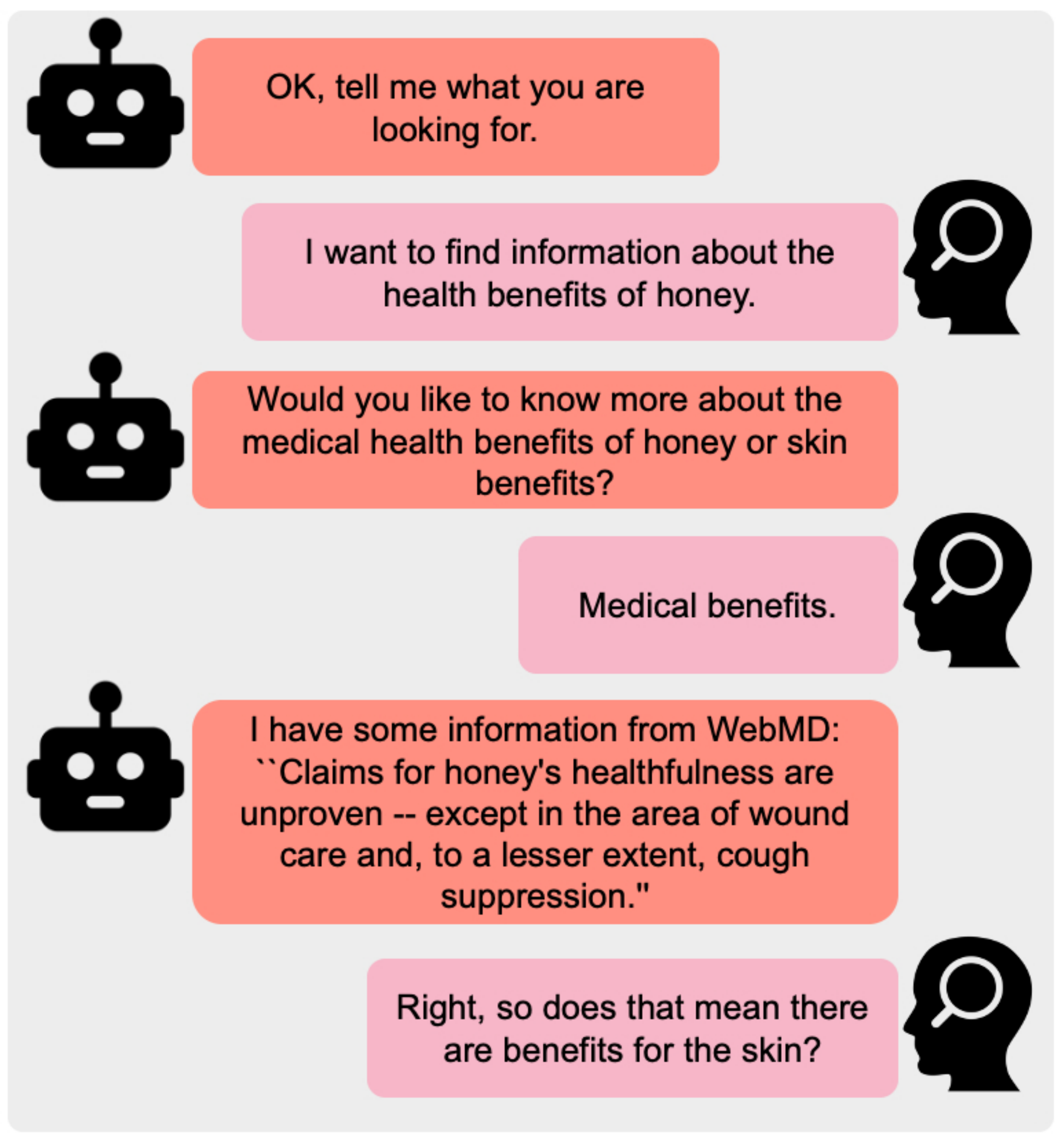}
\caption{Example conversation where the system asks an open-ended question in the opening utterance and a more specific question next.}
\label{fig:chapter3:example3}
\end{figure}

Independent of the kind of conversational interface, these interfaces are often considered from the perspective of \textit{initiative}. That is, to which degree does the system maintain an active role in the conversation~\citep{Zue:2000:IEEE, Mctear:2016:conversational, mctear2017rise}. Three different levels are often used to distinguish these, \ie, system-initiative, mixed-initiative, and user-initiative and are often used interchangeably with levels of control \textit{system}, \textit{user}, \textit{interchangeable}.
With \textit{system-initiative} applications, or system-directed dialogue, the computer takes control over the sequences in the conversation and which information needs to be exchanged. The aim of the system is to elicit information from the user to provide relevant details back to the user. 
This can be done by asking open-ended questions, such as seen in the first utterance in Figure~\ref{fig:chapter3:example3}, in which the system invites the user to provide information and then elicits further details (third utterance).

As seen in the example, in \textit{system-initiative} dialogues, the system takes the initiative to drive the conversation and the user only answers the system's queries. This strategy aims to constrain the user input or request variety, thus making the dialogues more efficient. However, this comes at a cost, with rigid and restricted conversations making the interactions less natural.

In the third user utterance, the user takes control of the dialogue by asking a question, turning the conversational interaction into a \textit{mixed-initiative} dialogue. Hence, both user and system now actively participate in addressing the information need through the interactive conversational paradigm. Thus, \textit{mixed-initiative} dialogues are known for a more natural exchange, however, more advanced ASR and language understanding are needed.

Lastly, \textit{user-initiated}, or user-directed dialogues, are conversations in which the user has complete control and can say anything to the system and the user always has the initiative. This means that the system will only respond to the user's requests. The disadvantage of this approach is that the user may find it challenging to understand the system's capabilities because the system will never suggest anything. Furthermore, dialogues with \textit{user-initiative} may lead to frustration from the user because the system is not a conversational \textit{partner} but rather only replies to requests. 

\section{Interface Limitations in Conversational Systems}
\label{sec:chapter3:interface limitations in conversational systems}

Even though conversational systems can have many advantages, such as enabling users or supporting natural language input, expression of multiple information needs in one turn, cross-platform compatibility and integration, and increasing engagement through personalization, many limitations need to be addressed.

For example, natural language input components, such as ASR and NLU, need to be optimized to handle the huge number of unknown and unexpected user inputs. 
Furthermore, conversational systems need to be optimized to handle \textit{non-explicit} information needs. For example, a user's tone of voice may imply that they want the conversational partner to do something, even though that need was not explicitly stated. Current CIS systems work reasonably well with narrow or factoid queries, however, they still have issues when the information need is more complex (\eg, multi-faceted) or has multiple information needs in one turn.

Besides the limitation of results presentation or output from the system discussed in Section~\ref{sec:chapter3:result presentation}, such as highlighting or bolding keywords, other more general limitations must be considered.
For example, GUIs should be carefully investigated before being directly translated into conversational or voice user interfaces. Even though many chatbots support menu-based interactions within the application, using buttons or menus will limit the benefits of natural language input. Furthermore, issues that already exist in GUIs are now passed on to conversational systems. As such, conversational systems now inherit the GUI experience devaluing the natural language advantage.

In addition to these existing output difficulties, speech-only conversational systems have distinct challenges. For example, simply reading out textual components or reading out lists has shown to be ineffective~\citep{Trippas:2015:sigir, Vtyurina:2020:ICTIR, Sahib:2015:Evaluating}. Indeed, the serial and transient nature of audio can challenge the users' ability to recall all information presented~\citep{dubiel:2020:conversational}. This exacerbates the difficulty of skimming audio and makes it challenging to present results while not overwhelming the user with information nor leaving them uncertain as to whether they have covered the information space~\citep{Trippas:2015:sigir, Trippas:2019:Thesis}. These CIS systems cannot maintain a lengthy information exchange or keep sufficient track of the context.
In addition, images and graphs are more challenging to be displayed and may need to be interpreted by the system to inform the user what is displayed~\citep{Trippas:2019:CHIIR}. Other limitations, such as the tone of voice or persona of the system interacting with existing stereotypes or biases of humans speaking in particular ways may plausibly both reinforce existing biases as well as cause systems to be perceived in particular ways \citep{nag:2020:gender}

Considerations must also be made for limitations of automatic speech recognition (ASR). For example, users' speech-input may include disfluencies or errors. Users may mispronounce words, use filler words such as "uhm" or "ah", or add extra pauses. They may also use words from other languages or made-up words and phrases (\eg, a made-up name for a personal music playlist). Furthermore, different speech variabilities such as patterns, dialects, age, gender, or speech impairments can impact ASR performance. For example, speaking faster or slower can have an impact on the acoustic models used for transcriptions~\citep{Benzeghiba:2007:Automatic}. Indeed, an apparent challenge for conversational systems is the barrier to recognize speech from a diverse population~\citep{Zue:2000:IEEE}. To make information more accessible and enable wide adaptation of conversational systems, including by people with cognitive or physical impairments is needed~\citep{Baldauf:2018:exploring, Derboven:2014:designing}. Beyond this, there has been very limited published work on the design of speech-only systems to consider users who are either hard of hearing or vision impaired.

\section{Summary}
\label{sec:chapter3:summary}
This section covered conversational interfaces, results presentation, different kinds of initiative in conversational systems, and interface limitations. 
We explained some of the essential building blocks for conversational interfaces and presented the central features such as natural language, conversational style, mixed-initiative, and context tracking.
This section also provided a historical context on existing conversational interfaces (\ie, SDS, VUIs, live chat support, and chatbots) and their application. We highlighted the differences between these various types of conversational interfaces with respect to CIS.
Since information presentation is key to transferring the found information to the users for CIS, we provided an overview of information presentation research. We described text-only presentations on desktops and small screens, speech-only presentations, and multi-modal presentations. In each part, we highlighted the limited research on \textit{how} and \textit{what} to present with interactions in CIS systems.
Lastly, we explained key elements regarding initiative in conversations and how different initiatives create different conversational styles, followed by a brief summary of limitations in conversational interfaces.

To conclude, different interfaces can support people in their information seeking journey differently. Thus, CIS systems can benefit from different interfaces, and models and evaluation methodologies for handling multiple interfaces will be needed. Understanding the impact of interactivity in CIS is crucial to make user-friendly systems. Different ways of understanding these conversational interactions are discussed in the next section.

\chapter{Understanding Conversational Interactions}
\label{chapter4}

A distinctive property of CIS systems is the ability of the systems to understand multi-turn interactions expressed in the form of natural language. This section discusses the varying approaches to conversational understanding required to follow the conversational structure, with a focus on information seeking conversations.

We give an example of a possible CIS conversation taken from the TREC Conversational Assistance Track (CAsT) Year 3 \citep{Dalton:2021:TREC} -- topic number 113. This illustrates coreference issues, dependence on previous user and system turns, and explicit feedback with topic-switching behavior. 

\begin{UserSystemUtterance}
    \item[User:] How do genes work?
    \item[System:] A gene is a short piece of DNA. Genes tell the body how to build specific proteins. There are about 20,000 genes in each cell of the human body. Together, they make up the blueprint for the human body and how it works. A person's genetic makeup is called a genotype... Many personal characteristics, such as height, are determined by more than 1 gene. However, some diseases, such as sickle cell anemia, can be caused by a change in a single gene. 
    \item[User:] What others are caused by a single change?
    \item[System:] The four types of (human) Genetic diseases are: (1) Single-gene/monogenic Genetic Diseases: In this category the starting point is a mutation/change in one gene... Almost 6000 single gene disorders are known and it is estimated that 1 of 200 newborns face a single gene genetic disorder... 
    \item[User:] What are the other types of diseases?
    \item[System:] (3) Chromosomal Genetic Diseases: Chromosomes are big DNA molecules composed from genes... Abnormalities in the structure, number (and not only) of the chromosomes can cause some of the most dangerous genetic disorders...
    \item[User:] You missed the second type.  What was that?
    \item[System:] A large genetic mistake typically occurs in the woman's egg, which may partially explain why older women are more likely to have babies with Down syndrome...
    \item[User:] That’s not what I wanted. How about recent developments in gene therapy to treat those defects?
    \item[System:]...
\end{UserSystemUtterance}

As illustrated above, CIS systems must ``understand'' a natural language utterance and represent the state of an information dialogue as it evolves. Second, they need to track the evolution of that state across turns and relate them to one another including detecting feedback and sub-topic switches. The fundamental mechanisms and representations for conversation understanding vary widely. The architecture may have an $implicit$ internal representation based upon the model's input (\eg, attention-based models that encode the conversation history). The model may also have an $explicit$ representation of the salient state that could include structured representations (entity or schema), lexical (word-based), or vector-based. 

This section begins by discussing the representation used within a single turn. It then discusses conversation modeling and history tracking for information and how this differs from previous work in dialogue state tracking. Next, it covers models of conversational discourse and discusses approaches that identify turn salience, extract contextual keywords, and construct vector representations. It provides an overview of core conversational tracking sub-tasks including (1) entity recognition, (2) query expansion, (3) salient term selection, and (4) conversational query rewriting (CQR). It concludes with a discussion of how these approaches continue to evolve beyond short conversations towards longer and multi-session conversations.



\section{Modeling within Turn State}
This subsection introduces the building block for multi-turn conversations --- the representation of the state for a single turn. Because CIS systems operate in an open-domain environment, they do not often use predefined domain state (frame) ontologies. At its most basic level, the state representation includes the utterance $text$; whether it is typed or from automatic voice transcription. The state representation for a single turn in CIS is contextualized with the history with implicit or explicit relationships between turns and concepts in the conversation. 

A widely adopted approach to conversational representation uses pre-trained language models with contextualized embeddings, particularly Transformer-based models \citep{Vaswani:2017:Transformers,Raffel:2020:jmlr}. These exhibit transfer learning capabilities that allow them to be fine-tuned for one or more conversational ranking or question answering (QA) tasks. For conversations, utterances may be encoded separately, compared, and possibly combined in a dense embedding space \citep{Khattab:2020:ColbertQA,Xiong:2021:ANCE,Prakash:2021:RANCE}.

Some early systems use $explicit$ structured annotations of the utterances from the output of an NLP system: part of speech information, dependency parse, semantic frame parsing (\eg, FrameNet~\citep{Baker:1998:FrameNet}), entity recognition and linking, semantic parsing to a logical representation, and others. However, pre-trained language models demonstrate key elements of these NLP pipelines including coreference resolution, entity recognition, and relations \citep{Tenney:2019:ACL}. As a result, approaches in many leading CIS benchmarks (\eg, CAsT \citep{Dalton:2019:TREC} and QuAC \citep{Choi:2018:emnlp}) do not explicitly use the output from an NLP system, but instead, rely on the models to handle these tasks implicitly.

Because of these advances, modern CIS does not often focus on explicit structured state tracking. Widely used CIS datasets do not contain labeled annotations of ground-truth conversational state, except in the form of manually disambiguated utterances to resolve phenomena like coreference. The focus is then on generating these automatically via tasks such as query rewriting. Currently, instead of component-wise evaluation of understanding elements the primary evaluation of effectiveness is primarily on extrinsic effectiveness in the overall end-to-end retrieval task. 

\tipbox{The key differentiating element for CIS compared with single-turn information seeking is the type of interaction and discourse structure.}

There are various proposed models of conversational structure in the literature. Structure in a conversation builds on the actions of the participants, namely the speech or dialogue acts. A common task is `dialogue act recognition' to label the utterances with the type of interaction \citep{Bunt:2017:DialogueAA} (\eg, INFORM, REQUEST, GREETING) that encodes how the current turn relates to previous ones explicitly. The definition of these act types and their usage varies widely. 

One model developed specifically for CIS by \citet{Azzopardi:2018:ConceptualizingAI} presents a model of conversational search evolution and includes a taxonomy of the user and system action space. A conversational information need evolves with a single turn being the Current Information Need (CIN), past turns with results as Past Information Needs (PINs), and an agent's model of the information space including a model of varying trajectories with Alternative Information Needs (AINs). The action space includes rich types of both user and system revealment of varying forms. The work of \citet{Ren:2021:SIGIR} refine this with a focus on conversational interaction with existing search engines, including explicit user intents (such as reveal, revise, chit-chat) and system actions (suggest clarifications, show results, chit-chat, \etc). 

Many of the current benchmark datasets have simplistic discourse with the user asking questions and the system returning answers of varying types. For example, the widely used QuAC \citep{Reddy:2019:tacl} conversational QA dataset contains three categories of dialogue act annotations for each turn, (1) continuation (follow up, maybe follow up, or don’t follow up), (2) affirmation (yes, no, or neither), and (3) answerability (answerable or no answer). Later developments to tackle challenges in this area include richer types of user revealment, feedback, and others \citep{Dalton:2021:TREC}. 
\tipbox{
The action space and intents vary widely according to the task and interface constraints. What makes CIS distinctive is the unique focus on satisfying a user's information need that may encompass short answers, long answers, and other rich types of interactions.
}

\section{Modeling Conversation History and Tracking State}
\label{sec:chapter4:Topic Tracking in Conversation}

Understanding a conversation is primarily concerned with organizing how a series of turns relate to one another. The relationships in CIS differ from previous work in search systems in that they often exhibit natural language phenomena that span turns -- \textit{coreference} (two or more expressions referring to the same thing) and \textit{ellipsis} (omitting words or topics implied by the context). It also requires handling informal language use and implicature. \citet{Dalton:2020:sigir} looked at the use of coreference of varying kinds -- anaphora, zero-anaphora (omission), and others. They find that compared with traditional NLP corpora (such as OntoNotes and CoNLL coreference) conversation information seeking has a higher rate of ellipsis and zero-anaphora, which are extremely rare in narrative text. More recently, \citet{Radlinski:2022:Subjectivity} looked at subjective language more broadly, arguing for different forms of subjective language requiring different treatment.

Informal conversational phenomena also include interpreting indirect answers in context \citep{Louis:2020:Indirect}. An example is: ``Would you like to get some dinner together?'' with a reply, ``I’d like to try the new Sushi place.'', which is an implicit affirmative that indirectly implies an answer. For voice-based applications, they must also handle noise because of disfluency removal and the noisy channel from speech-to-text transcription \citep{HassanAwadallah:2015:cikm}.

The use of an explicit structured state is widely adopted by task-oriented dialogue systems. Frame-based approaches model the dialogue state with structured domain-specific schemas that have intents (actions) and typed slots with values. Keeping track of this evolving state is a standard task, Dialogue State Tracking (DST), with long-running benchmarks in the Dialogue State Technology Challenge (DSTC) \citep{Williams:2016:DSTC}. These systems often support a fixed number of pre-defined domains with schemas; the widely used MultiWoz dataset  \citep{Budzianowski:2018:Multiwoz} has an ontology with twenty-five slots spanning seven domains. The largest, the Schema Guide Dialogue dataset \citep{Rastogi:2020:DSTC8} contains sixteen domains with an average of five intents per domain and 214 slots (2.5 per intent on average). In contrast, CIS most systems typically do not have pre-defined domains, intents, or slot representations. 

\tipbox{In contrast to task-oriented dialogue systems, CIS systems typically do not have pre-defined domains, intents, or slot representations.}

One exception to this is a proposed frame-like model that builds a structured representation (SR) of a turn with context entities, question entities, predicates, and expected answer types \citep{Christmann:2022:SIGIR}. Unlike the structured schemas from DST, these are loosely defined text values. The state of SRs evolves through a conversational flow graph. 

\section{Modeling Conversation Discourse}
\label{sec:chapter4:Modeling Conversation Discourse}
There are varying approaches to modeling the evolution of a conversational topic across turns. These leverage the natural language phenomena used in conversational dialogue. Automatic approaches look for topic shifts based on changes in coreferent mentions, shared noun phrases, and common patterns \citep{Mele:2020:SIGIR}. The realism of the conversational discourse varies widely among the conversational corpora based on their creation methodology. The TREC CAsT topics are inspired by informational sessions in web search \citep{Dalton:2020:sigir} but are also engineered to be challenging for trivial reformulation systems. Other datasets such as SQuAD and CoQA are derived from artificially created information needs. The widely used QuAC dataset is limited to discussing an information need about a single person with a bias towards entertainers \citep{Choi:2018:emnlp}. The result is that the discourse of conversations vary based on the type of information, the topic being discussed, the user task, and the modalities supported for interaction. Most of the aforementioned datasets, including TREC CAsT (2019), assume that the sequence of questions is fixed and is independent of the system's response, which is different from real interactions. Further, they assume that the only action the system can take is answering the questions and do not support mixed-initiative interactions where the system make take other actions. This is changing, with increased result dependence in CAsT 2021 \citep{Dalton:2021:TREC} and mixed-initiative sub-tasks in 2022 \citep{Dalton:2022:TREC}. It represents part of the larger trend towards greater dependence on previous system responses as well as richer types of system responses. 

In the following subsections, we discuss the evolution of approaches to modeling conversational history including how it is represented. We then break down history understanding sub-tasks and discuss each. Finally, we conclude by looking towards the evolution of conversations to longer, more complex tasks and information needs. 

\subsection{History Models}
Modeling the conversation history requires determining relevant parts of the history, how the history is encoded, and how the encoding is leveraged by the model. Approaches to modeling state across turns vary. 


A simple and widely used approach to modeling history is the simple heuristic to concatenate the \textit{last-$k$} turns.  The approaches vary in length and type of context appended. One example of this approach uses approximately the previous two turns \citep{Ohsugi:2019:ASB} -- the previous user utterances, system responses, or one of those two.  For conversational QA, a popular approach is to only append the previous answer as context \citep{Choi:2018:emnlp}. Similar heuristics that append the first turn and previous turn(s) of a conversation were also used in the first year of TREC CAsT \citep{Dalton:2019:TREC}. 

The most important feature in modeling history is the positional relationship between turns to capture common patterns of conversational discourse. In particular, in current datasets, most references refer to immediate or short-term contexts \citep{Chiang:2020:aaai}. \citet{Qu:2019:sigir} append encodings of the history but do not model position explicitly. Multiple threads of work improved on this by adding the relative position of previous answers \citep{Qu:2019:CIKM,Chen:2020:GraphFlow}. Beyond position, \citet{Qu:2019:CIKM} adds a History Attention Module that takes the encoded representations of sequences or tokens and learns the importance of the representations to the current answer. Analysis shows that models appear to be relying heavily on positional understanding more than on textual semantic relationships \citep{Chiang:2020:aaai}. 

A challenge for history models is that many of the existing benchmarks only exhibit simple discourse with strong local positional bias. As shown in CAsT, most dependencies are local, on directly preceding turns \citep{Dalton:2020:TREC}. This is evolving as CIS systems become more capable, with non-local dependencies increasing from 12\% of the turns in CAsT 2020 to 22\% in CAsT 2021.

\tipbox{Improved models of conversation history is an area for future work, particularly for long and complex conversations where appending short-term history does not adequately model the discourse. Behavioral analyses of existing models show that they rely heavily on short-term distance cues rather than deeper understanding.} 

\subsection{History Representation}
As mentioned above, a simple approach to model representation is to concatenate relevant turns to history in the order they appear. This creates an $explicit$ text-based representation for downstream tasks including query expansion and rewriting.  

\jeff{This section could be clearer and would benefit from using example figures from the papers similar to the tutorial.}
This may be performed implicitly through the latent state from a sequence model. Recurrent networks (such as LSTMs \citep{hochreiter:1997:lstm}) encode long-term conversation history dependencies via latent hidden states \citep{Yang:2017:neuir}. More recent neural language models based on Transformer architectures \citep{Vaswani:2017:Transformers}, \eg, BERT \citep{Devlin:2019:BERT}, use attention to latently encode relationships. A key consideration is how to encode turn structure (for example using separators) to indicate boundaries between previous user and system responses \citep{Reddy:2019:tacl, Qu:2020:SIGIR}. This may also be done in the model as in \citet{Qu:2019:sigir} using a separate embedding indicator to determine if an utterance is a part of a question (user) or answer (system). \citet{Chiang:2020:aaai} use a special word embedding indicator if a token is used in a previous answer. 

\citet{Gekhman:2022:TACL} extend the separator approach by modifying the input with prompt-based separators with positions. They study and compare this approach that modifies the input text with symbols compared with the other widely used approaches that modify the embedding layer for conversational question answering. They find that the simple prompt-based approach is more effective with new language models. Another key difference from previous work is that they append the answers in most recent first order in addition to explicit prompt labels for order. Future work might explore the impact of these positional modeling decisions further.

A separate vein of research creates an $explicit$ history model with a mechanism to integrate the representation. The FlowQA approach \citep{Huang:2019:iclr} introduces the concept of \textit{Flow} which generates an explicit latent representation of the previous context.  Modeling the conversation was subsequently evolved in the context of graph networks to model the flow of information as a graph using recurrent neural networks \citep{Chen:2020:GraphFlow}. \citet{Yeh:2019:FlowDelta} extended this to a Transformer architecture and makes the context flow dependence explicit. 

Following recent trends in retrieval approaches, the adoption of approximate nearest neighbor search applied to learned dense representations, also known as dense retrieval, and/or sparse representations resulted in a significant shift. In these representations, the query and history are combined into one or more vectors issued as queries to a dense retrieval system. \citet{Yu:2021:ConvDR} encoded the history representation with a dense vector that is \textit{learned} with a teacher-student model to mimic a dense representation of the manually rewritten query. The model for multiple turns uses composition with dense retrieval approaches similar to those in multi-hop QA \citep{Khattab:2021:Baleen}, but applied to a conversational context. The results from~\citet{Dalton:2021:TREC} include these as a baseline, and they are widely adopted by many of the top-performing teams in TREC CAsT \citep{Dalton:2022:TREC}. Most of the CIS systems, although still using a dense vector representation adopt simplistic history heuristics to create one or more representations that may also leverage words combined via fusion. Further, to maximize effectiveness most current models require an explicit text representation for further reranking (as discussed in the next section), although this is starting to change with effective learned sparse representations like Splade \citep{Formal:2021:SPLADE} being used in CAsT '22 \citep{Dalton:2022:TREC}. 

\section{Conversational Language Understanding Tasks}
\label{sec:chapter4:History Understanding Tasks}

Given a representation of conversation a key consideration is how to use elements of the history in the current turn to retrieve relevant information. There are varying and complementary approaches to address this problem. The tasks include unsupervised or supervised query expansion, generative query rewriting, identifying and tracking concepts and entities, identifying salient turns, and extractive or abstractive summarization.  

\subsection{Turn Salience}
This task involves explicitly modeling the relationship of turns in a dialogue to determine their relevance and relationship to the current turn. The CAsTUR dataset created by \citet{Aliannejadi:2020:chiir} adds turn salience data to the TREC CAsT 2019 dataset \citep{Dalton:2019:TREC}. The authors performed a detailed analysis of the dependencies. The resulting relation labels were used to train classifiers of turn salience \citep{Kumar:2020:EMNLP}. We note that subsequent iterations of CAsT in 2020 and 2021 \citep{Dalton:2020:TREC} include explicit dependence annotation labels by the topic creators with labels on the dependence on previous user utterances as well as previous results. The content of relevant turns can be used directly for multiple tasks including expansion and rewriting.

\subsection{Query Expansion}
In this section we discuss Conversational Query Expansion (CQE) including both unsupervised and supervised approaches. These augment the representation of the current turn with additional information from previous turns using a form of pseudo-relevance feedback \citep{Yang:2019:TREC,Mele:2020:SIGIR,Hashemi:2020:GT}.

\subsubsection{Unsupervised Approaches}
Work in this area started with heuristic approaches and unsupervised models. In TREC CAsT a simple heuristic baseline expansion approach was to expand with the first and previous turn in the conversation \citep{Clarke:2019:TREC}. These turns often represent an overall topic and the most recent (and therefore likely relevant) previous information. A mixture of feedback models \citep{Diaz:2006:sigir} can be used to combine feedback models across turns. However, these simple approaches are less effective when there is a sub-topic shift or there are non-relevant turns.

The expansion unit used varies, with some only using previous user turns and others using both user turns and system turns. The HQExp model proposed by \citet{Yang:2019:TREC} does both and leverages the combination of scores from a BERT model across past turns. This is an important model because it uses rules, but includes a model of topic shifts as well as query performance prediction.

Going beyond individual turns, some expansion approaches build a model explicit graphs and word networks that evolve. The Conversational Reasoning Over Word Networks (CROWN) \citep{Kaiser:2020:conversational} model is an unsupervised method for propagating relationships across turns based upon a network of words related by mutual information.

\subsubsection{Supervised Approaches}
Later work framed the task of expansion as a summarization task -- extractive or abstractive. These use supervised models to select or generate terms for use in query expansion. The Query Resolution by Term Classification (QuReTeC) model proposed by \citet{Voskarides:2020:SIGIR} models the task as a binary term classification, effectively performing term-level extractive summarization. In parallel and similar work, the Conversational Term Selection (CVT) method by \citet{Kumar:2020:EMNLP} frames the problem as a term extraction task but further applies the same extraction to pseudo-relevant results. These methods extend previous methods that extract key concepts from long verbose queries for web search \citep{Bendersky:2008:sigir} to a conversational language understanding task. 

The overall utility of both unsupervised and supervised expansion approaches is mixed, with many of the expansion approaches being outperformed by rewriting approaches \citep{Dalton:2019:TREC, Dalton:2020:TREC}, but turn and term salience is often complementary and a key part of an overall end-to-end effective system. 

\subsection{Conversational Query Rewriting}
Given a query and a dialogue history, the goal of Conversational Query Rewriting (CQR) is to generate a new query that contains the relevant context needed to rank relevant content in a single unambiguous representation. In a pipeline system with multiple passes of retrieval, this step is critical because it determines the effectiveness of both candidate passage retrieval as well as subsequent re-ranking. 

A widely adopted approach is to model the task as a sequence-to-sequence task~\citep{Sutskever:2014:seq2seq}. The task-oriented dialogue systems community used pointer-generator networks and multi-task learning to rewrite turns but they are limited to a handful of task domains \citep{Rastogi:2019:naacl}.  This approach rapidly evolved with pre-trained language models based on Transformer architectures \citep{Vaswani:2017:Transformers} and with evaluations on a Chinese dialogue dataset \citep{Su:2019:acl}. They showed that these architectures implicitly solve coreference resolution more effectively for the target task than previous state-of-the-art coreference models. 

Subsequent work by \citet{Vakulenko:2021:QuestionRe} in the TREC CAsT 2019 benchmark \citep{Dalton:2019:TREC} demonstrated the effectiveness of pre-trained models based on GPT-2, resulting in one of the top three best-performing automatic runs in that year. Subsequent work showed the model can generalize with relatively few examples, particularly when combined with weak supervision based on rules to handle omission and coreference \citep{Yu:2020:sigir}. Improvements continued to evolve by training the models with additional data spanning both CAsT and conversational QA datasets \citep{Elgohary:2019:EMNLP, Vakulenko:2021:QuestionRe}.


Improvements in this area continue with newer generations of sequence-to-sequence models (\eg, T5 \citep{Raffel:2020:jmlr}) based on larger corpora, increased model size, and refined objectives.  Additionally, recent work from the dialogue community \citep{Henderson:2020:EMNLP, Mehri:2020:DialoGLUE} demonstrated that pre-training and fine-tuning on conversational data provides significant gains for both task-oriented and chit-chat models over models pre-trained on general corpora only. It is common to train on public large-scale social media data, such as Reddit (heavily filtered), because of its size and diverse informal language \citep{Henderson:2020:EMNLP, Roller:2020:OpenChatbot}. 

CQR remains a active research area because high quality rewrites are more effective with current neural rankers trained on web search. There remains a significant gap, 20-40\% in CAsT between manual and automatic queries \citep{Dalton:2021:TREC,Dalton:2022:TREC}. An area for future work in these models is to handle the rich forms of discourse, like user clarification or feedback. 

CQR models track conversation state implicitly resulting in resolving ambiguity and missing information in a generative approach.  In contrast, an alternative model is to explicitly detect and track concepts as they evolve in a conversation.

\subsection{Entity Detection and Linking}
Tracking the evolution of concepts and entities used in a conversation explicitly is the task of Conversational Entity Detection and Linking. This includes tracking coreferent mentions, but also other forms of concept evolution. Due to the informal and personal nature of conversational discourse this task can be quite challenging.

Work on tracking entities across turns first appears in multi-turn factoid QA at TREC 2004 \citep{Voorhees:2004:TREC}. This evolved with the introduction of `related' questions that included anaphora and TREC 2005 with dependence on previous factoid responses \citep{Voorhees:2005:TREC}. A related line of research uses search session history to improve named entity recognition effectiveness in queries \citep{Du:2010:SIGIR}.  Approaches grounded in concepts and entities were widely used by Alexa Prize socialbot systems \citep{Ram:2018:Alexa} that allowed them to track topics and shifts across turns in similar ways as CIS systems.  

A key consideration for these systems is that they need to be able to identify general concepts, commonly referred to as Wikification \citep{Cucerzan:2007:EMNLP}. \citet{Joko:2021:SIGIR} studied the effect of entity linking on conversational queries and found that existing linking systems perform significantly worse on conversations than on other types of text. In follow-up work, they extend the REL entity linking system to conversations creating CREL \citep{Joko:2022:SIGIR}. To evaluate the model they create ConEL-2, an extension of the Wizard-of-Wikipedia dialogue dataset to add annotations of personal entity mentions and links. As systems and benchmarks evolve, we expect the importance of this area to grow and to address issues like personal bias \citep{Gerritse:2020:ICTIR}.

\section{Long and Multi-Session Conversations}
\label{sec:chapter4:Modeling Longer Conversations}

Many of the existing approaches discussed in the previous focus on a single relatively short conversation. For example, years one and two of TREC CAsT averaged between 8-10 turns \citep{Dalton:2019:TREC, Dalton:2020:TREC}, QuAC has fewer, with approximately 7 turns \citep{Choi:2018:emnlp}. A common heuristic based on the observation that many dependencies are local is to use only the three previous turns \citep{Mehri:2020:DialoGLUE}.

As conversations become longer, simple methods for modeling conversations break down. While there are new model variants for longer sequences \eg, Conformer~\citep{Mitra:2021:Conformer} and Longformer~\citep{Beltagy:2020:Longformer}, many widely used neural models, including those used for conversational query rewriting or term salience prediction are only capable of encoding a limited (short) context of a few hundred tokens. To address this, approaches that select conversational context to use, a task referred to as sentence selection in a dialogue \citep{Dinan:2019:convai2}. 

\subsection{Long Answer Dependence}
Another dimension of modeling conversations is understanding long responses. Much of the previous related work focused on tracking and reformulation mostly based on previous utterances (queries) with only limited result interaction \citep{Dalton:2019:TREC}. The structure of previous conversational QA tasks had limited reliance on immediate previous results \citep{Choi:2018:emnlp,Voorhees:2005:TREC}. This is because the responses given by the system are short factoid responses.   

\tipbox{In contrast to prior work on ConvQA with factoid responses, the broader scope of CIS systems has richer questions that require varied length responses. These may be one or more passages, a document, a multi-source summary, or even an entire search engine results page. These long answers make the overall conversational history much longer than typical chit-chat dialogues.}

Interactions with long results in later turns make language understanding significantly more challenging for CIS systems. They need to be able to understand references across a longer distance in more complex discourse. This remains a challenging area, where current neural approaches for conversation understanding struggle \citep{Dalton:2021:TREC}.

\subsection{Turn Retrieval and Recommendation}
\jeff{This is a bit awkwardly positioned; turn salience was earlier. Now we're doing retrieval much later.}
Similar to previously discussed work on turn salience, an alternative approach is to model finding relevant information from previous history as a ranking rather than classification task. Previous turns and responses are ranked for relevance to the current turn using the same sparse, dense, or neural ranking models \citep{Humeau:2020:ICLR} used in response ranking.  The evidence from previous turns may be encoded independently (or concatenated) \citep{Izacard:2020:FusionInDecoder} or fused \citep{Xiong:2020:Multi-hop} before being used in the target generative task.

Blenderbot from \citet{Xu:2021:goldfish} retrieve turns from past conversational sessions for additional historical context. The model retrieves sessions as a document and uses these as the context in the generation process. 

There are also clear connections to classic recommendation tasks here. Recommender systems often encode rich long-term sequences of interactions (which may be considered a ``conversation'') in a user model that is meant to summarize this sequence of interactions. Recent work has advocated representing such knowledge about users' needs in natural language \citep{Radlinski:2022:Scrutability}.

\jeff{add work here on the conversational knowledge FID model}

Finally, a possible area for future work might be to create summaries of turns or conversations, similar existing work on text compression \citep{Rae:2020:CompressiveTF}. 

\section{Summary}
This section reviewed conversational state-tracking approaches and models. We examined the fundamentals of modeling intra-turn states including vector representations, entity representations, and discourse classification. We discussed approaches to model conversational history and differentiating features of conversational search as contrasted with voice search or traditional text narratives, with a key differentiating feature being the wider use of implied context including indirect answers, zero-anaphora, and ellipsis.  We discussed long-range history models with many current approaches using a static window of context (last few turns) as well as dynamic turn salience, or attention-based models. Within this history, we examined key sub-tasks: entity recognition and linking, query expansion, query summarization, and query rewriting. The best-performing approach leverages multiple diverse techniques: rewriting, expansion, and reranking in a multi-stage pipeline \citep{Lin:2020:QueryRU}. An approach based upon both early and late fusion of multiple expansions and rewrites across both retrieval and reranking is currently the most effective \citep{Lin:2020:QueryRU,Lin:2020:MultiStage}. This indicates an opportunity for more unified approaches combining the different sub-components. Overall, this section suggested that notably challenging areas in understanding conversational interactions include result dependence on long responses as well as modeling long conversations, possibly spanning multiple sessions.  
\chapter{Response Ranking and Generation}
\label{sec:chapter5}
In this section, we discuss response ranking and generation used in conversational information seeking. The task of response ranking is selecting the relevant information item(s) for a turn in the conversation from the knowledge available to a conversational system. The types of methods are often categorized based on the type of conversational response provided: short answer (QA), longer single passage or document, automatically generated responses from extractive or abstractive summarization, and structured entities (products, restaurants, locations, movies, books, \etc). 

The evolution of ranking and generation is heavily influenced by the publicly available resources in this area. Early work in this area evolved existing QA datasets and models towards ones that include context. This includes single-turn QA or asynchronous discussions from Community Question Answering (CQA) on data including Reddit, StackExchange \citep{Penha:2019:MANtIS}, and Yahoo! Answers \citep{Hashemi:2019:ANTIQUE}. But going beyond context, conversational approaches evolve this towards interactive chat-like discussions that use different types of language patterns.  

\section{Short Answer Selection and Generation}
\label{sec:cqa}
This section covers an overview of Conversational QA (ConvQA) approaches, also referred to as Conversational Machine Comprehension in the NLP community. ConvQA often assumes that the question in each turn is answerable by a span of text within a particular passage (from a conversational retrieval model) and selects one or more spans of text from the passages. 
We begin by discussing traditional models, then more recent neural approaches, and end with recent work combining elements of retrieval and selection with end-to-end approaches.

The evolution of ConvQA follows advances in QA and machine comprehension. The adoption of deep neural models brought new interest in  the task and approaches. They are the building blocks for later ConvQA models. Early models are extractive and select one or more spans of text as the answer(s). These models have evolved to use generative sequence-to-sequence models. 

\subsection{Early Conversational QA Models}
Early ConvQA models started in the TREC 2004 QA Track \citep{Voorhees:2004:TREC, Voorhees:2005:TREC} with questions grouped into different series related to a single target entity (or event). Each question asks for more information about the target. This requires models to use previous questions in the sequence, mainly the first with the target. Unlike a dialogue or conversation, the questions did not mention answers (responses) from previous questions in the series, resulting in a limited discourse structure. Effective models \citep{Harabagiu:2005:lccruleqa} use straightforward and rule-based models, the response ranking methods did not leverage the multi-turn nature of the series.

A building block for later ConvQA models is extractive neural models for single-turn QA. Notable models include DrQA \citep{Chen:2017:ACL} and BiDAF \citep{Seo:2017:bidaf} that use Recurrent Neural Networks -- specifically bidirectional long short-term memory networks (Bi-LSTMs). The BiDAF++ QA model~\citep{Peters:2018:ELMO} includes self-attention and the use of pre-trained contextualized word vectors (ELMo). Later Pointer Generator Networks \citep{See:2017:GetTT} extended these by supporting copying spans from an input context in the decoder. These models and related datasets are extractive QA and do not focus significantly on ranking the input text. They are also not conversational, although as we discuss next they were later adapted to encode conversational context.

The shift from QA to ConvQA for these models required the development of new benchmark datasets. The Question Answering in Context (QuAC) dataset \citep{Choi:2018:emnlp} is one of the early ones. The baseline model on the dataset was BiDAF++, the state-of-the-art QA model at the time. To adapt it for ConvQA, the conversational history was appended (as described in Section~\ref{chapter4}) and was referred to as `BiDAF++ with k-Context'. This model appends previous \textit{k} (1-3) answers (their contextual embeddings) as context, along with the question turn number. We note that the QuAC dataset is limited to people entities, with a particular emphasis on entertainment.\footnote{See the datasheet description of QuAC for details including details of bias, \url{https://quac.ai/datasheet.pdf}}
Concurrent with QuAC, the CoQA benchmark \citep{Reddy:2019:tacl} was released with similar goals. Because of its crowdsourcing task setup, the extracts are shorter (2.7 words vs over 15 for QuAC). It includes conversational questions from seven diverse domains: children’s
stories, literature, middle and high school English
exams, news, Wikipedia, Reddit, and science. The CoQA baseline models were also similarly single-turn QA models adapted for conversation. They used BiDAF++ w/ k-Context. They also extended the DrQA model~\citep{Chen:2017:ACL} by including context history markers to separate turns, which outperforms the BiDAF model variants. These datasets and models are important because they represent the first steps towards a large-scale evaluation of ConvQA systems with models simply adapted from previous QA systems. 

One of the first steps towards new models explicitly designed for conversation are models that incorporate Flow (FlowQA) \citep{Huang:2019:iclr} to model the conversational dialogue.  Instead of appending history with a marker, they introduce a method that provides the model access to the full latent state used to answer the previous questions. This is a stack of two recurrent networks - one for each turn and one across turns. Their first model uses Bi-LSTMs to encode each turn and then processes each full turn representation linked with GRUs (for efficiency reasons). This represents a significant advancement over previous models that were extended to other types of networks including Transformers, discussed next.

\subsection{Conversational QA with Transformers}
The introduction of pre-trained language models based on Transformer architecture that supports transfer learning represents a significant shift for ConvQA systems. This subsection describes this evolution in approaches and challenges with these models.

Following the early success of Transformer-based models, such as BERT \citep{Devlin:2019:BERT} in QA tasks, these models were applied to ConvQA and yielded similarly impressive improvements. In many cases, the early work using Transformer approaches simply appended previous turn context with separators similar to previous extensions of BiDAF and DrQA. However, results show this has significant limitations. Naive approaches appending answer context degrade faster because of the limitations of the sequence input length \citep{Qu:2019:sigir}. To overcome these issues, \citet{Qu:2019:sigir} proposed the History Answer Embedding (HAE) model that uses BERT for ConvQA while modifying the representation to explicitly encode whether parts of the input are present in the previous history. On QuAC they found that this model outperforms BERT models that naively append the question or answer history, and is also more robust to appending longer conversations. In a different thread, \citet{Yeh:2019:FlowDelta} introduced the FlowDelta model that extends the previously discussed Flow model to use BERT for encoding, as well as changing the Flow loss to focus on the difference in Flow (Delta) across turns. They found that the proposed FlowDelta outperforms the previous Flow and BERT-based models. 

A long-standing top-performing system on the CoQA leaderboard is RoBERTa+AT+KD \citep{Ju:2019:coqatech}, an extractive model using a RoBERTa language model in combination with Adversarial Training (AT) that performs perturbation of the contextual embedding layer and Knowledge Distillation (KD) using a student-teacher setup. It ensembles nine models and has a post-processing step for the multiple-choice questions to match extracted spans to the target answer. Beyond the leaderboard, \citet{Staliunaite:2020:EMNLP} studied the behavior of BERT- and RoBERTa-based models on CoQA. They found that the key gain between the base models is that RoBERTa provides a better lexical representation. However, it does not capture more of the fundamental linguistic properties in ConvQA. To address these issues, they tested incorporating varying types of linguistic relationships in a multi-task model and combined the models in an ensemble. They found that incorporating the linguistic structure outperforms the base models. This indicates that the base representation of the language model is important for effectiveness and that there is an opportunity for models that incorporate more linguistic and conversational discourse structure.

\tipbox{
Note that the behavior of the current models for response ranking and generation in CIS is constrained by issues with current datasets and task formulation.
}

For example, an issue highlighted by \citet{Mandya:2020:LREC} is exposure bias: CoQA systems use gold answer labels for previous turns in both training and test time. As a result, CoQA evaluation sometimes overestimates the effectiveness of systems that have to rely on noisy previous predictions rather than human-written gold responses. They find this particularly problematic for longer conversations and longer questions. As discussed later, there is a similar phenomenon for conversational retrieval systems that perform conversational query rewriting. Systems that use manual query rewrites instead of predicted ones for earlier turns overestimate their effectiveness \citep{Gemmell:2020:TREC}. 

The ConvQA models and datasets (QuAC and CoQA) use a short pre-defined narrative of 200-400 tokens with the conversation focusing on one passage. As a result, the previously discussed ConvQA systems work well for extracting information from short passages with conversations grounded in a single paragraph. Further because of the way they were constructed, the headroom for generative models is very limited, approximately 5\% on CoQA \citep{Mandya:2020:LREC}. The next subsection covers more realistic models that include the retrieval of passages in the QA process.

\subsection{Open Retrieval Conversational QA}
\label{chapter2:OR-ConvQA}
This subsection discusses approaches that incorporate retrieval into the ConvQA task. This is referred to as open retrieval ConvQA (OR-ConvQA) or end-to-end ConvQA. The distinguishing feature of these models is that they operate on a large corpus of passage content and rank the passages used in the conversation. A common architecture for these systems is that they consist of two components - a \textit{Retriever} and a \textit{Reader}. The Retriever takes the conversational context and uses it to identify candidate passages. The Reader takes the context and candidates (text) and produces an answer. The base retrieval systems are effectively the Conversational Passage Retrieval long answer systems discussed below in Section~\ref{sec:chapter5:Long Answer Ranking} combined with a QA reader model to extract or generate the answer. A key challenge is that the existing ConvQA benchmarks are not designed for open retrieval QA and that current conversational passage retrieval benchmarks do not have short answer annotations. As a result, recent efforts \citep{Qu:2020:SIGIR,Gao:2021:OpenRetrievalCMR, Ren:2020:ConversationsWS} adapted and extended the datasets to bridge this gap. The first of these by \citet{Qu:2020:SIGIR} extended QuAC to incorporate passage retrieval over Wikipedia, creating the OR-QuAC dataset. To do this a synthetic query representing the information needed is created by providing the Wikipedia title and first paragraph with the initial question that is rewritten to be unambiguous. 

Recent developments in dense retrieval are also being applied to OR-ConvQA. \citet{Qu:2020:SIGIR} performed retrieval using a dot-product of a query history representation (previous k queries) and a passage that is based upon a learned query and passage encodings using ALBERT~\citep{Lan:2020:ALBERT}, a lite BERT representation. One of the novel contributions is the multi-task training objective where the retriever, a BERT-based cross-attention reranker, and a BERT-based reader are trained concurrently to avoid issues of error propagation. Another contribution is that it uses a distribution over candidate answers. One potential issue is that for training the model, golden query rewrites are used rather than employing a noisy query rewriter. This approach was further extended and improved upon by leveraging distant supervision to handle the free-form responses more effectively \citep{Qu:2021:WeaklySupervisedOC}.

One of the large-scale efforts in OR-ConvQA is the development of the Question Rewriting in Conversational Context dataset \citep{Anantha:2021:QReCC}. For a baseline, they used a BERTserini passage retriever combined with a BERT-large reader model. They found that a key factor in the success of reader models that leverage retrieval is incorporating the passage relevance score into the reader model \citep{Anantha:2021:QReCC}. Recent results by \citet{Tredici:2021:CIKM} demonstrate that different representations should be used for retrieving and reading models.  
One missing aspect from these is the existing models and datasets didn't exhibit topic switching. The TopiOCQA dataset is a large-scale OR-ConvQA dataset \citep{Adlakha:2021:TACL} that includes topic switching behavior. They start with seed questions from the Natural Questions QA dataset and traverse topics in Wikipedia. A leading approach on this dataset is a variation of Fusion-in-Decoder \citep{Izacard:2020:FusionInDecoder} extended to dialogue \citep{Wu:2021:EMNLP}.
 
Recently, mirroring a trend in QA there is increased attention to ConvQA over heterogeneous sources that combine text, tables, and entity KGs. \citet{Christmann:2022:SIGIR} propose a new heterogeneous conversational dataset (ConvMIX) and pipeline called CONVINSE to perform the task. One key difference in their proposed approach is a variant of conversational rewriting that instead of predicting a natural language utterance generates a frame-like representation ``intent-explicit structured
representation'' (SRs) whose nodes and sub-graphs are connected across turns in a graph. 

\subsection{Response Generation for Conversational QA}
Recent trends in QA increasingly have a focus on generative sequence-to-sequence models. These models are used to (1) perform generation and put the answer in the conversational context, and (2) to make the model more effective by generating responses from retrieved passages and past sessions. 

The first type focuses on the conversational natural language generation of answers, putting the answer into the natural conversational context and focusing on fluency. They follow the pattern of \textit{retrieve} and \textit{refine} \citep{Weston:2018:RetrieveRefine}. The initial results, retrieved from previous document collections or previous conversation responses, are used as the context that is refined during generation. The refining processing connects the answer to the dialogue and puts it into a natural language response form. An example of this is AnswerBART \citep{Peshterliev:2021:ConversationalAG}, which provides an end-to-end model that performs answer ranking, generation, and includes abstaining from answering when there is none. A novelty of this model is that it jointly learns passage reranking with the extraction task. A variation of this is treating generation as a ranking task. \citet{Baheti2020FluentRG} used syntactic patterns and templates to generate multiple candidate responses. This was combined with a GPT-2 based model that was pre-trained on Reddit conversations. These models focus primarily on fluency and putting the answer in context. 


The ability to incorporate long-term memory from past sessions is important for CIS systems.  The work from \citet{Shuster:2021:RetrievalAR} extended \citep{Lewis:2020:neurips} by incorporating the turn structure for knowledge-grounded conversations and they found this reduces model hallucination (i.e., producing factually invalid information), and results in a model that generalizes more effectively. Going beyond this, the work from \citet{Xu:2021:goldfish} extended the retrieval aspect to incorporate retrieval from past conversational sessions. The model retrieves sessions as a document and uses these as the context in the generation process. 
\jeff{There's probably some missing work here now.  Do we want to separate the fluency and generation out more? It feels a bit artificial to interleave these two.}

One limitation of many of these ConvQA approaches is that because the answers are short (even if they are put into a natural language utterance), they are usually simple factoid responses. As a result, the level of discussion in the conversation does not discuss aspects of the response and the ability to reference previous results in follow-up parts of the conversation is limited. The Question Rewriting in Conversational Context (QReCC) dataset from \citet{Anantha:2021:QReCC} is noteworthy because approximately 25\% of the answers are not simple extractions, but are human-generated paraphrases, possibly of multiple passages. Systems with these types of responses continue to evolve and represent an area for further work.

This section covered multiple threads of the evolution of these systems to use Transformer and attention-based architectures for ConvQA. They focus on improving the contextualized encoding (BERT vs RoBERTa), multi-task learning of discourse or token importance, stacking networks to capture cross-turn relationships, and approaches to make the models more robust using adversarial training and data augmentation. Recent work by \citet{Kim:2021:ACL} brought together generative conversational query rewriting using T5 in the QA process and showed that it outperforms more complex models that attempt to model both simultaneously. The models largely target factoid QA with most being extractive, possibly with minor adaptions for yes/no questions or multiple choice.  None of the existing ConvQA benchmarks are based on real user information needs (queries) with multiple results from retrieval. This represents an opportunity for new systems and methods to evolve towards more realistic tasks based upon real information needs.

\subsection{Conversational QA on Knowledge Graphs}
\label{sec:chapter5:ConvQA Knowledge Graphs}
Similar to parallel threads in question answering over unstructured text, ConvQA can also be performed on structured knowledge graphs (KGs) containing entities. This sub-area of conversational QA over a knowledge graph is called KG-ConvQA. These approaches allow conversational information seeking over structured data. Therefore, the nature of the questions they can answer is also structured and may involve logical operations including joins, aggregations, quantitative comparisons, and temporal references. KG-ConvQA systems may be partitioned into two distinct types. The first performs QA directly using the KG internally or traversing it using actions to produce an answer. The second type performs conversational semantic parsing and produces an executable logical structured query for producing the answer.


For the first type of KG-ConvQA systems, a neural sequence-to-sequence model is combined with a memory network to generate an answer. One of the first attempts to do this was done by \citet{Saha:2018:AAAI}, who introduced the Complex Sequential QA (CSQA) dataset and baseline model. A baseline for KG-ConvQA is HRED+KVmem, which combines a base conversational recurrent neural network (RNN) model, HRED, with a key-value memory network for modeling the KG, and finally an RNN decoder to generate the answer. This baseline model works well for many categories of questions but struggles with quantitative and comparative reasoning. 

Another approach, CONVEX, proposed by \citet{Christmann:2019:CIKM} starts from a seed entity and performs actions to traverse the graph to identify an answer. To handle the conversational evolution, CONVEX maintains a dynamic sub-graph that changes with each conversational turn using look-ahead, weighting, and pruning techniques to limit the graph size. This is effective because
traversing the  graph on the evaluation benchmark CONVQUESTIONS finds answers that are relatively close 
(no more than five edges away from the seed entity that starts the conversion).

The dynamic sub-graph approach was extended by \citet{Kaiser:2021:SIGIR} with their model, CONQUER. It uses reinforcement learning to select graph traversal actions. CONQUER maintains a set of context entities from which the agents traverse the graph. Their model uses a policy network that uses weak supervision from a fine-tuned BERT model. One of the key differences from previous work is that the model also predicts if the query is a reformulation. This is built on an extension of the CONVQUESTIONS dataset that adds manual reformulations when the baseline system produces incorrect answers. The results show that CONQUER outperforms CONVEX, demonstrating that its reformulation and policy network outperform the previous sub-graph tracking approach, particularly when there is implicit feedback from reformulation for wrong answers. The later PRALINE model learns graph traversal using contrastive learning that models the dialogue and possible KG paths in a joint space \citep{Kacupaj:2022:CIKM}.

The second direction taken to address this task is based on conversational semantic parsing. Instead of generating an answer, these approaches generate structured responses from a grammar. \citet{Guo:2018:Neurips} propose the Dialog-to-Action (D2A) model that builds on a GRU sequence-to-sequence model with a question and context from interaction history and outputs an action sequence from a predefined grammar. The dialogue history is managed in the action space. In contrast to the earlier HRED+KVmem model, the D2A model is much more effective, particularly for queries requiring reasoning. 

Subsequent approaches improve upon the semantic parsing quality by incorporating entity recognition and disambiguation in the semantic parsing process with multi-task learning. For instance, \citet{Shen:2019:EMNLP} presented the Multi-task Semantic Parsing (MaSP) model, performing both entity typing and coreference resolution together with semantic parsing. A subsequent multi-task model is CARTON (Context trAnsformeR sTacked pOinter Networks) \citep{Plepi:2021:ESWC}, with an encoder and decoder model to model the conversational representations. A series of three stacked pointer networks focus on the logical form needed for execution (types, predicates, and entities). 

A later approach using Transformers with multi-task learning and graph attention (LASAGNE) by \citet{Kacupaj:2021:EACL} built on this semantic parsing approach leveraging a graph attention network. It has a grammar-guided Transformer model to generate logical forms as well as a sub-model that learns correlations between predicates and entity types to avoid spurious logical forms. LASAGNE appears to outperform CARTON across most categories. However, CARTON performs better on coreference and quantitative reasoning.  They perform ranking on the KG by selecting the relevant entity and are implicit in the semantic parses produced by the model.  

The work from \citet{Marion:2021:EMNLP} generates a hierarchical JSON-like logical form that is KG executable. They used an Object-Aware Transformer that includes entity linking. They highlight that the CSQA approaches often use a base gold seed entity and only require coreference to the previous turn.  The results demonstrate strong effectiveness across multiple datasets using pre-trained encoder-decoder models.

\tipbox{The focus of most of the KG-ConvQA models is traversing the graph for structured comparison. The conversational structure, such as ellipsis and dependency support is limited in current models.}

\section{Conversational Long Answer Ranking}
\label{sec:chapter5:Long Answer Ranking}
This subsection discusses open-domain conversational long answer retrieval, sometimes referred to as ConvPR (for Passage Ranking). Analogous to the previous distinction between ConvQA and OR-ConvQA (see Section~\ref{chapter2:OR-ConvQA}), this subsection distinguishes between ConvPR and OR-ConvPR. ConvPR focuses on conversational passage reranking from a closed set of responses. In contrast, OR-ConvPR includes full retrieval over a corpus of passages in the ranking step. The questions may require one or more long answers to sufficiently answer the questions. This class of responses covers work on Ubuntu/Quora, MSDialog, AliMe, TREC CAsT, and similar corpora. 

The task of ConvPR has a rich history that builds on response retrieval and selection from discussion forums. These models have a long history in retrieval-based chatbots, see \citep{Tao:2021:RetrievalBasedDialoguesSurvey} for details. For the ConvPR task, the Deep Attention Matching Network (DAM) \citep{Zhou:2018:DAM} encodes each turn with a transformer model and combines them with a matching network and a final 3D convolutional network that incorporates the history. The intent-aware ranking model from \citet{Yang:2020:WWW} extends this model by adding explicit conversation intents. The encoding is similar to DAM, but it also produces a vector representing the user intent. This represents dialogue discourse types specific to CIS and includes: asking a question, clarification, elaboration on details, and both positive and negative feedback. The encoded turns are combined with the intent classification using a weighted attention model and aggregated into a matching tensor. Similar to DAM, the result is used in a final two-layer 3D-CNN model to rerank the candidate responses. 

One of the fundamental aspects of the effectiveness of any ConvPR model is the language model used in the encoding. Many of the encodings used are off-the-shelf language models, but an alternative is to perform a step of model fine-tuning with the language modeling objective on conversational corpora. Current leading approaches in chatbots and similar use models are trained on heavily filtered and curated conversations from web forums like Reddit. For example, the ConveRT model \citep{Henderson:2020:EMNLP} fine-tunes a BERT-based model on Reddit discussions and applies the resulting model to the task of response selection. This pre-training objective results in significant gains on Ubuntu DSTC7 and the AmazonQA response selection tasks. It is also widely used as a pre-training objective for dialogue system models.  In contrast to the intent-aware model, these do not use pre-defined intents and instead learn common discourse patterns directly from the text. 

In contrast to the previous ConvPR models, the OR-ConvPR models must retrieve and optionally rerank from large passage corpora. As a result, a current pattern exemplified by many CAsT systems is a pipeline with specialized modules. This includes modules that focus on understanding the context, as described in Section~\ref{chapter4}, that include conversational question rewriting and expansion across turns. These are then used with neural ranking models for passage retrieval. For more information on neural ranking models, we refer the readers to the recent survey articles~\citep{Mitra:2018:NeuIR,Guo:2020:NRM,Lin:2020:BERTandBeyond}. This architecture allows existing components trained on large existing datasets for query expansion, rewriting, and ConvPR to be used in the open retrieval context. 

It is common to use a multi-stage cascaded architecture for OR-ConvPR tasks. One of the prototypical multi-stage systems that perform this is developed by \citet{Lin:2020:MultiStage}. A core building block of their approach is Historical Query Expansion (HQE) which generates expanded queries based on the dialogue history using a sequence-to-sequence model. The conversational query rewriting is a standard T5 model trained on QuAC/Canard. One aspect is that the system additionally performs rank fusion to combine multiple query interpretations and formulations. This fusion can be performed early (in initial retrieval) or late (in reranking) and they find that fusion in early retrieval is critical for getting sufficient candidate passages in the pipeline for later reranking. 

Instead of the multi-stage cascade architecture, an alternative is end-to-end approaches based upon dense retrieval, sometimes referred to as Conversational Dense Retrieval (ConvDR) \citep{Yu:2021:ConvDR}. The representations of the query and document encodings vary and may include ANCE, TCT-Colbert \citep{Lin:2021:EMNP}, and others. The distinguishing feature is that retrieval and conversation are encoded with dense vectors rather than an explicit word-based query. This avoids explicit rewriting and instead builds a vector-based representation for retrieval directly. This approach can also be applied to OR-ConvQA. This continues to be an active area of research with few-shot approaches that rely on a multi-stage learning process including data augmentation, curriculum learning, and multi-task learning \citep{Mao:2022:SIGIR}. These elements are important to reduce noise and improve overall effectiveness. There are also attempts at zero-shot approaches \citep{Krasakis:2022:SIGIR} that can approach few-shot model effectiveness in some cases. There is also work demonstrating that efficiency in conversational dense retrieval process can be optimized to achieve fast latency by leveraging topical relatedness in the conversation \citep{Frieder:2022:Caching}. Although not (yet) as effective as the best complex pipeline systems incorporating explicit rewriting, they are rapidly improving.

\section{Long-Form Response Generation for CIS}
\label{sec:chapter5:Long Form Response Generation}
\jeff{Refactor this section into three parts: extractive, retrieve and generate, and purely generative. Add discussion of chatgpt.}
The previous subsection discussed retrieval of (mostly) passage-level responses.  In contrast, a recent development is extractive or generative summarization of retrieved results appropriate to a conversational interface and in a conversational context.

One approach to handling multiple retrieved documents or passages for CIS is to combine them with extractive summarization approaches.  This is particularly important for summarizing long documents for CIS interfaces and interaction. A text editing approach is to keep, delete, or make small insertions. This approach is used by the LaserTagger \citep{Malmi:2019:EMNLP} and Felix \citep{Mallinson:2020:EMNLP} models. They leverage pre-trained Transformers trained with supervised data. They produce responses for CIS applications that are true to the document and add elements of fluency by putting them in a conversational form. 

Beyond extractive approaches, generative systems are evolving towards longer and more complex information responses.  Recently, these developments include pre-training of language models for generation on dialogue, such as Language Model for Dialogue Applications (LaMDA) that builds upon the previous Meena \citep{Adiwardana:2020:Meena} architecture based on Evolved Transformers \citep{so:2019:evolvedtransformer} and trained on social media data. 

Generative approaches for long answers are a significant open area of research for CIS. This area is particularly important as generated answers become longer and more complex. Year four of TREC CAsT included evaluation of generative responses \citep{Dalton:2022:TREC} with human crowdworkers that assessed relevance, naturalness, and conciseness. The most effective models used T5 and BART to generate abstractive summaries of the input passages. As summarizes and inputs become longer and more complex work there will need to be architectures like the Routing Transformer \citep{Krishna:2021:naacl} with dynamic attention routing to support longer sequences.

The most significant advance in this area is from ChatGPT\footnote{\url{https://openai.com/blog/chatgpt/}} by OpenAI. ChatGPT is a purely generative model that encodes all of its knowledge parametrically. Although it doesn't leverage search, the breadth and scope of its generation capability as well as its ability to generate long-form, fluent responses across diverse areas is remarkable. Although formal evaluation is limited, its generated significant press with its fluent responses that can succinctly summarize complex content and even pass challenging medical exams \citep{Kung:2022:ChatGPT}.

A key consideration for all of these generative models is their factual consistency and fidelity to the input passage (or corpus), with previous work showing that the degree to which the model uses the input varies \citep{Krishna:2021:naacl}. To address this for short answers, an early benchmark by \citet{Dziri:2021:BEGIN}, Benchmark for Evaluation of Grounded INteraction (BEGIN), uses generated responses from Wizard-of-Wikipedia~\citep{Dinan:2019:WizardOfWikipedia}. Further, the provenance of facts to source passages and attribution of information will become increasingly important. 

\jeff{Add a discussion on chatgpt}

\section{Procedural and Task-Oriented Ranking}
The previous subsections describe formulations of CIS response ranking that largely extend previous research from QA, retrieval, and recommendation to a conversational context. However, because CIS systems are often embedded or used in combination with task assistants, the types of information needs and tasks performed are more likely to be grounded in procedures and real-world tasks.  Information seeking is interleaved with task-oriented processes and structured dialogue actions, such as task navigation \citep{Ren:2021:SIGIR, Azzopardi:2018:ConceptualizingAI}. This subsection discusses multiple veins of work in these areas and their connection to CIS.

\subsection{Procedural Question Answering}
In Procedural QA, the task is to interact conversationally to determine outcomes based on complex processes represented in text documents. To address this task, \citet{Saeidi:2018:sharc} introduced the Shaping Answers with Rules through Conversation (ShARC) benchmark. It contains varied types of discourse and natural language inference required within it. The procedures come from conversations on complex regulatory decisions. Because they are vague, the model must generate clarifying questions and understand the complex rule structures encoded in documents. Instead of providing excerpts like a typical QA task, the goal is to use rules in the text and the conversational responses to infer a yes/no answer. Similar to the evolution of other QA systems, a baseline model for this task includes a conversational BiDAF model for encoding history which is then combined with a natural language inference model, such as the Decomposed Attention Model (DAM) \citep{Parikh:2016:DAM} for interpreting rules. 

Subsequent work \citep{Gao:2020:Discern} focused on segmenting documents into elementary discourse units (EDUs) which are tracked through the conversation. Going further, recent work built on this by explicitly modeling the conversational structure using Graph Convolutional Networks (GCNs) \citep{ouyang:2021:aclfindings}. The results show that using both explicit and implicit graph representations allows the model to  effectively address conversations with complex types of discourse structure. 
Mirroring the evolution of QA towards open retrieval, \citet{Gao:2021:OpenRetrievalCMR} extended the ShARC conversational entailment task by adding rule retrieval, creating OR-ShARC. In this task, systems must first search a knowledge base of rule texts with context from the user and scenario (although it is limited to rule texts used in the original ShARC benchmark). It uses a basic TF-IDF retriever achieving over 95\% recall in the top 20 rules; approximately the top five rules are used with a recall of over 90\%. These are used in a RoBERTa machine comprehension system that also leverages inter-sentence Transformer layers to combine evidence. It is noteworthy that systems capable of reading multiple passages in the top-k retrieved results, \eg, \citep{Dehghani:2019:QA}, can be more effective than systems that only use the top (often golden) rule.

\subsection{Task-Oriented Information Seeking}
Task-based virtual assistants perform tasks in the world. They are largely separate from CIS systems. Recently, there is a trend towards systems and models capable of both: A critical aspect of CIS is that information seeking is occurring within an explicit task context with domains and intents.  It may start with conversational search to find an appropriate agent or task to execute (for example, finding a recipe to cook) and then unfold as the task is performed. This may involve answering procedural questions grounded in the task execution, questions requiring external knowledge for QA, and other types of information needs. The CIS should also respond to changes in the task execution environment. From the dialogue community, this task was proposed and evaluated as part of the DSTC9 challenge in the Beyond Domain APIs track \citep{Kim:2020:DSTCUnstructured}.  \jeff{Add descriptions of QA systems and behavior.}

\jeff{This could be updated to add more details summarizing key aspects of the challenge and relevance CIS findings.}
The recent Amazon Alexa Prize TaskBot Challenge \citep{Gottardi:Taskbot:2022} introduced the challenge of using multi-modal conversation to solve real-world tasks. This challenge includes conversational task retrieval and refinement, task-oriented QA, and conversational procedural instruction responses. Further, because interactions are multi-modal (including voice and screen), the responses may include images and videos in response to the information need. In practice, this means that elements of a dialogue system to navigate the task are interleaved with task-specific question answering and open-domain question answering. Additionally, the goal is also implicitly to select responses that are natural and engaging for the user with elements of social chat related to the task.

The winning approach \citep{GrillBot2022} during the first iteration of TaskBot challenge focused on automatic creation of TaskGraphs -- a dynamic graph unifying steps, requirements, and curated domain knowledge enabling detailed contextual explanations and adaptable task execution. They showed offline creation and enrichment of TaskGraphs, potentially with the help of large language models, can reduce the system's complexity in navigating through the steps and responding to user's requests, leading to a more efficient and effective TaskBot. Several participating teams found that the system's ability in finding relevant instructions plays a key role in the overall TaskBot performance \citep{TacoBot2022,MarunaBot2022}. This competition also demonstrated a successful use of visual content in conversational systems. \citet{TwizBot2022} successfully took advantage of  visual interactions and proposed a multimodal curiosity-exploration task guiding assistant to improve user experience by  potentially reducing the cognitive load on the user.

\section{Conversational Recommendation}
\label{sec:chapter5:recsys}
\hamed{focus on ranking} \hamed{relate to sec 2} \hamed{Relate to sec 6}
Traditionally, recommender systems mainly exploit historical user-item interactions for predicting user preferences. This has led to the development of collaborative filtering methods which are at the core of effective real-world recommendation engines. Other recommendation models, such as content-based and demographic filtering, have also been studied and showed promising results mostly for cold-start users and items. All of these approaches provide users with little control over the recommended list. For instance, users often cannot ask for a revised recommendation list based on their preferences. Conversational recommender systems address this limitation. During a human-machine dialogue, the system can elicit the current user preferences, provide explanations for the recommended items, and/or take feedback from users for recommendation refinement. 

Interactive and conversational recommender systems have been studied for several years~\citep{Thompson:2004:personalised,Mirzadeh:2005:ConvRecSys,Mahmood:2009:ConvRecSys,Blanco:2013:ConvRecSys}. Due to the potential real-world applications, conversational recommender systems have recently attracted considerable attention. Most efforts in this domain focus on preference elicitation by asking questions from users. \citet{Christakopoulou:2016:KDD} studied this task and proposed a conversational model based on probabilistic matrix factorization for restaurant recommendation. They proposed to initialize the conversational recommendation model's parameters by training the model on offline historical data and updating the parameters while the users interact with the system through online learning. They focused on question selection from a question bank during online interactions for preference elicitation. This approach was later revisited by \citet{Zhang:2018:cikm} who used multi-memory neural networks for template-based question generation. They unified conversational search and recommendation and trained their model based on item reviews in the e-commerce domain. In more detail, they extracted attribute-value pairs mentioned by users about items in their reviews, and train a model that generates attribute-based questions based on the attributes. Besides the explicit attribute-value pairs, implicit knowledge learned by pre-trained large language models can also be used for preference elicitation in recommendation \citep{Penha:2020:ProbingBERT4RecSys}. 
Preference elicitation in conversation can be improved by conditioning the dialogue on the user profile. To this aim, \citet{Li:2022:UserConvRec} proposed a multi-aspect user modeling approach that uses historical conversational interactions collected from look-alike users to go beyond the current dialogue session.

More recently, the applications of conversational interactions have been extended to bundle recommendation problems, where a \textit{set} of items is recommended to a user. Bundle recommendation largely suffers from data sparsity and the interactive nature of conversations would help the recommender system to collect more feedback and overcome this issue. Based on this idea, \citet{He:2022:BundleMCR} proposed Bundle MCR which models bundle recommendation as a Markov Decision Process with multiple agents, for user modeling, consultation, and feedback handling in bundle contexts. Additionally, \citet{Leszczynski2022} studied conversational music playlist recommendation which is another example of bundle recommendation tasks.

Another line of research focuses on modeling conversational recommendation using reinforcement learning (RL). \citet{Sun:2018:ConvRecSys} developed an early interactive RL-based recommendation model that can take two actions: (1) selecting an attribute (or facet) for preference elicitation, or (2) making a personalized recommendation. They simply used a two-layer fully-connected neural network as the policy network and defined the reward function based on the recommendation quality at every timestep during the dialogue. They demonstrated the benefits of conversational recommendation via both offline and online experimentation. This approach was later improved by modeling conversational recommendation using an Actor-Critic framework \citep{Montazeralghaem:2021:ActorCritic} as well as improving user and item representations based on implicit feedback \citep{Hu:2022}.

Recently, \citet{Lei2020EAR} introduced the \textit{Estimation-Action-Reflection} (EAR) framework for conversational recommendation. This framework unifies the following three fundamental problems in conversational recommendation: (1) what questions to ask, (2) when to recommend items, and (3) how to adapt to the users' online preferences. Another approach to conversational recommendation is to exploit multi-armed bandit solutions which have shown promising results in sequential and interactive recommendation. \citet{Zhang:2020:ConvContextualBandit} followed this path and proposed conversational contextual bandit. Later on, \citet{Li:2021:ConTS} improves this model by introducing the Conversational Thompson Sampling (ConTS) model. ConTS builds upon multi-armed bandit and models items and attributes jointly. This enables the model to compute the exploration-exploitation trade-off between preference elicitation and item recommendation automatically.

An interesting research direction in conversational recommender systems is producing responses that explain the rationale behind the recommendations \citep{Volokhin:2022}. This will help users to engage with the conversational system to provide more feedback and express their opinion. \citet{Li:2022:self-bot-play-recsys} developed a self-supervised bot play approach that learns to produce such explanations through reasoning and demonstrated that it can go beyond user simulations and can also work well in the wild. 

Popularity bias has always been an important challenge in recommender systems, especially collaborative filtering models \citep{Ricci:2010:RecSysHandbook}. \citet{Lin:2022:PopularityBiasConvRec} recently explored the correlation between popularity bias and exposure rate, success rate, and conversational utility in a conversational recommendation setting. They proposed a three-stage de-biasing framework and demonstrated that reducing the impact of popularity bias improves the overall conversational recommendation quality.

For more information on conversational recommender systems, we refer the reader to the recent survey on the topic \citep{Jannach:2022:ConvRecSysSurvey}.

\section{Summary}

This section focused on core conversational response ranking. The models started with ConvQA, with basic extractive factoid QA with context naively appended that operated in a closed environment. These evolved towards variants that are more realistic by incorporating retrieval from a corpus (OR-ConvQA), including incorporating multiple results and their retrieval scores as well as other forms of external memory, including past turns or conversations. 

As the retrieval task evolved towards longer and exploratory responses (OR-ConvPR and OR-ConvDR), the systems evolved to be complex pipelines that required query rewriting, query expansion, dense retrieval, multi-pass re-ranking, and result fusion. However, the ranking components are still largely separated and trained on external datasets specific to those tasks.

Later, models evolved to include conversational models of richer types of responses, including entities (KG-ConvQA), as well as ones that are longer and more natural.  Longer and more complex responses support richer types of result dependency and more natural conversations. This includes generating responses from one or more retrieved sources. Most of the effective ranking and generation models build upon pre-trained language models. The effectiveness of the models varies depending on their lexical representations and training to support linguistic and conversational structure. The most effective ones include additional fine-tuning with a language modeling objective on the target conversational data before final task-specific training. For ranking models, there is a common pattern of having a model for a single turn and then incorporating evidence across turns by stacking models to capture conversational context (\eg, Flow or 3D-CNNs). 

Finally, the section covered response ranking for structured prediction in the form of task-oriented dialogues and recommendations. Beyond these, the ranking tasks and models will continue to evolve to include richer types of responses and to support more realistic and complex information seeking tasks.
\chapter{Mixed-Initiative Interactions}
\label{sec:mixed_init}

Most approaches to human-computer interactions with intelligent systems are either controlled by a person or the system (i.e., user- or system-initiative). For example, in current search engines, users always initiate the interaction by submitting a query and the search engine responds with a result page. Therefore, search engines are user-initiative systems. 
That being said, developing intelligent systems that support \emph{mixed-initiative interactions} has always been desired. \citet{Allen:1999:MII} believed that development of mixed-initiative intelligent systems will ultimately revolutionize the world of computing. Mixed-initiative interactions in dialogue systems have been explored since the 1980s~\citep{Kitano:1991:MID,Novick:1988:MID,Walker:1990:MID}. Early attempts to build systems that support mixed-initiative interactions include the LookOut system~\citep{Horvitz:1999:PMU} for scheduling and meeting management in Microsoft Outlook, Clippit\footnote{\url{https://en.wikipedia.org/wiki/Office_Assistant}} for assisting users in Microsoft Office, and TRIPS~\citep{Ferguson:1998:TRIPS} for assisting users in problem solving and planning.

\citet{Horvitz:1999:PMU} identified 12 principles that systems with mixed-initiative user interfaces must follow. They are listed in Table~\ref{tab:chapter6:mixe_init_principles}. 

\tipbox{Mixed-initiative interactions should be taken at the right time in the light of cost, benefit, and uncertainties.} 

Many factors mentioned in these principles can impact cost and benefit of interactions. In addition, systems with mixed-initiative interactions should put the user at the center and allow efficient invocation and termination. Systems with mixed-initiative interactions are expected to memorize past interactions and continuously learn by observation. Based on these principles, conversational systems by nature raise the opportunity of mixed-initiative interactions.

\begin{table}[t]
    \centering
    \caption{Principles of mixed-initiative user interfaces by \citet{Horvitz:1999:PMU}.}
    \begin{tabular}{ll}\toprule
        \textbf{Principle} \\\midrule
        1. Providing genuine value \\
        2. Considering uncertainty about user intents \\
        3. Considering the user status in the timing of services \\
        4. Inferring ideal action in light of cost, benefit, and uncertainties \\
        5. Employing dialogue with users to resolve key uncertainties \\
        6. Allowing efficient direct invocation and termination \\
        7. Minimizing the cost of poor guesses about action and timing \\
        8. Scoping precision of service to match uncertainty in goals \\
        9. Providing mechanisms for efficient result refinement \\
        10. Employing socially appropriate behaviors  \\
        11. Maintaining working memory of past interactions \\
        12. Continuing to learn by observing  \\
        \bottomrule
    \end{tabular}
    
    \label{tab:chapter6:mixe_init_principles}
\end{table}

\citet{Allen:1999:MII} defined four levels of mixed-initiative interactions in the context of dialogue systems, as follows:
\begin{enumerate}
    \item \textbf{Unsolicited reporting:} An agent notifies others of critical information as it arises. For example, an agent may constantly monitor the progress for the plan under development. In this case, the agent can notify the other agents (e.g., user) if the plan changes. 
    
    \item \textbf{Subdialogue initiation:} An agent initiates subdialogues to clarify, correct, and so on. For example, in a dialogue between a user and a system, the system may ask a question to clarify the user's intent. Since the system asks the question and the user answers the question, and this may be repeated for multiple turns, the system has temporarily taken the initiative until the issue is resolved. This is why it is called subdialogue initiation.
    
    \item \textbf{Fixed subtask initiation:} An agent takes initiative to solve predefined subtasks. 
    In this case, the agent can take initiative to ask questions and complete the subtask. Once the subtask is completed, initiative reverts to the user.
    
    \item \textbf{Negotiated mixed-initiative:} Agents coordinate and negotiate with other agents to determine initiative. This is mainly defined for multi-agent systems in which agents decide whether they are qualified to complete a task or it should be left for other agents.
\end{enumerate}

When it comes to (pro-active) open-domain conversational information seeking, some of these mixed-initiative levels remain valid.  Mixed-initiative interactions in the context of CIS have been relatively less explored, but are nevertheless identified as critical components of a CIS system~\citep{Radlinski:2017:CHIIR, Trippas:2018:Informing,Aliannejadi:2019:sigir,Wadhwa:2021:SystemInitiative,Wu:2022:MixedInit}. \citet{Vakulenko:2021:TOIS} conducted a large-scale analysis of 16 publicly available dialogue datasets and established close relations between conversational information seeking and other dialogue systems. Clarification and preference elicitation are the two areas related to mixed-initiative interactions that have attracted considerable attentions in recent years. Therefore, in the rest of this section, we first review the role of agents in initiating a conversation (Section~\ref{sec:mixed_init:initiating}), and continue with discussing methods for generating, analyzing, and evaluating clarification in conversational search (Section~\ref{sec:mixed_init:clarification}). We further summarize preference elicitation in conversational recommendation (Section~\ref{sec:mixed_init:pref_elicit}), and  finally discuss how the user and system can be involved in mixed-initiative interactions with the goal of providing feedback (Section~\ref{sec:mixed_init:feedback}).





\section{System-Initiative Information Seeking Conversations}
\label{sec:mixed_init:initiating}

Typically, users initiate the interaction with a conversational system, for example by clicking or touching a link or button, by using pre-defined voice commands such as ``Alexa'' or ``OK Google'', or by asking a question or submitting an action request. In mixed-initiative conversational systems, the agent is also able to initiate the conversation. This is also called a system-initiative (or agent-initiative) conversation. Making a recommendation is perhaps the most common scenario for initiating an interaction by the system. 
For example, a CIS system can initiate a conversation by recommending an item based on the situational context of the user (e.g., location and time) and their preferences.
Note that this is different from many conversational recommendation settings, where users first submit a request about the item they are looking for, \eg,~\citep{Sun:2018:ConvRecSys,Zhang:2018:cikm}. 
Joint modeling of search and recommendation~\citep{Zamani:2018:DESIRES,Zamani:2020:JSR} is a step towards developing mixed-initiative search and recommendation systems. However, initiating a conversation by the system is not limited to recommendation. 
For instance, \citet{Avula:2020:Wizard} developed a system for conducting wizard-of-oz experiments to study system-initiative interactions during conversational collaborative search. This system can be integrated into collaborative discussion tools, such as Slack.\footnote{\url{https://slack.com/}} 
In this system, while a group of users are performing a collaborative search task, another user (who plays the role of wizard) can intervene and provide additional information.
Although little progress has been made in this area, there is a great potential for systems to initiate conversations based on context and engage with users or even get feedback. For instance, assume a user drives to a restaurant using a mapping application. When it has access to the context, a CIS system could initiate a conversation when the user is driving back, by asking about their experience at the restaurant. This could potentially lead to improving the user experience with the conversational system, collecting feedback on the restaurant, and also collecting information on the user's preferences for improving the user profile. As another example, if a user is struggling with completing a task, a CIS system can be automatically triggered to start the conversation with the user, hear their complaints, and help them complete the task. Related to this line of research, \citet{Rosset:2020:WWW} studied how a system can lead a conversation while users are searching for or exploring a topic. They formulated the problem as a conversational question suggestion task and demonstrated its impact by presenting the question suggestions in search engine result pages. 

\tipbox{
Initiating a conversation by the system can be risky and it may annoy users and hurt user satisfaction and trust. For instance, in some situations, a user may not be interested in engaging in a conversation, and thus predicting opportune moments for conversation initiation is an important part of developing system-initiative CIS systems. Therefore, whether and when to initiate a conversation are the key decisions a mixed-initiative CIS system should make. 
}

\begin{figure}[t]
    \centering
    \includegraphics[width=1.1\textwidth]{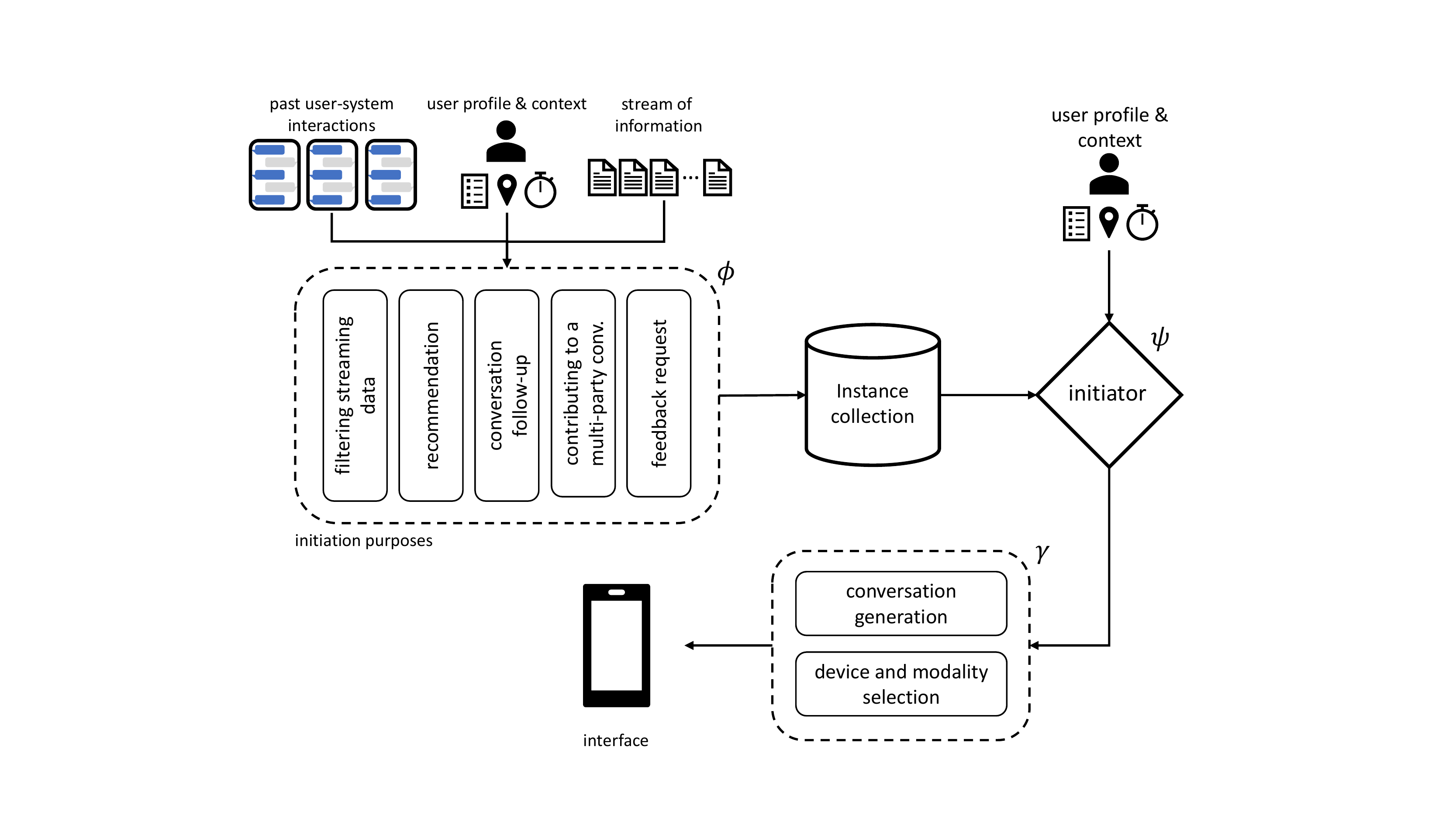}
    \caption{A generic pipeline for conversation initiation in CIS systems by \citet{Wadhwa:2021:SystemInitiative}.}
    \label{fig:chapter6:systeminit}
\end{figure}

\citet{Wadhwa:2021:SystemInitiative} studied system-initiative CIS systems and discussed their challenges and opportunities. They introduced a taxonomy of system-initiative CIS systems by defining three orthogonal dimensions: (1) initiation moment (\emph{when} to initiate a conversation), (2) initiation purpose (\emph{why} to initiate a conversation), and (3) initiation means (\emph{how} to initiative a conversation). They further identified five different purposes for initiating conversations for CIS systems, some of which have been mentioned above: (1) filtering streaming information, (2) context-aware recommendation, (3) following up a past user-system conversation, (4) contributing to a multi-party human conversation, and (5) requesting feedback from users. Based on this taxonomy and conversation initiation purposes, they introduced a generic pipeline that is depicted in Figure~\ref{fig:chapter6:systeminit}. According to this pipeline, several algorithms are constantly monitoring the user's situation (user context) and the stream of generated information to produce conversation initiation instances. These instances are stored in a database which is constantly monitored by a conversation initiator component. Based on the situation, the initiator may select one of the initiation instances. Then, a fluent conversation will be initiated. For more information on this architecture, we refer the reader to \citet{Wadhwa:2021:SystemInitiative}.

\section{Clarification in Information Seeking Conversations}
\label{sec:mixed_init:clarification}
Clarification is defined as ``an explanation or more details that makes something clear or easier to understand.''\footnote{\url{https://dictionary.cambridge.org/us/dictionary/english/clarification}} In information seeking systems, it is often used to clarify the user's information need or user's intent, and it can be in any form. For instance, relevance feedback is one form of clarification that is provided by the user. In mixed-initiative interactions, systems can take initiative to ask for clarification. This is why asking for clarification has been identified as a necessary component in developing ideal CIS systems~\citep{Radlinski:2017:CHIIR,Aliannejadi:2019:sigir,Anand:2020:Dagstuhl,Zamani:2020:GCQ,trippas2020towards}. As pointed out earlier, subdialogue initiation is one of the four levels of mixed-initiative interactions in conversational systems, which involves \emph{asking a clarification}. In a study of mixed-initiative collaborative planning in human conversations, clarification accounts for 27\% of interactions, more than any other type of mixed-initiative interactions~\citep{Allen:1999:MII}.
A conversational agent can ask a clarifying question to resolve ambiguity, to prevent potential errors, and in general to clarify user's requests and responses. 
Clarification may happen in multiple levels for various purposes. \citet{Stoyanchev:2014:ClarifySpeech} used clarification for resolving ambiguity and uncertainty in speech recognition, while \citet{Aliannejadi:2019:sigir} used clarification to identify query intent in a conversational search setting. Besides CIS systems, asking clarifying questions has been explored in various tasks. For instance, \citet{Rao:2018:acl} used clarification for identifying missing information in a passage, such as community question answering posts. \citet{Trienes:2019:ecir} identified the community question answering posts that require clarification. Subsequent work by \citet{Tavakoli:2021:Clarification} studied properties of clarification in community question answering websites based on user responses. Asking clarifying questions has also been studied in the context of task-oriented dialogue systems which are mostly closed-domain~\citep{Krum:2005:DialogueClar,Rieser:2005:ClarImplication}. In the following  subsections, we mostly focus on query intent clarification which is the most relevant type of clarification for information seeking systems.



\subsection{A Taxonomy of Clarification Types}
\label{sec:mixed_init:clarification:taxonomy}
In the context of information seeking systems, clarification has been studied in both synchronous and asynchronous information seeking scenarios. 

For instance, \citet{Braslavski:2017:ClarQ} studied clarifications asked in community question answering (CQA) websites as an example of asynchronous human-human information seeking conversations. They derived a taxonomy of clarification types for questions asked in CQA websites. The clarification types and their examples are reported in Table~\ref{tab:chapter6:cqa_clarq_taxonomy}. 

\begin{table}[t]
    \centering
    \caption{A taxonomy of clarification types for questions asked in CQA websites by \citet{Braslavski:2017:ClarQ}.}
    \begin{tabular}{ll}\toprule
        \textbf{Clarification Type} & \textbf{Example} \\\midrule
        More Information & What OS are you using? \\
        Check & Are you on a 64-bit system? \\
        Reason & What is the reason you want a drip pan? \\
        General & Can you add more details to this question? \\
        Selection & Are you using latex or oil based Kilz? \\
        Experience & Have you tried to update video card drivers? \\\bottomrule
    \end{tabular}
    \label{tab:chapter6:cqa_clarq_taxonomy}
\end{table}

Later on, \citet{Zamani:2020:GCQ} studied clarification in open-domain search systems by analyzing a large-scale query reformulation data collected from a commercial web search engine. This resulted in a clarification taxonomy for open-domain information seeking queries. Their taxonomy consists of four main categories and a number of subcategories as follows:
\begin{itemize}[leftmargin=*]
    \item \textbf{Disambiguation:} some queries (or part of the queries) are ambiguous and could refer to different concepts or entities. Clarifying questions can be used to disambiguate the query intent.
    
    \item \textbf{Preference:} Besides disambiguation, a clarifying question can help identify a more precise information need. Four major subcategories of preference clarifications are:
    \begin{itemize}
        \item Personal information (``for whom''):  personal information, such as gender, age, language, and expertise, can limit the search space.
        \item Spatial information (``where''): spatial information is also reflected in reformulations in many cases.
        \item Temporal information (``when''): some queries have a temporal aspect which can be clarified by the system.
        \item Purpose (``for what purpose''): if the answer to a query depends on the purpose of user, a clarifying question can seek the purpose. For example, a user searching for ``screwdrivers'' may be interested in screwdrivers for different kinds of screws in different sizes, depending on the user's purpose.
    \end{itemize}
    \item \textbf{Topic:} In case of broad topics, the system can ask for more information about the exact need of the user. This would narrow down the search space and would potentially lead to more accurate results. Topic clarification includes:
    \begin{itemize}
        \item Sub-topic information: The user might be interested in a specific sub-topic of the query.
        \item Event or news: based on an event or breaking news, many users often search for a topic related to the news, while the query may have different meanings out of the context of that event or news.
    \end{itemize}
    \item \textbf{Comparison:}  Comparing a topic or entity with another one may help the user find the information they need.
\end{itemize}

Note that clarifying the information need of a user may lie in multiple categories in this taxonomy. As mentioned earlier, this taxonomy was obtained based on web search query logs. Therefore, it can be considered as a taxonomy for open-domain information seeking queries. However, there may be other domain-specific types of clarification that are not easily recognizable in web search query logs. For more information on this taxonomy, we refer the reader to \citet{Zamani:2020:GCQ}. 

For all clarifying questions, we note that it is also essential to consider a system's need for specific information, with particular attention to personal or private information. As an example, while personal information such as gender or age may help a CIS system better answer a particular information need, is it clear to the user why this is being asked? Is it clear how this information will be processed and/or recorded? What would be the effect should the user decline to answer this question? While there are commonly accepted UI affordances for visual search systems (such as an asterix for required fields and hover-over information tags to provide background on questions), such affordances rarely exist in verbal modalities.

\subsection{Generating Clarifying Questions}
\label{sec:mixed_init:clarification:generation}


There exist three categories of solutions for generating clarifying questions: (1) selecting and filling out pre-defined question templates, (2) selecting and editing a clarifying question, (3) generating clarifying questions based on sequence-to-sequence modeling by maximizing the likelihood of generating the questions in a training set, and (4) generating clarifying questions by maximizing a clarification utility. In the following subsections, we briefly discuss solutions from each of these categories. 

\subsubsection{Template-based Slot Filling Models}
Template-based slot filling is the simplest approach for asking a clarification. In this approach, a small set of question templates is first defined. The templates are task- and domain-dependent. For instance, \citet{Coden:2015:AorB} simply used the question template ``Did you mean \_\_\_ or \_\_\_?'' for entity disambiguation. The question template ``Did you mean \_\_\_?'' has been widely used by various commercial search engines, such as Bing and Google, to clarify misspelling. \citet{Zamani:2020:GCQ} listed a handful of question templates for search clarification. The question templates can be as generic as ``What would you like to know about \_\_\_?''. However, more specific questions, such as ``What \_\_\_ are you using?'' or ``Who are you shopping for?'' would be desired in most scenarios. 


Once the question templates are defined, the task is to select one of the templates and fill it out. The template selection can be as simple as a rule-based algorithm or modeled as a machine learning problem, either as a multi-class classification or a learning to rank task. Similarly, rule-based solutions can be used to fill out the templates. For example, a substring of the user request or its entity type obtained from a knowledge base can be used to fill out some templates. Machine learning solutions are often preferred due to their superior performance for filling out the templates. Slot filling is not specific to clarification. A number of slot filling models used in task-oriented dialogue systems can be employed in clarification as well \citep{Wu:2019:TaskOrientedDialoge,Budzianowski:2019:GPTforDialogue,Zhao:2019:RLforTaskOrientedDialogue}. 

\subsubsection{Sequence Editing Models}
Another category of approaches for generating clarifying questions is based on selecting a clarifying question and editing it based on the conversation context. For instance, \citet{Liu:2021:LearningToAsk} proposed a Reinforcement Iterative Sequence Editing (RISE) framework that minimizes the Levenshtein distance between the model's output and ground truth questions through explicit editing actions. In more detail, the authors used BERT2BERT \citep{Rothe:2020:BERT2BERT} to implement the policy network in RISE and used a variant of Markov Decision Process (MDP) for optimization, in which the reward function is defined as the Levenshtein distance obtained by each action compared to the last iteration. RISE is able to pay attention to tokens that are related to conversational characteristics. Therefore, this approach is able to produce questions with coreferences to the conversation history. The idea of retrieve-and-edit has also been explored in the context of generating structured output, \eg, programming code \citep{Hashimoto:2018:Ret_and_edit}. Similar ideas can potentially be applied to this category of clarification generation models.

\subsubsection{Sequence-to-Sequence Models}

\filip{You might also want to reference \citep{Kostric:2021:Questions}, who transform review text using a large language model, to clarify user needs by asking questions how a searched for recommendation will be used.}

As discussed in~\citet{Rao:2019:naacl,Zamani:2020:GCQ}, generating clarifying questions can be seen as a sequence generation task, in which the inputs are the query $q$ and the context $c$ and the output is a clarifying question $q^*$. The context here may refer to the query context, e.g., short- and long-term search or conversation history~\citep{Bennett:2012:MIS} and situational context~\citep{Zamani:2017:WWW}, or some additional knowledge about the query, such as query aspects. Sequence-to-sequence models, including seq2seq~\citep{Sutskever:2014:seq2seq} and the Transformer encoder-decoder architecture~\citep{Vaswani:2017:Transformers}, can be adopted and extended to address this task.

Sequence-to-sequence models consist of at least one encoder and one decoder neural network. The encoder model $E$ takes the query $q$ and the corresponding context $c$ and learns a representation $v$ for the input tokens. The decoder model $D$ uses the encoder's outputs and generates a sequence of tokens, i.e., a clarifying question. The training objective is to maximize the likelihood of generating the clarification $q^*$ by the decoder. 
This maximum likelihood objective is equivalent with minimizing the cross-entropy loss. 

Once the model is trained, it is autoregressively to generate the clarification at inference time. 
This decoding step can be achieved using beam search, its variants, or in the most simplest case, generating the clarification token by token until observing an \texttt{end} token. For more detail on sequence-to-sequence modeling, we refer the reader to~\citet{Sutskever:2014:seq2seq,Vaswani:2017:Transformers}. 

It is widely known that training text generation models by maximizing likelihood of generating a ground truth output will result in frequent generation of the most common outputs. Thus, the models often suffer from generating diverse outputs. This has been addressed using different techniques, such as unlikelihood training~\citep{Welleck:2020:unlikelihood} and $F^2$-Softmax \citep{Choi:2020:F2Softmax}. Clarification utility maximization (next subsection) also implicitly addresses this issue.

\subsubsection{Clarification Utility Maximization Models}
An alternative to the presented sequence-to-sequence models that maximize the likelihood of generating clarification observed in the training set is clarification utility maximization models. The intuition is to generate a question that best clarifies the user information need, while there is no notion of clarification in the training objective of sequence-to-sequence models. 

In this approach, the goal is to maximize a clarification utility function $U$ that measures the likelihood of clarifying the user information need or a similar objective. For instance, \citet{Rao:2019:naacl} estimated the information value of the possible answer that a user may give to the generated clarification as a utility function. \citet{Zamani:2020:GCQ} estimated the likelihood of covering all information needs observed in the query logs based on the past interactions. 

The clarification utility functions are often non-differentiable, which prevents us from using gradient descent based optimization. Therefore, clarification generation can be modeled as a reinforcement learning task whose reward function is computed based upon the clarification utility function $U$. The REINFORCE algorithm~\citep{Williams:1992:REINFORCE} can then be used for learning the clarification generation model. It has been shown that using the models that are pre-trained using maximum likelihood training for the REINFORCE algorithm can lead to more effective and more robust outcomes. This approach is called Mixed Incremental Cross-Entropy Reinforce (MIXER)~\citep{Ranzato:2016:MIXER}. For more information, we refer the reader to~\citet{Rao:2019:naacl,Zamani:2020:GCQ}.

\subsection{Selecting Clarifying Questions}
Clarifying question generation models can be evaluated using human annotations or online experimentation. However, both of these approaches are time consuming and are not always available. On the other hand, offline evaluation based on text matching metrics, such as BLEU~\citep{Papineni:2002:BLEU} and ROUGE~\citep{Lin:2004:ROUGE}, are not reliable for clarification generation models. Therefore, due to the challenges in offline evaluation of clarifying question generation models, \citet{Aliannejadi:2019:sigir} introduced the task of \emph{selecting} clarifying questions from a set of candidate human- or machine-generated clarifying questions. The authors created and released the Qulac dataset, consisting of over 10K human-generated (through crowdsourcing) question-answer pairs for 198 topics associated with the TREC Web Track 2009-2012. An alternative dataset is MIMICS~\citep{Zamani:2020:MIMICS} that contains over 450K unique real queries and machine-generated clarifying questions along with user engagement signals (\ie, clickthrough rate). The more recent MIMICS-Duo dataset \citep{Tavakoli:2022:MIMICS-Duo} enables both online and offline evaluation of clarifying question selection tasks.

Baseline models that use a combination of contextual representations of the query and clarifying questions (\eg, BERT) and query performance prediction indicators (\eg, standard deviation of retrieval scores) demonstrate the best performance on clarification selection tasks on Qulac~\citep{Aliannejadi:2019:sigir}. \citet{Zamani:2020:ASQ} showed that the clarifying question selecting model can benefit from query reformulation data sampled from search engine query logs.
Subsequent work by \citet{Hashemi:2020:GT} proposed Guided Transformer, an extension to the Transformer architecture that uses external information sources (\eg, pseudo-relevant documents) for learning better representations for clarifying questions. This model significantly improves upon the baseline models for clarification selection tasks. Specifically, they showed that the model performs well for clarifications with short negative responses. Subsequently, \citet{Bi:2021:CQNeg} focused on a BERT-based model for clarification selection based on negative feedback.  This model works well for document retrieval when clarifying questions are asked. \citet{Kumar:2020:ClarificationNLI} looked at clarification selection as a special case of natural language inference (NLI), where both the post and the most relevant clarification question point to a shared latent piece of information or context.
Both clarifying question generation and selection tasks are still active areas of research in both the IR and NLP communities.

\subsection{User Interactions with Clarification}
The way users interact with clarification can reveal information on the clarification quality. For example, user engagement with clarifying questions can be studied as a proxy to measure clarification quality. \citet{Zamani:2020:ASQ} studied how users interact with clarifying questions in a web search engine. They found out that more specific questions have a higher chance to engage users. They showed that the majority of engagement comes for one of two reasons: (1) high ambiguity in the search queries with many resolutions, and (2) ambiguity but where there is a dominant ``assumed'' intent by users where they only realize the ambiguity after issuing the query. Interestingly, users are more likely to interact with clarification in case of faceted queries in comparison with ambiguous queries. Note that the user interface may affect these findings. For instance, in the web search interface with ten blue links, users can simply skip a clarification and directly interact with the retrieved web pages. However, this may not be possible in a conversational search system with a speech-only interface. Therefore, besides generating high-quality clarifying questions, (spoken) CIS systems should make a (binary) decision at every step on whether to ask a clarifying question or to show the result list or answer. \citet{Wang:2021:RiskClarification} addressed this issue by developing a risk-aware model that learns this decision-making policy via reinforcement learning. Their model considers the common answers to each clarification in order to minimize the risk of asking low-quality or out-of-scope clarifications. The model enables the CIS system to decide about asking a clarification with different levels of user tolerance. 

In a separate line of research, \citet{Tavakoli:2021:Clarification} studied user interactions with clarifying questions in asynchronous conversations. They focused on user interactions in community question answering websites, e.g., StackExchange.\footnote{\url{https://stackexchange.com/}} To study user interactions, they categorized clarifying questions to three categories: (1) clarifications that have been answered by the Asker (the person who submitted the questions/post), (2) clarifications that have been answered but not by the Asker, and (3) clarifications that are left unanswered. They found that clarifications with the goal of disambiguation account for the majority of clarifying questions and they are very likely to be answered by the Asker. On the other hand, clarifications with the goal of confirmation are more likely to be left unanswered. For more analysis on user interactions with clarification in asynchronous information seeking conversations, refer to~\citet{Tavakoli:2021:Clarification}. 


\section{Preference Elicitation in Conversational Recommendation}
\label{sec:mixed_init:pref_elicit}
\hamed{link to section 5}
Preference elicitation in conversational recommender systems forms another type of mixed-initiative interactions. Typically, recommender systems create a user profile or user representation based on the user's past interactions (e.g., click)~\citep{Jannach:2018:ImplicitFB,Oard:1998:ImplicitFB} and/or her explicit feedback on items using ratings and reviews~\citep{Resnick:1994:GroupLens,Ricci:2010:RecSysHandbook}. Conversational systems enable recommendation engines to ask for user preferences in a natural language dialogue. This creates a significant opportunity for the system to learn more about the current context of the user, and how their preferences at this point in time may differ from their preferences in general. \citet{Christakopoulou:2016:KDD} studied the task of conversational recommender systems by focusing on preference elicitation in a closed-domain scenario, like restaurant recommendation. They observed 25\% improvements over a static model by asking only two questions. Following their work, \citet{Sun:2018:ConvRecSys} proposed a reinforcement learning model for preference elicitation by asking questions about item facets in a closed-domain setting, i.e., restaurant recommendation. \citet{Zhang:2018:cikm} focused on a broader domain by automatically extracting user preferences about item facets from user reviews on an online e-commerce website. They showed that multi-memory networks can be successfully used for asking questions about item facets in their setting. \citet{Sepliarskaia:2018:PrefElicit} used a static questionnaire to ask questions from users in the context of movie and book recommendation. They studied different optimization strategies for the task with a focus on cold-start users. In this work, user responses to the system questions are automatically generated and may be different from real-world settings. To mitigate this issue, \citet{Radlinski:2019:PrefElicit} conducted a crowdsourcing experiment with a wizard-of-oz setting, where a crowdworker plays the role of user and another person (i.e., assistant) plays the role of the system. They introduced a ``coached'' preference elicitation scenario, where the assistant avoids prompting the user with specific terminology. 

The mentioned methods ask questions about items and item attributes for preference elicitation. In case of incomplete information on item attributes, \citet{Zhao:2022:PrefElicitKnowledge} proposed a knowledge-aware preference elicitation model. Moreover, users may not be able to answer all questions about item attributes especially if they have limited knowledge. More recently, \citet{Kostric:2021:Questions} proposed to address this issue by asking questions about item usage, which is related to ``purpose'' in the clarification taxonomy presented in Section~\ref{sec:mixed_init:clarification:taxonomy}. 
Preference elicitation in recommendation is tightly coupled with the design of conversational recommender systems. Refer to Section~\ref{sec:chapter5:recsys} for further information.


\section{Mixed-Initiative Feedback}
\label{sec:mixed_init:feedback}
The system can take advantage of mixed-initiative interaction to get feedback from users and even give feedback to them. For instance, in the middle (or at the end) of a dialogue in a conversational recommender system, the system can ask for explicit feedback from the users. Existing systems often have a static pre-defined questionnaire that will automatically be triggered after a conversation ends. For instance, the Alexa Prize Challenge~\citep{Ram:2018:Alexa} has sought explicit rating feedback from users upon the completion of the conversation and used average ratings for evaluating the participant teams. This simple approach can be further improved by asking context-aware questions for feedback and making natural language interactions within the conversation. 

Mixed-initiative feedback can be also relevant to the concept of ``grounding as relevance feedback'' introduced by \citet{trippas2020towards}. Grounding is defined as discourse for the creation of mutual knowledge and beliefs. The authors demonstrated grounding actions in a spoken conversational search data, such as providing indirect feedback by reciting their interpretation of the results. This grounding process can potentially enable CIS systems to better understand a user's awareness of the results or information space. 

As mentioned earlier, mixed-initiative interaction can be used to give feedback to the users. As an emerging application, users may not directly know how to effectively use the system. Hence, the system can take advantage of this opportunity to educate users on the system capabilities. Educating users on interacting with CIS systems has been relatively unexplored.


\begin{figure}[t]
    \centering
    \includegraphics[width=0.6\textwidth]{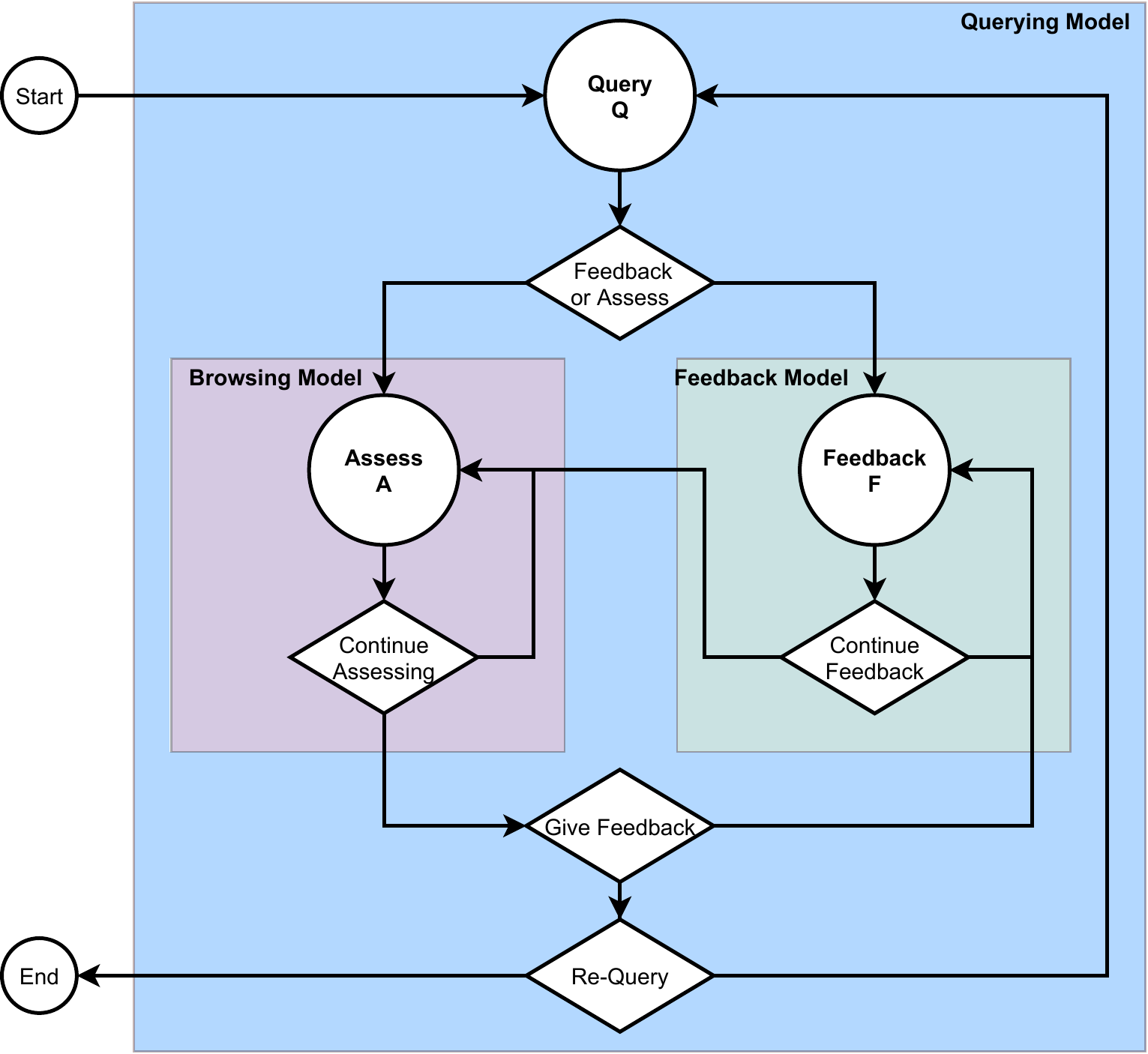}
    \caption{A user model of mixed-initiative conversational search proposed by \citet{Aliannejadi:2021:AMI} which is composed of three sub-components the Querying, Browsing and Feedback Models. Diamonds represent user decision points, while circles represent the action/turn taken.}
    \label{fig:chapter6:usermodel}
\end{figure}

\section{Modeling Mixed-Initiative Strategies}
The CIS system needs to make decision on what action to take at each timestamp and mixed-initiative interactions significantly increase the number of options, resulting in a complex decision making problem. Thus, formulating, modeling, measuring, and simulating mixed-initiative information seeking conversations is quite important. \citet{Aliannejadi:2021:AMI} proposed a user model for mixed-initiative conversational search that consists of three major phases: querying, feedback (\ie, mixed-initiative), and browsing (\ie, assessing search results). This user model is shown in Figure~\ref{fig:chapter6:usermodel}. Based on this user model, they considered two extreme cases. (1) Feedback First, where the system first asks for multiple feedback (\eg, clarification) once the user submits the query and then presents the results, (2) Feedback After, where the results are shown first and then unsatisfied users can provide feedback to refine the search results. To measure each conversation they rely on gain to cost ratio, where gain is defined by the relevant documents assessed by the user and cost is defined by the time the user spent on each conversation. Note that the definition of gain and cost can be simply revisited, if needed. Through extensive simulations by modeling the gain to cost ratio, \citet{Aliannejadi:2021:AMI} provided guidelines for taking mixed-initiative interactions in different situations, for example for patient and impatient users. Such modeling is later extended by proposing an economic model of conversation search \citep{Azzopardi:2022:EconomicConvSearch}. This theoretical framework for conversational search can provide insights to guide and inform the development of conversational search agents.

\section{Summary}
In this section, we discussed the opportunities and challenges that mixed-initiative interactions bring to CIS systems. We drew connections with mixed-initiative user interfaces and mixed-initiative interactions in dialogue systems. We discussed system-initiative CIS and reviewed different purposes for conversation initiation. 
We also provided an overview of clarification in CIS systems and how a clarifying question can be generated or selected to identify the user's intent. We briefly reviewed preference elicitation and demonstrated its connections with intent clarification. We finished by showing how systems can get feedback from and give feedback to the users through mixed-initiative interactions.

Overall, understanding mixed-initiative interactions and initiating conversations have been identified as a key part of CIS research. Clarification, as a form of mixed-initiative interaction, has been studied quite extensively. However, other forms of mixed-initiative interactions require further significant efforts. Evaluating mixed-initiative CIS systems is another under-explored yet important research area. 

\chapter{Evaluating CIS Systems}
\label{chapter7:evaluation}

Evaluation of conversational information seeking systems continues to be a rapidly evolving research area due to unique challenges of assessing the quality of \emph{conversations}, and the parallel difficulty in creating benchmark datasets.

In contrast to non-conversational information seeking settings, the multi-turn nature of conversations requires evaluations to model long-term state, and consider \emph{what} information is conveyed, \emph{when} the information is conveyed, as well as \emph{how} this communication happens. All these are dependent on \emph{why} a user is engaging in a conversational interaction in the first place (as opposed to non-conversational alternatives). The same conversation may be considered of high or of low quality depending on context: For example, if a user is in a rush or not, or if the user requires high confidence in the conclusion or not.


\section{Categorizing Evaluation Approaches}

There are a number of ways that CIS evaluation may be presented. We structure Section~\ref{chapter7:evaluation} by evaluation modality: Offline or online evaluation, and sub-types of these modalities.

However, evaluation approaches can be broken down in other ways (see Chapter 4.2 of \citep{Anand:2020:Dagstuhl}). We summarize some here as researchers in CIS may find some of the specific micro-evaluation or user-centric questions particularly pertinent to the research questions being asked in a given setting.

For example, individual components of conversations can be evaluated at a micro-level, leading to a catalogue of micro-evaluation techniques including \emph{How well does the system predict the dialogue act of a given utterance? How well does the system predict the user's goals and sub-goals? Can the system identify terms in statements to fill slots in a structured search query? How well does the system select responses from a set of candidates? How well does the system answer individual questions?} 
As we shall see, such an evaluation approach has the benefit that these questions lend themselves well to traditional information retrieval evaluation approaches. A major drawback, however, is that high performance on micro-level metrics does not necessarily translate into a CIS system being effective for satisfying users' needs.

An alternative is to break down by evaluation approaches in a user-centric manner: \emph{Does the user trust the system? What is the cognitive load of interactions? How fluent and efficient is the system in communication in general?}
Within the context of a particular information need, one can seek metrics to evaluate based on properties such as \emph{Is the user satisfied with the outcome of the conversation? How much effort and/or time was required to satisfy the information need? Is the information need ultimately resolved? Was the user frustrated in the process?} For such metrics, subjectivity is a common concern. Additionally, while such evaluation does assess the overall quality of a CIS system, such metrics are particularly difficult to optimize.

\tipbox{Individual components of CIS systems can be evaluated at a micro-level. A major drawback, however, is that high performance on micro-level metrics does not necessarily translate into overall user satisfaction. Thus, an alternative is end-to-end user-centric evaluation methodologies.}


\begin{sidewaystable}
    \centering
     \resizebox{\linewidth}{!}{%
    \begin{tabularx}{\linewidth}{p{4cm}lp{1.7cm}p{4.3cm}l}\toprule
        \textbf{Dataset} & \textbf{Domain} & \textbf{Task} & \textbf{Construction} & \textbf{Scale} \\\midrule
        {CAsT 2019 - 2022  \citep{Dalton:2019:TREC}} & open domain & ConvPR & {questions written by organizers \& passage pooling} & 100+ conversations \\
        {CoQA  \citep{Reddy:2019:tacl}} & seven domains & ConvQA & wizard-of-oz & 1K+ conversations \\
        {QuAC  \citep{Choi:2018:emnlp}} & people & ConvQA & wizard-of-oz & 10K+ conversations \\
        {MISC  \citep{thomas2017misc}} & open domain & CIS & {spoken human conversations} & 10+ conversations \\
        {Qulac  \citep{Aliannejadi:2019:sigir}} & open domain & CIS clarification & crowdsourcing & 10K+ clarifications \\
        {MIMICS  \citep{Zamani:2020:MIMICS}} & open domain & CIS clarification & search logs & 100K+ clarifications \\
        {RiDial  \citep{li2018redial}} & movies & ConvRec & wizard-of-oz & 10K+ conversations \\ \bottomrule
    \end{tabularx}
    }
    \caption{Examples of notable datasets for various conversational information seeking tasks: conversational passage retrieval (ConvPR), conversational question answering (ConvQA), clarification in conversational search, and conversational recommendation (ConvRec). There are many other datasets that are not listed here.}
    
    \label{tab:datasets}
\end{sidewaystable}

\section{Offline Evaluation}
\label{sec:chapter7:offline evaluation}

As a staple of information retrieval evaluation, offline evaluation permits reproducible evaluations that can reliably compare different systems. We start with a discussion of some of the existing datasets commonly used to evaluate CIS systems. Following a summary of each category of dataset, we present open challenges with respect to offline evaluation of CIS tasks.


\subsection{Conversational Datasets}

Conversational datasets are transcripts of actual conversations that have occurred between two or more parties, either as part of natural information seeking or through a role-play conversation exercise. Table~\ref{tab:datasets} reports a few notable CIS datasets.

We begin by observing that some conversational datasets are \emph{synchronous} (\eg,~\citep{Budzianowski:2018:Multiwoz}), while others are \emph{asynchronous} (such as datasets derived from Reddit~\citep{Henderson:2019:Repository}). Although, in principle, the content of these can be similar, subtle timing effects can lead to meaningful practical differences. For instance, asynchronous conversations may contain fewer disfluencies and unintentional errors as participants take time to consider their utterances \citep{Serban:2018:Corpora}. 
Asynchronicity also makes it possible to carry out time-consuming tasks such as consulting external sources between conversational turns. Of particular importance to studies of mixed initiative, the role of initiative and conversational turn taking is very different in synchronous and asynchronous conversations \citep{Gibson:2009:Intercultural,Boye:2000:Asynchronous}.

An example of a widely used conversational dataset is Multi-WOZ \citep{Budzianowski:2018:Multiwoz}. Consisting of synchronous naturalistic task-oriented dialogues designed to simulate a possible conversation between a tourist and information agent, it focuses on recommendation tasks with well-defined slots and values. To create these, one person is presented with search criteria, while a second (``wizard'') has access to a search system that allows them to identify recommendations that satisfy the ``user's'' constraints. However, by presenting such specific requirements that perfectly match the wizard's known fields, it may be argued that the conversations can be somewhat unnatural. The TaskMaster dataset \citep{Byrne:2019:Taskmaster} generalizes on the Multi-WOZ idea, with dialogues around making orders and setting up appointments, such as ordering a pizza or creating an auto repair appointment.
In addition to synchronous wizard-of-oz dialogues similar to those from Multi-WOZ, the authors also include asynchronous self-dialogues where a single person types both sides of a conversation, focusing on given needs. To make the conversations more natural, the authors also instructed raters to intentionally include understanding errors and other types of dialogue glitches, with some conversations created to be intentionally unsuccessful.
This type of dataset is predominantly used for the evaluation of slot-filling algorithms. As an alternative to task-oriented dialogues, \citet{Radlinski:2019:PrefElicit} presented Coached Conversational Preference Elicitation, intending to obtain realistic synchronous dialogues by instructing a ``wizard'' to simply motivate a ``user'' to describe their preferences, without setting a detailed goal for either.

Another category of conversational datasets is used for conversational question answering \citep{Iyyer:2017:Sequential,Choi:2018:emnlp,Reddy:2019:tacl} or TREC CAsT Track~\citep{Dalton:2019:TREC,Dalton:2020:TREC}. Here the major challenge addressed is co-reference resolution, evaluating the systems ability to answer questions in sequence, particularly when a given question may refer to earlier questions or their answers (for example, ``Who won the superbowl?'' followed by ``Who is their quarterback?''). Such dialogues can be sampled from search engine interactions, known answers, or manually constructed.

Two more types of conversational datasets are commonly used in developing CIS systems. Asynchronous discussions on a given topic, often from the Reddit forum (for example, \citep{Henderson:2019:Repository, Qu:2018:sigir,Qu:2019:chiir}), are often used to model open-ended conversations. 
As a massive corpus of free-form dialogues, these exchanges can be used to train and evaluate conversational agents with a goal of responding reasonably to any utterance on any topic without an assumption of a particular task. 
Of course, it is important to note in the context of web forums that careful attention must be paid to the representativeness of the authors of the corpus being used. For instance, training or evaluating CIS systems based on a forum with a particular type of contributor may lead to bias in a CIS system evaluation, and may lead to undesirable conversational behaviors being learned if they mirror the behavior of the authors who contributed to that forum. For instance, language ranging from microaggressions to insults or worse is often observed \citep{Bagga:2021:You}. For this reason, the use of massive web corpora must be done with care.
Other formus, like Slack, can similarly be used \citep{Sabei:2022:SCC} to observe asynchronous communication. To obtain open-ended synchronous conversations with higher quality than may be expected in an open forum, transcripts of movie and television dialogues are frequently used \citep{Muller:2013:StatisticalMT,Henderson:2019:Repository}.

There are numerous challenges in creating and using conversational datasets for offline evaluation. One of the key challenges is that the motivation of the participants can greatly influence the dialogues observed. In a wizard-of-oz setting, if the wizard is provided with a particular interface to obtain answers for user requests, this is likely to influence their utterances \citep{Radlinski:2019:PrefElicit}. If the user is given detailed instructions, especially if these do not align with the person's actual interests, this again can result in unnatural dialogue \citep{Serban:2018:Corpora}. If several wizard-of-oz datasets are used together for evaluation, they may uncover slight differences in the study setup impacting the conversations \citep{Trippas:2019:biirrr}. Moreover, if users are asked to complete predefined tasks, there is a risk that they do not approach these tasks as someone who actually \emph{wants} to perform that task \citep{Serban:2018:Corpora}. For example, suppose a user is tasked with purchasing something under a given price. A real user may exhibit certain flexibility regarding the price, or may ask questions relating to value for money, rather than solely around price -- and examples of realistic behavior around pricing may end up missing from the collected corpus.
A second major challenge in evaluation with offline datasets lies in how the datasets are interpreted. Where dialogues are taken to contain \emph{correct} responses in a particular context, they can suffer from false negatives: A perfectly capable system may be judged to perform poorly when it is simply performing the task \emph{differently} \citep{Finch:2020:Towards,Zhang:2020:Evaluating,Sekulic:2022:Evaluating}. 

\subsection{Single-Step Datasets}


As a step towards fully conversational systems, a number of challenges have been proposed to address the necessary sub-tasks. Here we refer to them as single-step datasets, as the focus is on a single step within the many that a conversational system must perform. We note that they do not focus on \emph{single dialogue turns} (as is the case with Conversational QA datasets), but even more fundamental steps of information processing. 

One recent example is generating the natural text from structured information to describe a particular search result, as the conversational equivalent of search snippet generation \citep{Turpin:2007:Fast}. For instance, suppose a conversational agent needs to explain a specific restaurant to a user, showing how it satisfies their request. The agent may possess rich structured information about the restaurant -- its name, address, the type of food offered, pricing information, and other key attributes. However, just presenting these facets of information to the user may not be suitable.
The End-to-End NLG Challenge \citep{Dusek:2018:ACL} produced a dataset mapping a set of attributes to natural language descriptions, allowing a challenge for generating text from structured information -- a critical single step of many CIS systems.

A second example where single-step datasets are used is for applications where generating full text is unnecessary. This common task treats conversational information seeking as the ranking of possible (existing) responses that an agent could give at a particular time. For instance, \citet{Yang:2018:sigir} described datasets derived from transcripts of past technical support dialogues: They assume that for any given user utterance, the system should select from previous agent utterances (as most technical support problems are not novel). Such specialized single-step datasets will address this single-turn ranking problem.

As a third example, when an agent asks a question, it must be able to interpret the user's answers. Taking the seemingly simple case of yes/no questions, a user may answer indirectly. For instance, if an agent asks if a user would be interested in an evening activity, the user may say ``I'd prefer to go to bed'' rather than simply ``no''. The Circa dataset \citep{Louis:2020:Indirect} was developed to contain natural questions and answers to train and evaluate reliable answer interpretation by CIS systems. The approach used multiple phases of crowdworker tasks first to develop natural questions and then, in turn, natural answers while attempting to minimize bias and maximize the diversity and naturalness of answers.

\subsection{Simulated Users}


A recent alternative to static conversational datasets is relying on simulators \citep{ie:2019:recsim,Aliannejadi:2021:AMI,Salle:2021:Studying,Erbacher:2022:SimulationClarification}. For instance, \citet{Zhang:2020:Evaluating} argued that a simulator ``should enable to compute an automatic assessment of the agent such that it is predictive of its performance with real users''. In this way, rather than evaluating with a fixed dataset, an agent could be assessed dynamically against a (fixed) simulator to obtain the benefits of effective offline evaluation. As another example, \citet{Sekulic:2022:Evaluating} develop a simulator capable of answering clarifying questions posed by a CIS system. Both these recent works showed a high correlation between simulation-based evaluation and an online evaluation approach. Simulation also addresses challenges in fixed datasets, particularly relating to user privacy \citep{Slokom:2018:Synthetic,Hawking:2020:Simulating}.

Although long studied in information seeking in general, this is a relatively new methodology in the context of CIS. As such, it has been the subject of two recent workshops \citep{balog:2021:sim4ir, Ekstrand:2021:Simurec}. These identified particular open challenges: Developing increasingly realistic user simulators, and making simulators easier to share. It was observed that one particularly pertinent still open question is ``\emph{how realistic simulators can be, or indeed should be}'' noting that simulations need only correlate well with other approaches \citep{balog:2021:sim4ir}. For instance, \citet{Zhang:2022:Simulating} considered how to design simulators to reformulate their utterances when a conversational agent fails to understand them similarly to how human do. As such, the general problem evaluation/validation of simulators itself is also an open area to ensure simulation-based evaluation is valid. 

\subsection{Datasets Beyond the Text}

Several authors have considered evaluating CIS tasks beyond simply the text of interactions between a user and a CIS system. Typically this involves additional annotation of the conversational dialogue to indicate relevant aspects, although it can also involve other content modalities in addition to the conversation.

One example is the annotation of the high-level role of individual utterances. This may be at the level of sentences within a conversation (annotated as to whether they are asking a question, sharing an opinion, thanking, or so forth) \citep{yu:2019:MIDAS}, or may be at the level of the high-level structure of conversations as in the case of sub-goal or sub-task prediction. Alternatively, user-centric metrics may be annotated, such as indicators of customer frustration at specific points in customer service conversations \citep{Oraby:2017:Customer}.
Note that these evaluation annotations are in contrast and complementary to datasets which have been annotated to investigate \textbf{how} interactions between the user and CIS system are structured~\citep{Vakulenko:2021:TOIS, trippas2020towards}. A key challenge in such datasets is ensuring that the (indirect) labels produced by raters agree with the (direct) opinion of actual participants. Promisingly, \citet{Fu:2022:Paradigm} recently studied this question and found that it is possible to collect labels where there is a fair agreement between direct and indirect assessments at least in terms of user satisfaction. 

A fundamentally different type of CIS dataset involves multiple modalities. The conversation may include both text, images, or gestures to illustrate the user's need in a recommendation setting \citep{nie:2019:multimodal,deldjoo:2021:SIGIR}, or even include navigation within a virtual or physical environment as part of the conversational task \citep{ku:2020:room}.


\section{Online Evaluation}
\label{sec:chapter7:online evaluation}

In contrast to offline evaluation, CIS systems may also be evaluated \emph{online}: deploying a system that real users interact with, dynamically obtaining user utterances and the system's responses.

\tipbox{Online evaluation allows systems to be evaluated much more robustly, as the consequences of earlier system actions can be seen in how users respond, which in turn determines what options the system has and how these are handled. In this way, online evaluations are much more predictive of real-world system performance, and is more likely to identify limitations in current solutions.}

Online evaluation can be done in one of two ways: (1) a lab or crowdsourcing study, or (2) a real-world study.

\subsection{Lab or Crowdsourced Studies}
\label{sec:chapter7:crowdsourced}

It is often desirable to evaluate components of a system that is not end-to-end complete (such as when developing specific aspects of a CIS system), or where it is necessary to control certain conditions (such as when performance for specific use cases is of particular interest). In this situation, paid users or volunteers are often employed.

For instance, \citet{Christakopoulou:2016:KDD} studied different approaches for eliciting user preferences in a restaurant recommendation setting. As the authors' goal was to assess how well different ways of asking questions efficiently established users' interests, the authors chose to perform a lab study. Participants were presented with preference questions that a conversational system might want to ask. The results were used to inform algorithms for learning about users interests. This type of evaluation was appropriate as the information could not be collected through an offline corpus (as rating data in offline studies is usually incomplete), nor in a real-world system (as preference elicitation studied here is but one part of the overall CIS recommendation challenge).

Similarly, \citet{Aliannejadi:2019:sigir} introduced a crowdsourced approach for evaluating clarification question selection. They started with a variety of queries, crowdsourced a collection of possible clarifying questions, then collected possible answers to these questions. Despite simplifying assumptions, the approach allowed a clarifying question selection model to be evaluated based on the retrieval performance, giving possible answers to the system's potential questions. For the same task, \citet{Zamani:2020:MIMICS} provided guidelines for manual annotation of clarifying questions and their candidate answers based on their fluency, grammar, usefulness for clarification, comprehensiveness, coverage, understandability, diversity, coherency, and so forth.

Evaluating a different aspect of CIS behavior, \citet{Balog:2020:Measuring} studied the role of explanations in recommendation tasks. As one may expect explanations of results presented to be part of CIS, the authors focused on assessing what constitutes a valuable explanation. Using a non-conversational approach, crowdworkers were first asked to express their preferences in a given domain. They were then presented with recommendations along with explanations. These explanations were assessed using a focused questionnaire addressing different reactions the participants may have to the explanations.

As another example, \citet{Jiang:2015:www} recruited participants to complete specific tasks using a well established CIS agent, including specific search tasks. After the tasks were complete, participants were asked to answer specific questions about their experiences. Based on the answers to these questions and a record of the participants' interactions with the CIS system, the authors developed an automated approach for predicting satisfaction and natural language understanding.

As these examples show, controlled studies can allow investigation of the performance of particular aspects of CIS. A detailed treatment of designing user studies for interactive IR systems is presented by \citet{Kelly:2009:MEI}.

\subsection{Real-World Studies}

When a CIS system is complete and a fully realistic evaluation of the users' overall experience is desired, a real-world study is the gold standard. This involves observing actual natural interactions between users and the CIS system, particularly with users motivated by relevant information needs. The key difference between such studies and lab or crowdsourced studies described in Section~\ref{sec:chapter7:crowdsourced} above is that of \emph{motivation}. Specifically, in real-world studies the user comes with their own rich needs (which may or may not be clear to the user from the start), and they may be satisfied or dissatisfied with any aspect of a CIS system. They may choose to engage with a system, or simply leave if some aspect of performance is poor --- or perhaps just become distracted by something outside the system designers' control. Given sufficient scale, the conclusions of such an evaluation are most likely to generalize to other users with other needs and in other contexts. 

The key consideration is that while on one hand users bringing their own information needs leads to more realistic interactions, on the other such an evaluation depends on actual interactions with only limited feedback usually available. As an example of such a study, \citet{Park:2020:Hybrid} presented a study of a commercial CIS agent where, for some intents (such as asking for weather), the agent asked for user feedback. In particular, the agent asked users ``Did I answer your question?''. Responses to this question were used to assess the quality of the end-to-end CIS system.
A similar approach is used in the Alexa Prize Challenge \citep{Ram:2018:Alexa}. Here, real users may request to interact with a conversational system. At the end of the interaction, the user is asked to rate their experience. Such online evaluation can assess the quality of the conversational abilities of the system according to predetermined criteria (here, user-declared satisfaction, and level of engagement based on time spent).


\section{Metrics}
\label{sec:chapter7:metrics}


Having considered evaluation approaches, here we briefly discuss an essential aspect of CIS evaluation separately, namely that of metrics. While a complete treatment of metrics suitable for conversational information seeking is beyond our scope, we provide a high-level overview of the metric types used in different cases, and some of the appropriate considerations that are required when determining the right ones. We refer the reader to~\citet{Liu:2021:Metrics} for a more extended treatment of conversational systems' single-turn and multi-turn metrics.

\subsection{Metrics for Individual Steps}

At individual steps, it is possible to evaluate whether the system understood a user's utterance, whether the search system respected a constraint, or whether a system utterance was fluent among other things. Most often, these aspects can be measured with metrics that can be computed offline.

As an example, we take conversational question answering (ConvQA), discussed in depth in Section \ref{sec:cqa}. Often used to assess clarification approaches in the NLP community~\citep{Rao:2019:naacl}, common metrics include BLEU~\citep{Papineni:2002:BLEU}, ROUGE~\citep{Lin:2004:ROUGE}, and METEOR~\citep{Banerjee:2005:METEOR}. At a high level, these metrics match the similarity between a given string and reference strings. While effective for some applications, these metrics do not correlate highly with user satisfaction in conversational systems~\citep{Liu:2016:EMNLP}. More recently, machine learned metrics have achieved significantly higher correlation with manual human ratings for such language tasks~\citep{Ma:2018:Results,Sellam:2020:BLEURT}.

When assessing the relevance of recommendations that terminate a conversational exchange, classic information retrieval metrics are used~\citep{Croft:2010:Search}. For instance, Normalized Discounted Cumulative Gain (nDCG), Mean Reciprocal Rank (MRR), and Precision are often used to assess if recommended items match information needs of users given a particular user representation, \eg,~\citep{Christakopoulou:2016:KDD}, if a system is able to rank possible clarifying question, \eg,~\citep{Aliannejadi:2019:sigir}, or if a system accurately provides answers to specific requests, \eg,~\citep{Christmann:2019:CIKM}. As with language metrics, such metrics do not necessarily agree with user experience of an end-to-end system \citep{Jiang:2016:session}.

As an example of more nuanced refinements of relevance in a conversational setting, consider work by \citet{Rashkin:2021:Measuring}. Here, the authors propose a metric that assesses whether a CIS system only presents verifiable information, rather than hallucinated or factually unverifiable information.

\subsection{Metrics for End-To-End Evaluation}

An essential characteristic of conversational information seeking systems is the multi-turn nature of conversations. As such, it is vital that evaluation considers an end-to-end interaction. For example, consider catastrophic failure in the middle of a long conversation, where an agent may lose the state after a user has provided significant information to a CIS system. \citet{Kiseleva:2016:sigir} showed how one failure in a more extended conversation often leads to dissatisfaction. This can happen even if the vast majority of individual conversational steps are successful.

The richness of conversational interactions thus means that CIS systems can be assessed along many different dimensions. Trivially, one may consider whether users were successful at their task~\citep{Chuklin:2018:prosody, dubiel:2018:investigating} or achieved success quickly~\citep{thomas:2018:CHIIR, Trippas:2017:chiir}. Despite this, a shorter time to success is not necessarily sufficient. For instance, in a non-conversational recommendation setting, \citet{Schnabel:2016:Shortlists} showed that more successful recommendations may be obtained using systems that require more prolonged user interactions,  leading to overall higher user satisfaction. In conversational settings, a system may trade-off long-term and short-term utility \citep{Radlinski:2017:CHIIR}. It is important to note that it is also possible to succeed while leaving users frustrated, as studied by \citet{Feild:2010:Frustration}. A particular end-to-end evaluation approach was recently presented by \citet{Lipani:2021:Evaluating}, based on the flow of different subtopics within a conversation. 

Two other classes of metrics are critical to consider. First, \emph{trust} between the user and a CIS system. For example, \citet{daronnat:2020:impact} studied how trust affects users satisfaction. Trust usually requires factuality. It has been noted that some modern neural conversational systems can produce utterances that are false (often termed hallucinations). A detailed treatment of hallucination, and references to further work, can be found in \citet{Shuster:2021:RetrievalAR}. Trust may also be affected by explanations being incorporated into CIS systems. Explainable AI is, in general, an extremely active and critical area of study \citep{Adadi:2018:Explainable}. In a conversational recommendation setting, explanations have recently received attention as well, for example, see \citep{Chen:2020:Towards,Balog:2020:Measuring}. 

The second critical concept to consider in CIS systems is that of \emph{fairness}. While often not treated as a key metric of effectiveness, many researchers have recognized this as a principal desirable aspect of AI systems in general and recommendation systems in particular. A CIS system that provides recommendations in the course of a conversation, for instance, may aim to do so in a fair manner. Thus biases that may be present within the conversational system warrant careful consideration. We refer interested readers to \citet{Beutel:2019:Fairness,Ge:2021:Fairness} and their citations for definitions, approaches and relevant metrics.




\section{Summary}

This section presented an overview of key concepts in the evaluation of conversational information seeking systems. We provided an overview of offline as well as online evaluation techniques, discussing common methodologies in both cases. Benefits and drawbacks of the two high-level approaches were discussed. Finally, we provided an overview of common metrics used to evaluate CIS systems, as well as references to broader topics that should be considered when measuring the performance of CIS systems, such as trust and fairness.
\chapter{Conclusions and Open Research Directions}
\label{chapter9:Conclusions}

This survey aimed to provide an overview of conversational information seeking (CIS) research, summarizing current research and presenting an introduction to researchers new to this area. We addressed CIS from both a user- and system-centred approach, aiming not to single out one view but provide a holistic overview. CIS could be naively approached as a straightforward pipeline of all components such as user input (\eg, automatic speech recognition), which transcribes the spoken query as input, information retrieval, which identifies and retrieves the relevant items to the query, or information visualization, which summarizes and presents the found information to the user.
However, many more components are needed to make CIS truly useful in solving diverse information needs, including features that can capture and utilize interaction and preference history, adapt results presentations to the user's need or context, and track the conversation flow in long-term representations, and interact with external systems. Indeed, we argue that the interconnectedness of all the CIS building blocks makes them intrinsically interrelated, meaning they should be investigated beyond the sum of the parts. Furthermore, we show that CIS is more than system evaluation, and retrieval effectiveness requires a broad range of techniques. 

CIS is a new interaction paradigm beyond the basic query-response approach. This means that existing knowledge and assumptions of traditional IR should be challenged, reviewed, and expanded. Furthermore, CIS research aims to investigate and develop systems that users use and perceive as genuinely helpful, which means taking actions as well as returning information. The more users interact with CIS systems across diverse tasks and contexts the use cases and types of support the systems can provided will evolve and advance. As such, creating more usable CIS systems will help users adopt and adapt conversational and interactive methods to search for and utilize information.

Current research often makes simplifying assumptions about user interactions and system capabilities. Given these assumptions, this monograph showed that large-scale pre-trained language models have many applications in developing different parts of CIS systems that deal with natural language, \eg,~conversational search, question answering, preference elicitation, and clarification. However, deciding about the interaction type, modality, initiative, explanation, \emph{etc.}~involves many components that must work cooperatively with such models for optimal understanding and generation. 

We provided an overview of evaluation methodologies for CIS research. Due to the interactive nature of CIS systems, developing reusable datasets and user simulation strategies for model training and offline evaluation is incredibly important and challenging. Again, most existing benchmarks and evaluation methodologies make numerous simplifications to the CIS tasks.
Currently, online evaluation and collecting human annotations are the most robust and reliable approaches for evaluating CIS systems, although simulation is also gaining popularity. 

It can be challenging to negotiate all the different components of CIS, being ethical and rigorous in the research while maintaining a vision of an information system that does not hinder access to information. We hope that the overview of the broad range of research topics within CIS reflects the various research disciplines that should be part of the conversation studying and developing CIS.

\section{Open Research Directions}
\label{chapter8}

Many open questions and directions of research have been mentioned throughout this monograph. In this section, we bring many of them together with the aim of providing a unified starting point for researchers and graduate students currently investigating conversational information seeking.
While not intended to be exhaustive, we believe these critical areas for future work are particularly likely to have a profound impact on the field. The content of this section can be seen as complementary to directions suggested by the recent Dagstuhl Seminar on Conversational Search~\citep{Anand:2020:Dagstuhl} and the SWIRL 2018 report~\citep{Culpepper:2018:swirl}.

Although some of these topics could be grouped under multiple headings, we divide this section into four main topics, 
(1) \textit{modeling and producing conversational interactions}, which covers the foundation of conversational systems to understand and produce user-system interactions and the information transfer between them,
(2) \textit{result presentation} with different interaction modality and devices,
(3) \textit{types of conversational tasks} that are mostly under-explored,
and (4) \textit{measuring interaction success and evaluation}, focusing on interactivity, ethics and privacy in conversational systems, and lastly, looking at evaluation as a more extensive and broader topic than measuring success.

\subsection{Modeling and Producing Conversational Interactions}
\label{subsec:chapter8:Interactivity}
Interactivity, the process of two or more agents (human or machine) working together, is a crucial characteristic of information seeking. Modeling interactions and deciding the following action or interaction is at the core of CIS research. In this context, although much research has been devoted recently to \textbf{mixed-initiative interactions}, most mixed-initiative strategies have not been fully explored. In fact, our understanding of when a system can take the initiative without disrupting the flow of natural information seeking conversation needs significant further exploration. We believe that systems should more accurately identify opportune moments to initiate the conversation, introduce new topics, or support disambiguation. Similarly, the ability for systems to model \textbf{uncertainty} in user needs (including due to the ambiguity of language) requires further study to effectively and efficiently clarify needs. We argue that supporting all these interactions will enhance the user experience, enable improved information seeking interactions, and thus positively impact this collaborative process.

\textbf{Natural language understanding}, to understand the input from the user (\eg, queries or feedback) needs to be further optimized. This includes the ability of the system to understand complex ideas and concepts from a user's utterance. Furthermore, understanding short, incomplete, or ambiguous queries is still challenging for existing systems.

On top of the aforementioned open research directions for interactions, \textbf{long-term conversational interactions} may need specialized attention. In general, when investigating CIS, it is often assumed that the user is interacting with the system \emph{only} at the time of information need. However, supporting users in long-term information needs, be it multi-session tasks or the ability for a conversation to be continued and repeated much later, need further research. This implies that the history and memory of conversations may be stored and used in future user-system interactions. 
Thus, further research needs to be done on how users want this \textit{memory} to work, including \textbf{privacy and transparency} of what is stored and how the system retrieves and identifies relevant past interactions responsibly.

\subsection{Result Presentation}
\label{subsec:chapter8:Result Presentation}
Presenting results that the user can incorporate into their personal ``knowledge space'', and how the user interacts with them, can be seen as part of a broader challenge of information transfer. Result presentation has not received commensurate attention in the CIS research community relative to its impact. This includes \emph{what} information needs to be presented and \emph{how}.
For example, how can result presentations be optimized with personalization? Can CIS systems use the user's context (\eg, user's location or search history)? Can particular summarization or visualization techniques present results in a concise and easy-to-understand manner?

Furthermore, with the increased interest in \textbf{multi-modal and cross-device CIS}, further research on when, how, and on which device users want to receive information is crucial. Questions such as how CIS systems can/should use \textbf{sensor data} to optimize result presentation is an open problem (\eg, if a user is close to a screen, instead of using a smart speaker, should the information be presented visually?). As part of result presentation, further research on interactions between multiple devices will be pivotal. Thus, research on including more user context to predict how users will interact with the available devices is warranted.

\subsection{Types of Conversational Information Seeking Tasks}
\label{subsec:chapter8:Types of Conversational Information Seeking Tasks}
Many users will have different reasons for why they engage in CIS tasks, with these varying based on the time, context and social situation of their information need. Supporting each user's goals means recognizing these differences. For instance, users interacting with a CIS may choose this search mode to seek advice, look for a detailed summary of a complex topic, or verify a fact.
Developing CIS systems that can \textbf{integrate different kinds of information seeking tasks} and produce a human-like dialogue needs to be better understood. Further, different scenarios or settings may require distinct forms of interaction. For instance, searching for information in enterprise settings contrasts with ``everyday'' search. Conversations may also be structured differently, depending, for instance, on the number of actors in the CIS process, thus making \textbf{collaborative CIS} an essential topic for further exploration.

There are particular challenges for \textbf{domain-specific CIS systems}. Imagine research for a medical-specific system, it may be hard to find researchers with expertise in the particular medical domain and CIS. From a system point of view, it may be challenging to obtain datasets or resources within the medical domain to train and evaluate the CIS systems, this can be because there is hardly any data available or for ethical reasons. Consequently, the lack of data may hinder understanding the specific terminology or language and information seeking tasks. Furthermore, depending on \textbf{who} the end-user is (\ie, a medical professional or a layperson), the system may need to generate different responses addressing different levels of the domain-specific language.

\subsection{Measuring Interaction Success and Evaluation}
\label{subsec:chapter8:Measuring Success}

We distinguish measuring success from evaluation to emphasize the importance of interactivity in CIS. Even though interactivity has always been a major part of information seeking, interactivity becomes even more critical with the paradigm shift from the basic query-response approach to CIS. For example, further research is needed to investigate a more robust \textbf{definition of success in CIS} across different user populations, contexts, and modalities. The CIS interaction model affects the tradeoff between relevance, effort, trust, confidence in the correctness of a result, and the ability to understand the sources of information presented. Thus highlighting the difficulty of defining success since it is changeable depending on the context or modality. 
Currently, there is a lack of evaluation standards defining what constitutes a good conversational search result highlighting the need for a comprehensive benchmark enabling the performance evaluation of CIS systems.
Furthermore, that success may be incredibly personal, and metrics are only helpful when measuring what is desirable for a particular user. As such, further research on \textbf{personalized evaluation} of CIS is needed.

Another factor in measuring interaction success includes tracking how well a user has understood the system and vice versa. This measurement enables systems to increase their capability to explain answers identified and increase confidence in successful information transfer. As already mentioned in Section~\ref{subsec:chapter8:Types of Conversational Information Seeking Tasks}, this kind of conversation calibration can help with the \textbf{transparency} of how systems are
(1)~retrieving, presenting, or augmenting information, 
(2)~handling references to past user-system interactions beyond the current session, or
(3)~aiming to mitigate potential biases generated during the CIS process.

Furthermore, much more work is needed to evaluate the entire CIS process beyond measuring interaction success. Indeed, evaluating any highly interactive process in which users are involved is challenging. Recently, many reusable test sets have been created for CIS tasks. However, many of these sets simplify assumptions about user behaviors and system capabilities. For example, many datasets do not include whether a conversation has resulted in the user achieving their goal or satisfying their information need. Since a conversation is dynamic, evaluating the interaction at particular time points is challenging without users' input. However, including user judgments for each interaction is time-consuming and expensive. Efforts to optimize human evaluation while creating new datasets can dramatically impact CIS research.
This means that ongoing efforts on \textbf{creating datasets that enable training and offline evaluation} are needed. However, to compare datasets and ensure researchers are producing more useful corpora in general, \textbf{dataset protocols} are desirable.
We observe that the ongoing trend in \textbf{simulating users} could be helpful here.

Tools for dataset creation and evaluation may also benefit from further research efforts. For instance, many researchers build their wizard-of-oz frameworks with limited reuse capabilities. Other tools which are more intuitive to use, similar to WYSIWYG (What You See Is What You Get) website builders to test particular interactions, may accelerate all research in CIS.

\chapter*{Acknowledgments}
We would like to thank the researchers who contributed directly or indirectly to the field of conversational information seeking. 
We are also grateful to our colleagues, especially W. Bruce Croft, Reena Jana, and David Reitter, who gave us invaluable suggestions and helped us better position this monograph.
We also thank the anonymous reviewers and the editors of Foundations and Trends in Information Retrieval, especially Maarten de Rijke for his support throughout the process.

This work was supported in part by the Center for Intelligent Information Retrieval, in part by NSF grant number 2143434, in part by the Office of Naval Research contract number N000142212688, in part by an Alexa Prize grant from Amazon, and in part by an Engineering and Physical Sciences Research Council fellowship titled ``Neural Conversational Information Seeking Assistant'' grant number EP/V025708/1. Any opinions, findings and conclusions or recommendations expressed in this material are those of the authors and do not necessarily reflect those of the sponsors.

\appendix
\chapter{Historical Context}
\label{appendixA}
In this appendix, we briefly provide a historical context to information retrieval and dialogue systems research related to conversational information seeking systems. Readers that are not familiar with early IR research are especially encouraged to read this appendix.


\section{Interactive Information Retrieval Background}
\label{appendixA:IIR}



Conversational information seeking systems have roots in interactive information retrieval (IIR) research.
The study of interaction has a long history in information retrieval research, starting in the 1960s~\citep{Kelly:2013:IIRReview}. Much of the earlier research studied how users interacted with intermediaries (\eg, librarians) during information seeking dialogues but this rapidly shifted to studying how users interacted with operational retrieval systems, including proposals for how to improve the interaction. Information retrieval systems based on this research were also implemented. 
\citet{Belkin:1980:ASK} studied the concept of Anomalous States of Knowledge (ASK) of users of IR systems and discussed the importance of multi-turn interactions to help user formulate their needs and help systems successfully retrieve relevant information \citep{Belkin:1986:ASKRetrieval}.
\citet{Brooks1983} studied information seeking dialogues between a user and an intermediary and introduced a annotation coding scheme for discourse analysis of the dialogues.

\begin{figure}
    \centering
    \includegraphics[width=\textwidth]{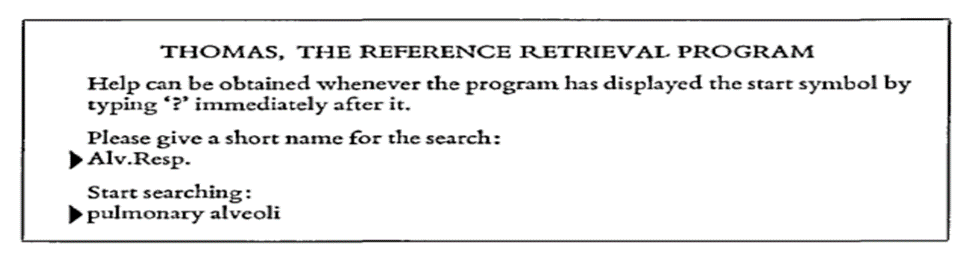}
    \includegraphics[width=\textwidth]{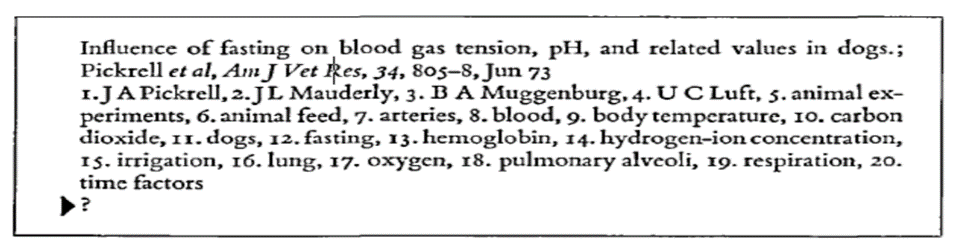}
    \includegraphics[width=\textwidth]{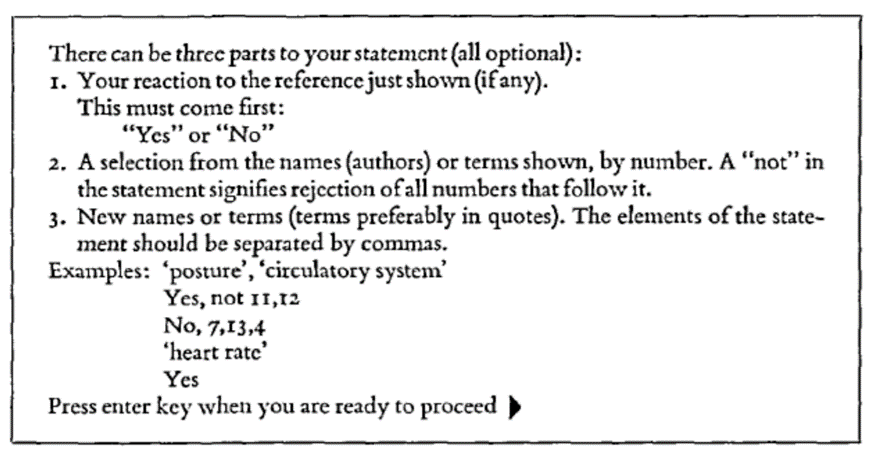}
    \caption{Snapshots from the THOMAS system \citep{oddy1977information}.}
    \label{fig:appendixA:thomas}
\end{figure}

\begin{figure}
    \centering
    \includegraphics[width=\textwidth]{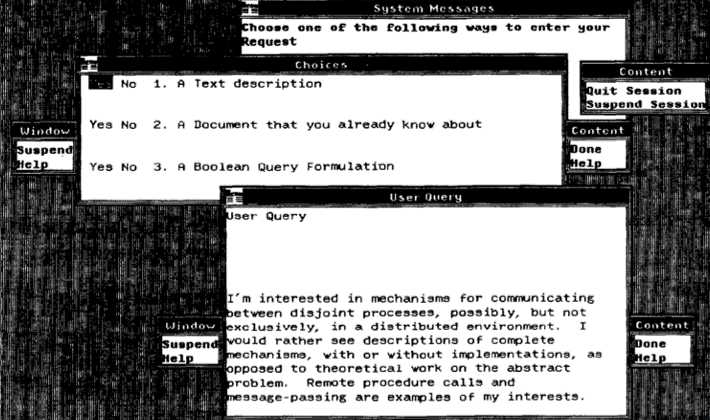}
    \includegraphics[width=\textwidth]{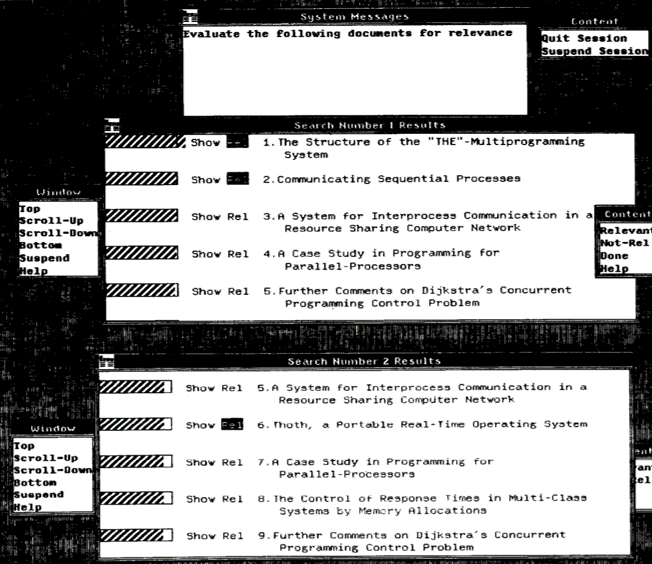}
    \caption{Snapshots from the I\textsuperscript{3}R system \citep{Croft:1987:I3R}.}
    \label{fig:appendixA:I3R}
\end{figure}

\citet{oddy1977information} developed an interactive information retrieval system with rule-based dialogue interactions in 1977, called THOMAS. Example snapshots of user interactions with THOMAS are presented in Figure~\ref{fig:appendixA:thomas}. As shown in the figure, THOMAS includes a few pre-defined interaction types. Even though THOMAS handles a sequence of interactions, it does not model users which is essential for IIR systems. \citet{Croft:1987:I3R} closed this gap by proposing the I\textsuperscript{3}R system -- the first IIR system with a user modeling component. I\textsuperscript{3}R uses a mixture of experts architecture. It assists users by accepting Boolean queries, typical text queries, and documents (query by examples). It enables users to provide explicit relevance feedback to the system. Example snapshots of user interactions with I\textsuperscript{3}R is presented in Figure~\ref{fig:appendixA:I3R}. Later on, \citet{belkin1995cases} focused on user interactions with IIR systems and characterized information seeking strategies for interactive IR, offering users choices in a search session based on case-based reasoning. They defined a multi-dimensional space of information seeking strategies and applied their model to the MERIT system, a prototype IIR system that implements these multi-dimensional design principles.

Since the development of web search engines, research has mostly focused heavily on understanding user interaction with search engines based on an analysis of the search logs available to commercial search engine providers, \eg, see~\citet{Dumais2014LogAnalysis,Buscher2009LoggingWorkshop,Teevan2007Refinding,Murray2007LogAnalysis}.
Since then, explicit modeling of information seeking dialogues or conversations with the aim of improving the effectiveness of retrieval has not been a focus of research until recently. Among them, session search is perhaps the closest research area to CIS (see Section~\ref{appendixA:session}).


\section{Formal Modeling of IIR Systems}
\label{appendixA:formal}
The proposition that IR systems are fundamentally interactive and should be evaluated from the users' perspective is not new~\citep{Kelly:2009:MEI}. This has been highlighted by many pioneers in the field since the 1960s~\citep{Cleverdon:1968:Performance,Salton:1970:IIREvaluation}. However, today's search engines are mostly based on algorithms designed for retrieving documents for a single query. A main reason for this is due to the complexity of IIR modeling and evaluation. Recently, there has been some promising progress in formal modeling of IIR problems, including the probability ranking principle for IIR~\citep{Fuhr:2008:PRPIIR}, the economics models for IIR~\citep{Azzopardi:2011:EconomicsIIR}, the game theoretic framework for IR~\citep{Zhai:2016:GameTheoryIR}, and the interface card model~\citep{Zhang:2015:CardInterface}. Conversational information seeking is an emerging application of IIR systems and many of the developed IIR models and evaluation methodologies can be extended to CIS systems too. For further reading on approaches for designing and evaluating IIR systems, we refer the reader to the comprehensive survey by~\citet{Kelly:2009:MEI} and the tutorial by~\citet{Zhai:2020:IIRTutorial}.


\section{Session-based Information Retrieval}
\label{appendixA:session}
One can put information retrieval tasks in context based on the user's short-term history \citep{Bennett:2012:MIS}, their long-term history \citep{Keenoy:2003:Personalisation}, or their situation \citep{Zamani:2017:WWW}. 
\tipbox{Short-term history is often formulated by the user interactions with the search engine in a short period of time (\eg, a few minutes), referred to as a \emph{search session}. Sessions are different from conversations in that one can pick up and continue a past conversation, while this is not possible in sessions.}

Interactions in a session include past queries, retrieved documents, and clicked documents. Therefore, a session can be considered as a period consisting of all interactions for the same information need \citep{Shen:2005:CIR}. However, this is a strong assumption. In reality, sessions are complex and they are not all alike. Some sessions contain various interactions and query reformulations for a single information need, while some other sessions may involve a series of related simple tasks. Therefore, sessions should be treated differently. This makes modeling search sessions challenging. Existing methods oftentimes relax the assumptions. For instance, \citet{Shen:2005:CIR} assumed that all queries in a session represent the same information need and proposed a model based on the language modeling framework~\citep{Ponte:1998:QL} for session search tasks. In more detail, they provide a more accurate query language model by interpolating the distribution estimated from the current query, with the ones estimated from the past queries and clicked documents. \citet{Bennett:2012:MIS} introduced a learning to rank approach for session search and defined a number of features that can be used for improving the session search performance in web search. TREC Session Track \citep{Carterette:2016:ERO} focused on the development of query formulation during a search session and improving retrieval performance by incorporating knowledge of the session context. Session information can also be used for a number of other information retrieval tasks, such as query suggestion~\citep{Sordoni:2015:QuerySuggestion,Dehghani:2017:QuerySuggestion} and clarification~\citep{Zamani:2020:GCQ}.

Whole session evaluation of IR systems is also challenging. \citet{Jarvelin:2008:sDCG} proposed sDCG, an extension of the nDCG \citep{Jarvelin:2002:nDCG} metric to session search tasks. sDCG basically sums up the nDCG values of all the queries in the session and gives higher weight to the earlier queries. \citet{Kanoulas:2011:nsDCG} later introduced a normalized variation of sDCG, called nsDCG. \citet{Jiang:2016:session} conducted a user study to measure the correlation between these metrics and user's opinion. They found that nsDCG has a significant yet weak correlation with the user metrics. They also showed that user's opinions are highly correlated with the performance of the worst and the last queries in the session. More recently, \citet{Lipani:2019:sRBP} proposed a user model for session search in which users at each step make a decision to assess the documents in the result list or submit a new query. This user model led to the development of the sRBP metric.

It is clear that session search provides a logical foundation for conversational search tasks, however, there are some fundamental differences that necessitates developing novel models and evaluation methodologies for the conversational search tasks. For instance, since most conversational systems are using limited-bandwidth interfaces, the underlying user models of the aforementioned metrics cannot be extended to conversational search. From the modeling perspective, the type of queries in conversational systems are closer to natural language compared to the session search tasks. In addition, unlike in session search, co-reference and ellipsis resolutions play a central role in conversational search. That being said, we believe that the rich history of IR research on session search would be sometimes quite useful in developing and evaluating conversational search systems.




\section{Exploratory Search}
\label{appendixA:exploratory}
A significant research effort in interactive IR has focused on \emph{exploratory search} tasks. Exploratory search is an information retrieval task in which the user is unfamiliar with the search task, unsure about the goal, or even unsure about how to complete the task. 
Users engage in exploratory search with the aim of learning about and exploring a topic -- as opposed to known-item/look-up tasks in which users are focused on finding a fact or answering a specific question.  Exploratory search refers to a broad set of real-world search tasks that involve learning, investigation, planning, discovery, aggregation, and synthesis~\citep{Marchionini:2006:ESF}.  Exploratory search tasks can be generally categorized as (1) exploratory browsing and (2) focused searching~\citep{White:2017:ES}. Previous work on exploratory search has examined interface features to support users with query refinement and filtering (\eg, faceted search)~\citep{Hearst:2006:Faceted}; tools to help gather and synthesize information~\citep{Morris:2008:SearchBar,Donato:2010,Hearst:2013}; and tools to support collaboration~\citep{Golovchinsky:2009:ColabIS}.

\tipbox{Natural language conversation is a convenient way for exploratory search tasks. In many exploratory search tasks, users experience difficulties describing their information needs using accurate keyword queries. This is mainly due to a misconception of the topics and/or the document collection. Information seeking conversations would be the natural solution for this problem as natural language conversation is perhaps the most convenient way of human communication and users can express their exploratory search needs quite easily.} 

Interestingly, many conversations in the TREC CAsT Tracks \citep{Dalton:2019:TREC,Dalton:2020:TREC} are basically addressing exploratory information seeking tasks through natural language conversation. 



\section{Dialogue Systems}
\label{appendixA:dialogue}
CIS is also related to dialogue systems. Many concepts used in developing CIS systems were also explored in the context of dialogue systems. Dialogue systems, or conversational agents, refer to computer systems that are intended to converse with humans through natural language. That being said, dialogue systems are not limited to natural language interactions and can benefit from one or more of text, speech, graphics, haptics, gestures, and other modalities. Dialogue systems are mainly categorized as either chatbots (a.k.a. chit-chat dialogue systems) or task-oriented dialogue systems. The former is designed to mimic human conversations mostly for entertainment, while the latter is developed to help the user accomplish a task, \eg, hotel reservation. Task-oriented dialogues are closer to CIS yet with fundamental differences. 

Designing and developing dialogue systems require a deep understanding of human conversation. Therefore, the dialogue community spent considerable efforts on extracting and modeling conversations. \citet{Jurafsky:2021:SLP} reviewed these properties in detail. For instance, \emph{turn} is perhaps the simplest property -- a dialogue is a sequence of turns, each a single contribution from one speaker. \emph{Dialogue acts} is another important property -- each dialogue utterance is a type of action performed by the speaker. Different modules in real-world dialogue systems are designed because of this property, such as dialogue act classification. \emph{Grounding} is yet another property of dialogues -- acknowledging that dialogue participants understand each other. \emph{Initiative} is the final property we review here. As mentioned in Section~\ref{sec:mixed_init}, it is common in human conversations for initiative to shift back and forth between the participants. For example, in response to a question, a participant can ask for a clarification instead of immediately answering the question. Such interactions are called mixed-initiative. For learning more about dialogue properties and detailed explanations, refer to  \citep[Chapter~24]{Jurafsky:2021:SLP} and \citep[Chapter~3]{Mctear:2016:conversational}.

Dialogue systems have been studied for decades. ELIZA is an early chatbot developed by \citet{Weizenbaum:1966:ELIZA} in the 1960s. It is a rule-based dialogue system designed to simulate a Rogerian psychologist. It involves drawing the patient out by reflecting patient's statements back at them. It selects the best match rule for every utterance (regular expression matching) and uses it for producing the next utterance. PARRY is an updated version of ELIZA developed by \citet{Colby:1971:PARRY} with a clinical psychology focus, used to study schizophrenia. Besides regular expressions, PARRY models fear and anger and uses these variables to generate utterances. It was the first known system to pass the Turing test, meaning that psychologists could not distinguish its outputs from transcripts of interviews with real paranoids~\citep{Colby:1972:PARRYTuringTest}. 

Another successful implementation of dialogue systems in early years was done by the SHRDLU system~\citep{Winograd:1972:SHRDLU}. SHRDLU provides a natural language interface to a virtual space filled with different blocks. Therefore, SHRDLU users could select and move objects in the virtual space. Given the few number of object types, the action space and vocabulary in SHRDLU conversations are highly limited. The AT\&T How May I Help You? (HMIHY) system \citep{Gorin:1997:HMIHY} is also a notable example of dialogue systems developed in the 1990s. HMIHY involved speech recognition, named entity extraction, and intent classification with the goal of call routing. It used a wizard-of-oz approach for data collection and training. It also implemented an active learning algorithm for language understanding. 

Dialogue research was later accelerated by the DARPA Communicator Program. For instance, \citet{Xu:2000} developed a language modeling framework for dialogue systems during the Communicator Program. It was designed to support the creation of speech-enabled interfaces that scale across modalities, from speech-only to interfaces that include graphics, maps, pointing and gesture. Recent chatbot systems often use large-scale language models, such as GPT-3~\citep{Brown:2020:GPT3}, in addition to corpus-based approaches that retrieve information from an external corpus in order to produce more sensible utterances. 

For task-oriented dialogue systems, \citet{Bobrow:1977:GUS} introduced the GUS architecture in the 1970s. GUS is a frame-based architecture for dialogue systems, where a frame is a kind of knowledge structure representing the information and intention that the system can extract from the user utterances. Thus, frames consist of many slots and dialogue systems need to extract and generate the values of these slots based on the conversation. Architectures similar to or inspired by GUS are still used in real dialogue systems. An alternative to such a modular architecture is end-to-end dialogue systems that do not explicitly model slots and are based on text generation models. We refer the reader to \citet[Chapter~4]{gao:2019:FnTIR} for recent advances on task-oriented dialogue systems using neural models.

Evaluating dialogue systems is a challenging and widely explored topic. N-grams matching metrics, such as BLEU~\citep{Papineni:2002:BLEU} and ROUGE~\citep{Lin:2004:ROUGE}, have been used for dialogue system evaluation. Semantic similarity-based metrics, such as BERT-Score~\citep{Zhang:2020:BERTScore}, have also been used. However, research shows that these metrics have several shortcomings~\citep{Liu:2016:EMNLP}. Using human annotators to evaluate the output of the system and/or using implicit or explicit feedback provided by real users are perhaps the most reliable forms of evaluation for dialogue systems. The PARADISE framework~\citep{Walker:1997:PARADISE} for measure overall system success. Developing and evaluating dialogue systems are still active areas of research, we refer the reader to \citet{Finch:2020:Towards} for recent work.

\section{Summary}
In this appendix, we briefly reviewed decades of research related to systems and formal models for interactive information retrieval systems, exploratory search, and dialogue systems. Even though the natural language nature of interaction in CIS makes it more complex and many simplifying assumptions made by prior work on IIR cannot be overlooked in the context of CIS systems, many of the concepts that have been developed for IIR can be directly applied to or extended to CIS tasks. Same argument holds for past research on dialogue systems that has been briefly reviewed in the last subsection. Therefore, instead of re-inventing the wheel for various problems in CIS systems, we urge the reader to have a thorough review of the rich literature on IIR and dialogue research, some of which are pointed out in this appendix.






\backmatter  

\printbibliography

\end{document}